# Spin injection and detection in a Si-based ferromagnetic tunnel junction: A theoretical model based on the band diagram and experimental demonstration

Baisen Yu[1], Shoichi Sato[1, 2], Masaaki Tanaka[1, 2, 3], and Ryosho Nakane[1, 3, 4]

[1]*Deptartment of Electrical Engineering and Information Systems, The University of Tokyo,
7-3-1 Hongo, Bunkyo-ku, Tokyo 113-8656, Japan*
[2]*Center for Spintronics Research Network (CSRN), The University of Tokyo,
7-3-1 Hongo, Bunkyo-ku, Tokyo 113-8656, Japan*
[3]*Institute for Nano Quantum Information Electronics, The University of Tokyo,
4-6-1 Meguro-ku, Tokyo 153-8505, Japan*
[4]*System Design Lab (d.lab), The University of Tokyo,
7-3-1 Hongo, Bunkyo-ku, Tokyo 113-8656, Japan*

**Abstract**
We have experimentally and theoretically investigated the spin injection/detection polarization in a Si-based ferromagnetic tunnel junction with an amorphous MgO layer, and demonstrated that the experimental features of the spin polarization in a wide bias range can be well explained using our theoretical model based on the band diagram of the junction and the direct tunneling mechanism. It is shown that the spin polarization originates from the band diagrams of the ferromagnetic Fe layer and $n^+$-Si channel in the junction, while the spin selectivity of the MgO tunnel barrier is not necessary. Besides, we clarified the mechanism of the reduction in spin polarization when the bias is high and nonlinear properties are prominent, where the widely-used spin injection/detection model proposed by Valet and Fert is not applicable. The dominant mechanism of such reduction is found to be "spin accumulation saturation (SAS)" at the $n^+$-Si interface in contact with the MgO layer as the bias is increased in the spin extraction geometry, which is inevitable in semiconductor-based ferromagnetic tunnel junctions. We performed numerical calculations on a two-terminal spin transport device with a $n^+$-Si channel using the junction properties extracted from the experiments, and revealed that the magnetoresistance (MR) ratio is suppressed mainly by SAS in a higher bias range. Furthermore, we proposed methods for improving the MR ratio in two-terminal spin transport devices. Our experiments and theoretical model provide a deep understanding of the spin injection/detection phenomena in semiconductor-based spin transport devices, toward the realization of high performance under reasonably high bias conditions for practical use.

## I. Introduction

Toward next-generation mobile electronics with low power consumption, ones of the promising devices are Si-based spin transport devices that have nonvolatile and reconfigurable spin-related functions [1-6]. Especially, Si-based spin metal-oxide-semiconductor field-effect transistors (spin MOSFETs) [7-12] are highly attractive, because their spin-functional transistor characteristics can be used for various energy-efficient circuits by being directly integrated with the complementary MOS (CMOS) electronics on a Si platform. The basic spin-functional output characteristics in such devices originate from the spin-valve signal that is a resistance change when the relative magnetization configuration between the ferromagnetic source and drain electrodes is changed from parallel to antiparallel magnetization states, and vice versa. To achieve a high spin-valve signal required for practical use, the most important issue is to clarify the spin transport physics in device elements: Injection of spin-polarized electrons from the ferromagnetic source electrode to the Si channel (spin injection), transport of spin-polarized electrons through the Si channel (spin transport), and the detection of spin polarized electrons by voltage measurements during the extraction of electrons from the Si channel to the ferromagnetic drain electrode (spin detection). So far, we have progressively revealed the physical mechanism of spin transport through Si two-dimensional (2D) channels in spin MOSFETs with source/drain ferromagnetic Fe/MgO/Si tunnel junctions and a SiO$_2$/Si gate stack [6,11,12,15]. The effective spin diffusion length was found to be a few tens µm at room temperature with the aid of the spin drift effect. Thus, Si 2D channels have great potential for practical use.

On the other hand, there have been a lot of studies on spin injection and detection using multi-terminal spin devices with a ferromagnetic metal (FM) / insulator (I) / semiconductor (SC) tunnel junction for the source/drain and an $n^+$-Si channel or a Si-2D channel [13,14,16,17,18,43]. After the measurement of spin signals, such as the spin-valve signal and Hanle signal, the spin polarization $P_\text{S} = \sqrt{P_\text{inj}P_\text{det}}$ was mainly estimated based on the spin diffusion model, where $P_\text{inj}$ is spin injection polarization and $P_\text{det}$ is spin detection polarization. The physical origins of $P_\text{inj}$ and $P_{\text{det}}$ were studied with a theoretical model first developed by Valet and Fert for metallic heterostructures including ferromagnetic metal (FM) and nonmagnetic metal (NM), under the linear response approximation [19-22]. Hereafter, we refer to this theoretical model as the V-F model, which predicts that both $P_{\text{inj}}$ and $P_{\text{det}}$ are determined by the spin asymmetric factor $\gamma$, i.e., $P_{\text{inj}} = P_{\text{det}} = \gamma$. However, the V-F model cannot fully explain the following experimental phenomena: (A1) the nonlinear relationship between the current bias and spin injection polarization [14,16, 23,26], (A2) the nonlinear relationship between the current bias and spin detection polarization [14,16, 23,26], and (A3) disappearance of spin signals in the spin injection geometry [16,23]. This is probably because the V-F model does not take into account the following three realistic properties: (B1) nonlinear $I-V$ characteristics arising from the electron tunneling process through the source/drain ferromagnetic tunnel junctions, (B2) semiconductor properties of the Si channel, such as band bending near the I/Si interface, and (B3) the band offset between FM, I, and Si in the junction.

There has been an attempt to study the phenomena (A1)−(A3) using a spin device with four ferromagnetic Fe/MgO/$n^+$-Si junctions and a $n^+$-Si channel [16]: They developed a nonlocal measurement method that allows the voltage measurement of spin valve signals by one Fe/MgO/$n^+$-Si junction biased with a constant voltage while another junction supplies a spin current under a constant current [16,24,25]. The purpose of such measurement method is to study $P_{\text{inj}}$ and $P_{\text{det}}$ under a controlled junction voltage. For the analysis of experimental spin signals, a theoretical model based on the V-F model was derived by combining with the property (B1). Indeed, their model prediction reproduced the phenomenon (A2) in a small junction voltage range, but it was neither able to reproduce the phenomenon (A2) in a higher junction voltage range nor the other phenomena (A1) and (A3).

The purpose of this study is to clarify the detailed physical mechanism on the spin injection and detection through the analyses of experimental spin signals with our original theoretical model that takes into account more realistic properties of Si-based magnetic tunnel junctions. First, our theoretical model without $\gamma$ is derived based on the band diagram of a FM/I/Si tunnel junction, in which we take into account the properties (B1)−(B3). Next, spin-valve signals are measured with four-terminal devices with a Fe/Mg/MgO/SiO$_x$/$n^+$-Si junctions [14] and a $n^+$-Si channel, using the same nonlocal method developed in refs. [16,24,25]. Then, $P_{\text{inj}}$ and $P_{\text{det}}$ in a wide range of the junction bias voltage are experimentally and theoretically investigated. Both $P_{\text{inj}}$ and $P_{\text{det}}$ are found to highly depend on the junction bias voltage, particularly in higher junction voltage ranges where nonlinear electrical characteristics appear, because of the electron tunneling mechanism. In addition, it is found from our theoretical model that excessive spin accumulation in $n^+$-Si leads to reduction in spin current. The feature is analogous to the conductivity mismatch problem in the V-F model, but its origin is unique for semiconductor-based ferromagnetic tunnel junctions. Finally, we discuss how we can enhance the spin-valve signal in terms of the device design.

## II. Theoretical model

We present our theoretical model using the basic concept of the V-F model that is widely used in the research field of semiconductor-based spin transport devices. The motivation for our theoretical model construction is to incorporate the realistic properties of FM/I/SC junctions so that their experimental spin signals are reproduced under a wide range of bias conditions.

First, we briefly review the V-F model to confirm the definitions of $P_{\text{inj}}$ and $P_{\text{det}}$, and point out the problems that are due to the lack of the properties of spin injector/detector tunnel junctions. Next, we derive our theoretical model with taking into account the electronic band diagram and a general tunnel current model.

## II-A. Basic model without band-diagram-based tunneling current model

The V-F model was established as a theoretical model for electron spin transport in the diffusive transport regime when electrons travel through ferromagnetic layered structures, including FM/NM, FM/I/NM, FM/SC, and FM/I/SC [19-22]. There are two different spin injecting processes depending on the electron flow direction determined by a bias applied to the structure: Spin injection means electron flowing from FM to SC, and spin extraction means electron flowing from SC to FM. It is worth noting that the bias applied to the structure is infinitesimally small under the linear response approximation.

Consider a one-dimensional FM/I/SC structure with a thickness $t$ of the I layer and infinite thicknesses of both the FM and SC layers, as shown in Fig. 1(a), where the origin of the $z$ axis is located in the center of the I layer, $J_+(z)$ and $J_-(z)$ are the tunneling currents for the up-spin and down-spin electrons, and they satisfy continuous boundary conditions $J_\pm(t/2) = J_\pm(-t/2)$, i.e., spin relaxation events inside the I layer are neglected. Hereafter, $J_+(t/2)$ and $J_-(t/2)$ are denoted by $J_+$ and $J_-$, respectively, because we will focus on the spin injection polarization $P_{\text{inj}}$ and the extraction polarization $P_{\text{ext}}$, which are determined by the tunneling currents at the I / SC interface ($z = t/2$) and FM/I interface ($z$ $= -t/2$), respectively. The spin current $J_S$ and electron current $J$ at $z = t/2$ or $-t/2$ are expressed by

$$J_S = J_+ - J_-, \quad (1)$$
$$J = J_+ + J_-. \quad (2)$$

Under the spin injection geometry ($J > 0$), $P_{inj}$ at $z = t/2$ is given by

$$P_{inj} = \frac{J_S}{J} = \frac{J_+ - J_-}{J_+ + J_-}. \quad (3)$$

Under the spin extraction geometry ($J < 0$), $P_{ext}$ at $z = -t/2$ has the same form as Eq. (3). Hence, the same character $P_{inj}$ will be used for both $P_{inj}$ and $P_{ext}$.

Figure 1(b) shows the band diagram of a FM/I/SC structure in the spin injection geometry, where $\bar{\mu}^{FM}$ is the Fermi level of FM, $V$ is the voltage drop in the I layer, the up-spin and down-spin electrochemical potentials $\mu_+$ and $\mu_-$ are defined in SC, respectively, and the mean $\bar{\mu} = (\mu_+ + \mu_-)/2$ is also used. Here, $\Delta\mu = \mu_+ - \mu_-$ is called "spin accumulation" and it is generated by $J_S$ into SC having the spin resistance $r_{sr}$ [19-22]:

$$\Delta\mu = 2J_S r_{sr}, \quad (4)$$

where $r_{sr}$ is the product of the resistivity $\rho$ and spin diffusion length $\lambda_{sf}$. Figure 1(b) neglects $\Delta\mu$ in FM since FM has a much smaller $r_{sr}$ by several orders magnitude than SC [20-22]. The V-F model uses a spin-dependent boundary conditions $r_\pm = 2r_b(1 \mp \gamma)$ at the interfaces, where $r_+$ and $r_-$ represent the resistance-area products for up-spin and down-spin electron passes, respectively, $r_b$ is a constant resistance-area product of the I layer, and $\gamma$ is the constant spin selectivity of the I layer (spin asymmetric factor). Using the Ohm's law for the $J$–$V$ relationship, $J_\pm$ are expressed by:

$$J_\pm(V, \Delta\mu) = \frac{V_\pm}{r_\pm} = \frac{(V \mp \Delta\mu/2)(1 \pm \gamma)}{2r_b(1 - \gamma^2)}, \quad (5)$$

where $V_+$ and $V_-$ represent the voltage drops of the I layer along the up-spin and down-spin electron passes, respectively, and $V$ is the mean voltage drop over the I layer. The combination of Eqs. (3)-(5) leads to

$$P_{inj} = \frac{\gamma r_b}{r_b + r_{sr}} = \frac{\gamma}{1 + r_{sr}/r_b}. \quad (6)$$

It is important to note that $P_{inj}$ is independent of $V$. On the other hand, $P_{inj}$ depends on the relative magnitude between $r_b$ and $r_{sr}$: the maximum value $P_{inj} = \gamma$ under the condition $r_b \gg r_{sr}$, whereas $P_{inj} \sim 0$ under the condition $r_b \ll r_{sr}$. This is well-known as "the conductivity matching condition for $P_{inj}$".

Spin detection polarization $P_{det}$ is derived from a different consideration. Figure 1(c) shows a 4-terminal (4T) spin transport setup, in which $J_S$ generated at the L junction travels through the spin diffusion and then it is detected at the R junction having zero charge current [11,12,13,16,32,35,36]. Figures 1(d) and (e) show schematic band diagrams of the R junction when the relative magnetization configurations between the L and R electrodes are the parallel (P) and antiparallel (AP) magnetization states, respectively. When $\Delta\mu = 0$, $\bar{\mu}^{FM}$ is equal to $\bar{\mu}^{SC}$, as shown by the dashed line in Figs. 1(d) and (e). On the other hand, when $\Delta\mu > 0$ in Fig. 1(d), $\bar{\mu}^{FM}$ has $V_{det}$ against $\bar{\mu}^{SC}$ due to the spin asymmetry of the tunnel junction. In a similar manner, when $\Delta\mu < 0$ in Fig. 1(e), $\bar{\mu}^{FM}$ has $-V_{det}$ against $\bar{\mu}^{SC}$. The conversion efficiency from $\Delta\mu$ to $V_{det}$ is defined by $P_{det}$:

$$P_{det} = \frac{2\,\Delta V_{det}}{\Delta\mu}. \quad (7)$$

Since no current flows through the R junction in both P and AP magnetization states, $V_{det}$ is determined by solving the following equation:

$$J_R^P(V + V_{det}, \Delta\mu) = J_R^{AP}(V - V_{det}, -\Delta\mu) = 0, \quad (8)$$

where $J_R^P$ and $J_R^{AP}$ are the electron current densities in P and AP magnetization states, respectively. The combination of Eqs. (3),(4),(5), (7), and (8) leads to

$$P_{det} = \gamma. \quad (9)$$

Thus, $P_{det}$ is also independent of $V$ in the V-F model.

**II-B. Issues in the application of the basic V-F model**

There are some assumptions in the V-F model, which do not reflect the realistic properties of experimental tunnel junctions. The main problems are summarized as follows.

(I) Constant resistance-area product $r_b$

The insertion I layer between FM and SC, which corresponds to the insulating tunnel barrier, has a constant $r_b$ that is regarded as a conductive material with a high resistivity. This is because the V-F model uses the linear response approximation. However, the electron conduction in FM/I/SC structures contains the tunneling process through the I layer, which results in nonlinear *I–V* characteristics, i.e., the resistance is changed when the bias is changed in a wide range. The most experiments reported so far used biases higher than a few hundreds mV that are certainly beyond the linear *I–V* characteristics [13,14,16,23,26].

(II) Constant spin asymmetric factor $\gamma$

The constant $\gamma$ means bias-*independent* spin selectivity in the I layer. A constant $\gamma$ value is assumed without solid evidence in the analysis of experimental signals. However, the experiments reported so far have not shown any clear evidence that the I layer itself has such constant spin selectivity [14,23,26]. Rather, it was shown that the experimental $P_{\rm inj}$ decreases with increasing bias applied to the junction, namely, the spin selectivity is *dependent* on the bias [14,16,26-28]. Hence, the V-F model does not explain the features of spin signals experimentally obtained in tunnel junctions in a wide range of bias conditions.

(III) Boundary conditions of the electrochemical potentials

In the V-F model, the continuity conditions of the electrochemical potentials for the up-spin and down-spin electrons are defined by $\gamma$ at the FM/I and I/SC interfaces. However, such conditions are not applicable under high *V* bias conditions.

(IV) Band structure of SC and band bending at the SC/I interface

In the V-F model, the material difference between metals and SC is expressed only by resistivity since the bias is infinitesimally small. The Fermi energy $E_{\rm F}$ in $n^+$-SC is typically smaller than 100 meV [29], which is much smaller than $E_{\rm F}$ in metals, and this is important in the electron transport through a FM/I/SC junction under high bias ranges. Besides, the band bending occurs in $n^+$-SC. However, the V-F model does not take into account these material properties.

**II-C. Our theoretical model**

We construct a theoretical model for a Fe/MgO/$n^+$-Si tunnel junction whose structure is almost the same as that used in our experiments in Sec. III-A. As shown later, the experimental junction structure is Fe/MgO/SiO$_x$(0.2 nm)/$n^+$-Si, but there are unknown physical parameters, such as the band offset between Fe, $n^+$-Si, and the conduction band minimum of the series MgO/SiO$_x$ layers. Here, two assumptions are introduced to simplify the model junction structure. First, the MgO/SiO$_x$ tunnel barrier is regarded as a single MgO layer. Second, the single MgO layer has the same effective barrier height ($\Phi_0 = 1.15$ eV) with respect to the Fermi levels of Fe and $n^+$-Si at zero bias, which means that the flat band voltage of the junction is zero. The validity of these two assumptions will be verified using experimental results in Sec. IV-A.

In the following, we describe our approach to solve the problems (I)–(IV) described in Sec. II-B, focusing on the differences from other models [16,26-28].

(I) Tunneling process and properties of the electrode and channel

The previous model by another group [16] adopted a direct tunneling process using the WKB approximation with "specular transmission", in which all the wavevectors and energies of electrons are conserved through the tunneling [30]. Our model adopts a direct tunneling process using the WKB approximation with the diffusive tunneling process, in which only the energies of electrons are conserved through the tunneling.

(II) Spin selective conduction through the junction

Our model does not use $\gamma$. When an amorphous non-magnetic I layer is used for the tunnel barrier, it is expected that the spin selective conduction through the junction simply originates from the tunneling conduction and the electronic band structures of Fe, I, and Si.

(III) Boundary conditions of the electrochemical potentials

Under the two assumptions described above ((I) and (II)), the electrochemical potentials for up-spin and down-spin electrons are discontinuous at both sides of the I layer when the bias voltage is reasonably high. Immediately after the direct tunneling into final electron states at the interface, electrons at higher energies lose their energies to form the thermally equilibrium electron distribution that is defined by the Fermi level, the Fermi-Dirac distribution function, and temperature. During such energy relaxation, spin flip may occur, but we assume spin-conservative energy relaxation. Thus, the split of the electrochemical potentials is determined by the spin population at the final electron states in the tunneling process.

(IV) Semiconductor properties of $n^+$-Si

Our model takes into account $E_{\rm F}$ which is much smaller in $n^+$-Si than in metals, and the $n^+$-Si band bending both in the depletion mode (the spin injection geometry) and in the accumulation mode (the spin extraction geometry) [34].

Figures 2(a) and (b) show one-dimensional electron energy band diagrams of a Fe/MgO/$n^+$-Si junction in the spin injection and extraction geometries, respectively, where $E$ is the electron energy level, $\bar{\mu}^{Fe}$ and $\bar{\mu}^{Si}$ are the Fermi energies of Fe and Si, respectively, $E_C$ is the conduction band bottom of Si, and $E_2$ is the difference between $\bar{\mu}^{Si}$ and $E_C$. The upper line of the MgO barrier layer is the conduction band bottom and the $z$ axis is defined with the origin in the center of the MgO layer. The bias voltage $V$ has the positive and negative polarities in the spin injection and extraction geometries, respectively. Hereafter, $\bar{\mu}^{Fe}$ is defined as the zero potential. The flat band voltage is set at zero for simplification. In both geometries, $V_{MgO}$ is the voltage drop of the MgO layer and the band bending of $E_C$ is characterized by $\Phi_{Si}$ that is the energy difference between the surface and bulk region with the flat $E_C$, and thus $V = V_{MgO} + \Phi_{Si}$. The effective barrier height $\Phi_{eff}(E)$ for electrons at $E$ is defined by the energy difference between $E$ and the energy in the center ($z = 0$) of the MgO conduction band bottom, which is the same method in ref. [31]. For examples, $\Phi_{eff}(\bar{\mu}^{Fe})$ and $\Phi_{eff}(\bar{\mu}^{Si})$ are illustrated in Figs. 2(a) and (b), respectively.

Figures 2(c) and (d) show schematic illustrations of electron distributions in Figs. 2(a) and (b), respectively, where red and blue colors in Fe and Si represent the filled states of up-spin and down-spin electrons, respectively. Our model uses only a parabolic (s-like) band structure with a spin-splitting energy $\Delta$ of 1.6 eV in Fe [32], the energy range of electrons $E_1 = 1.75$ eV [32], and 24% spin polarization at $\bar{\mu}^{Fe}$. The assumption is reasonable, because the electrons in highly s-hybridized bands have a quite small effective mass and dominate the tunneling electron current [32,33]. The MgO tunnel barrier with a thickness $t_{MgO} = 1$ nm has a barrier height $\Phi_0 = 1.15$ eV from $\bar{\mu}^{Fe}$ and $\bar{\mu}^{Si}$ at zero bias $V = 0$ V (the flat band condition) [32]. In the $n^+$-Si, a parabolic conduction band structure with $E_2 = 100$ meV is assumed [29], and the split of the up-spin (+) and down-spin (−) electrochemical potentials corresponds to spin accumulation $\Delta\mu$ ($= \mu_+^{Si} - \mu_-^{Si}$): $\mu_+^{Si} > \mu_-^{Si}$ in Fig. 2(c) and $\mu_+^{Si} < \mu_-^{Si}$ in Fig. 2(d). The effective barrier heights for the up-spin (+) and down-spin (−) electrons are defined by $\Phi_+(E)$ and $\Phi_-(E)$, respectively, and $\Phi_+(\mu_+^{Si})$ and $\Phi_-(\mu_-^{Si})$ are illustrated in Fig. 2(d). On the other hand, under any biases, it is assumed that the electrochemical potential in the Fe remains unchanged at the equilibrium state, i.e., it does not split into up-spin and down-spin ones, owing to its much smaller spin resistance than that in Si [13,19-22]. Hence, only $\Phi_{eff}(E)$ is defined in Fe, that is, $\Phi_{eff}(E) = \Phi_+(E) = \Phi_-(E)$.

The direct tunneling currents for $J_+(V)$ and $J_-(V)$ are numerically calculated using the following formula:

$$J_\pm(V, \Delta\mu(V)) = \alpha \int_{-\infty}^{+\infty} g_{Fe}^\pm(E)\, g_{Si}(E-V)\, T(E,V) \left[f_{Fe}(E) - f_{Si}\left(E - V \pm \frac{\Delta\mu(V)}{2}\right)\right] dE$$

$$= A \int_{-\infty}^{+\infty} D_{Fe}^\pm(E)\, D_{Si}(E-V)\, T(E,V) \left[f_{Fe}(E) - f_{Si}\left(E - V \pm \frac{\Delta\mu(V)}{2}\right)\right] dE, \quad (10)$$

$$g_{Fe}^\pm(E) = \frac{1}{2\pi^2}\left(\frac{2m_{Fe}}{\hbar}\right)^{3/2} D_{Fe}^\pm(E),$$

$$g_{Si}(E) = \frac{1}{2\pi^2}\left(\frac{2m_{Si}}{\hbar}\right)^{3/2} D_{Si}(E),$$

where $V$ is the applied voltage bias to the junction, $\Delta\mu(V)$ is the spin accumulation at $V$, $D_{Si}(E-V) = \sqrt{E + E_2 - V}$ is the density of states (DOS) in $n^+$-Si, and $D_{Fe}^\pm = \sqrt{E + E_1 \pm \Delta/2}$ is DOSs for up-spin / down-spin (+/−) electrons in Fe, $T(E) = \exp\left(-2t_{MgO}\sqrt{2m_t(\Phi_{eff} - E)/\hbar^2}\right)$ is the tunneling probability based on a WKB model [31] with a tunneling effective mass $m_t$, $\Phi_{eff} = \Phi_0 - V_{MgO}/2$ is the effective barrier height [31], $f_{Fe}(E)$ and $f_{Si}(E-V\pm\Delta\mu/2)$ are the Fermi-Dirac distribution functions at 4 K for Fe and $n^+$-Si, respectively, and $\pm\Delta\mu/2$ stand for the energy split due to the spin accumulation in the up-spin / down-spin (+/−) bands. The constant $A$ has the unit of A/m$^2$/(eV)$^2$ and it includes the effective masses $m_{Fe}$ and $m_{Si}$ of electrons in Fe and Si, respectively. By fitting Eq. (10) to an experimental $J$–$V$ curve [see Sec. S1 in S.M. [44]], $A$ was estimated to be $7 \times 10^{11}$ A/m$^2$/(eV)$^2$. In this paper, $J$ ($= J_+ + J_-$) always represents the electron current. It should be emphasized that Eq. (10) is our original $J$–$V$ relationship for the FM/I/SC structure: Keys to spin-selective electron conduction are the contributions of both filled and empty electron states in the spin-dependent Fe band structure and different effective barrier heights for up-spin and down-spin electrons in Si when $\Delta\mu$ is significantly large. Our model is clearly different from the V-F model (Eq. (5)) and the models proposed by other groups [16,26-28]. An important feature is that Eq. (10) can be applied to FM/I/SC junctions having different $r_{sr}$ ($\geq 0$) values (different $\lambda_{sf}$ values) while the band diagram of the junction and electronic material properties are unchanged. In other words, we can study the spin transport properties through a FM/I/SC junction while only changing the spin transport property in SC. Such feature originates from the fact that continuous $\Delta\mu$ ($= \mu_+ - \mu_-$) at the I/SC interface is not required for the boundary condition, unlike the V-F model. Figures 3(a) and (b) are examples, which schematically show the band diagrams in the spin injection geometry ($V > 0$) with large and small $r_{sr}$ values, respectively, where $V$ is identical with each other. As can be seen, SC with smaller $r_{sr}$ has less $\Delta\mu$ than SC with larger $r_{sr}$. Comparison of the two cases $r_{sr} = 0$ and $r_{sr} \neq 0$ enable us to study the influence of $\Delta\mu$ on $P_{inj}$ and $P_{det}$, as will be described in Sec. IV.

One of the remarkable features is that Eq. (10) leads to nonlinear $\Delta\mu-V$ and $J_S-\Delta\mu$ relationships. To explain the effects of such nonlinearity, we schematically show $\Delta\mu-V$ and $J_S-\Delta\mu$ in the spin extraction geometry ($V < 0$) in Figs. 4(a) and (b), respectively, where red and blue curves are the relationships derived from our model and the V-F model, respectively. The linear $\Delta\mu-V$ and $J_S-\Delta\mu$ relationships in the V-F model come from the linear $J-V$ relation and Eq. (4), respectively. On the other hand, in our model, the nonlinear $\Delta\mu-V$ and $J_S-\Delta\mu$ relationships have the following features. As $V$ is increased, $\Delta\mu$ increases at first and saturates at a specific value $\Delta\mu_0$. In Fig. 4(b), according to such change in $\Delta\mu$, $J_S$ increases and then drastically decreases toward zero due to the asymptotic convergence of $\Delta\mu$ to $\Delta\mu_0$. Unlike the V-F model, $\Delta\mu$ in our model does not exhibit infinite growth but saturates toward a specific value as $V$ is increased. Hereafter, we refer to this unique nonlinear $\Delta\mu-V$ relationship as spin accumulation saturation (SAS). A brief explanation of SAS is as follows: At a high bias in the spin extraction geometry, $\Delta\mu$ in $n^+$-Si becomes large and the barrier height ($\Phi_-$) for the minority spin ($-$) electrons becomes lower than the barrier height ($\Phi_+$) for the majority spin ($+$) electrons, and thus the tunneling current $J_-$ increases and becomes comparable to $J_+$, resulting in a steep decrease in the spin current $J_S = J_+ - J_-$ (as will be shown later in Fig. 12). This is the main cause of SAS (Fig. 4). Owing to the steep decrease of $J_S$ in the $J_S-\Delta\mu$ curve in Fig. 4(b), $P_{inj}$ and $P_{det}$ decrease significantly, as will be shown with explaining physical origins in Sec. IV. In contrast, in the spin injection geometry ($V > 0$), SAS does not appear since the barrier height $\Phi_0 = 1.15$ eV is much larger than the magnitude of $\Delta\mu$ in $n^+$-Si (that is smaller than a few tens meV, as will be shown later in Fig. 9(b)) and $\Delta\mu$ does not make a significant difference between $\Phi_+$ and $\Phi_-$. Thus, $\Delta\mu$ does not saturate as $V$ is increased in the spin injection geometry.

## III Experimental methods and results
### III-A Device structure, measurement setup, and analysis methods

Figures 5(a) and (b) show the schematic cross-sectional and top views, respectively, of a device on a (001)-oriented silicon-on-insulator (SOI) substrate with a 200-nm-thick buried oxide layer. The L and R electrodes have ferromagnetic tunnel junctions (from the top to bottom) Mg(1nm) / Fe(4 nm) / Mg(1 nm) / amorphous-MgO(1 nm) / SiO$_x$(0.2 nm) / $n^+$-Si(001) and the topmost capping layers having Al, Pt, and Ta. Two reference L$_{REF}$ and R$_{REF}$ electrodes are located at the outside of the L and R, respectively. An $n^+$-Si channel has a thickness $t_{SOI} = 25$ nm and a phosphorus doping concentration $N_D \sim 1\times 10^{20}$ cm$^{-3}$, and the Cartesian coordinates are defined. The edge-rounded rectangular L and R electrodes have short sides $L_L = 0.7$ μm and $L_R = 2.0$ μm along the $y$ direction, respectively, and they have a long side 176 μm along the $x$ direction. The $n^+$-Si channel has a rectangle with a width $W_{ch} = 180$ μm along the $x$ direction and the channel length defined by the gap between the L and R electrodes is $L_{ch} = 0.7$ μm. The distance between the left sides of the L and L$_{REF}$ (the right sides of the R and R$_{REF}$) is 100 μm that is far longer than the spin diffusion length $\lambda_{sf} = 1$ μm at 4 K in $n^+$-Si [12,13,37-40]. The detailed fabrication processes of the ferromagnetic junction and device structure are described in refs. [12-14].

Figures 6(a) and (b) show experimental measurement setups named Setup-A and Setup-B, respectively, which are basically the same as those in refs. [16,24,25]. Spin-valve signals are measured at 4 K by detecting voltages $V_R$ and $V_L$ using Setup-A and Setup-B, respectively, while constant positive charge currents $I_L$ and $I_R$ are applied to the L−L$_{REF}$ and R−R$_{REF}$ circuits, respectively, and an external magnetic field $H$ along the $x$ direction (along the easy magnetization axis of the Fe electrodes) is applied and swept from +200 to −200 Oe, and vice versa. The positive and negative polarities of $I_L$ ($I_R$) correspond to the spin injection and detection geometries at the L junction (the R junction), respectively. In this study, we set $I_R = 0$ mA and −1.5 mA in Setup-A, $I_R = 20$ mA in Setup-B, respectively, and $I_L$ is varied from −20 to 20 mA.

The procedure of the analysis is as follows. First, a spin-valve signal is measured with specific $I_L$ and $I_R$ values and its maximum change $V^{SR}(I_L, I_R)$ in Setup-A and $V^{SL}(I_L, I_R)$ in Setup-B are estimated (Fig. 7(a) and (c)). After that, the product $P_{inj}P_{det}$ is estimated using the following equations:

In Setup-A

$$V^{SR}(I_L, I_R = 0 \text{ or} - 1.5 \text{ mA}) = P_{det}(I_R = 0 \text{ or} - 1.5 \text{ mA}) P_{inj}(I_L) I_L R_0 \exp\left(-\frac{L_{ch}}{\lambda_{sf}}\right), \tag{11}$$

In Setup-B

$$V^{SL}(I_L, I_R = 20 \text{ mA}) = P_{\text{det}}(I_L)\, P_{\text{inj}}(I_R = 20 \text{ mA}) I_R R_0 \exp\left(-\frac{L_{ch}}{\lambda_{\text{sf}}}\right). \tag{12}$$

In Eqs. (11) and (12), the effective channel spin resistance $R_0$ is given by [13,41]

$$R_0 = \frac{\rho}{W_{ch}}\frac{\lambda_{\text{sf}}}{t_{\text{SOI}}}\left\{\frac{\lambda_{\text{sf}}}{L_L}\left(1-\exp\left(-\frac{L_R}{\lambda_{sf}}\right)\right)\frac{\lambda_{\text{sf}}}{L_R}\left(1-\exp\left(-\frac{L_L}{\lambda_{\text{sf}}}\right)\right)\right\}, \tag{13}$$

where $\rho = 1\times10^{-5}$ $\Omega\cdot$m is the resistivity of the $n^+$-Si channel at 4 K [13,42,43]. In both setups, $P_{\text{inj}}P_{\text{det}}$ can be estimated from experiments. The advantage of combination use of both measurement setups is that Setup-A and -B can reveal different features as follows:

Setup-A:
The change in $P_{\text{inj}}P_{\text{det}}$ reflects the bias-dependent features of $P_{\text{inj}}(I_L)$, since $I_R$ is constant (0 or −1.5 mA) and $P_{\text{det}}$ is unchanged.

Setup-B:
The change in $P_{\text{inj}}P_{\text{det}}$ reflects the bias-dependent features of $P_{\text{det}}(I_L)$, since $I_R$ is constant (20 mA) and $P_{\text{inj}}$ is unchanged.

Next, the L junction voltage drop $V_{\text{JL}}$ for a specific $I_L$ value is estimated using the $V_{\text{JL}}-I_L$ characteristics measured in Setup-C in Fig. 6(c). Then, $P_{\text{inj}}P_{\text{det}}$ at a specific $I_L$ value is plotted as a function of $V_{\text{JL}}$ both for Setup-A and Setup-B: These $P_{\text{inj}}P_{\text{det}}$ $-V_{\text{JL}}$ plots will be analyzed based on the band diagrams under a specific junction voltage $V_{\text{JL}}$ in Sec. IV. Finally, the shapes of the $P_{\text{inj}}P_{\text{det}}$ $-V_{\text{JL}}$ plots in Setup-A and Setup-B are compared with $P_{\text{det}}-V_{\text{JL}}$ and $P_{\text{inj}}-V_{\text{JL}}$ plots predicted by our theoretical model with Eq. (10), respectively. It is noteworthy that we focus on the shapes of the plots because the absolute values of $P_{\text{inj}}$ and $P_{\text{det}}$ cannot be separately estimated from the experiments alone unlike theoretical predictions, therefore, comparison of the experimental results with the theoretical model is necessary.

**III-B. Experimental results**

Figure 7(a) shows $\Delta V_R$ as a function of magnetic field $H$ measured with various $I_L$ values (= −20 - 20 mA) and $I_R = 0$ mA in Setup-A, where blue, green, orange, and red curves represent the signals obtained with $I_L = -20, -10, 10,$ and 20 mA, respectively. The signal shapes clearly indicate the spin-valve signals, reflecting the change of the parallel and antiparallel magnetization states in the Fe electrodes. These data confirm the spin transport through the Si channel between the L and R electrodes. The estimation method of $V^{SR}(I_L, I_R)$ is as follows: First, the signal base is determined by subtracting a linear line fitting in higher $H$, and then, the maximum signal changes in the positive and negative $H$ ranges are averaged. Figure 7(b) summarizes $V^{SR}(I_L, I_R)$ estimated in Setup-A plotted as a function of $I_L$, where red and green open dots are the values estimated for $I_R = 0$ mA and $I_R = -1.5$ mA, respectively. As $I_L$ is increased from zero in the negative bias (the spin extraction geometry), $|V^{SR}(I_L, I_R)|$ nonlinearly increases while the rate of change gradually decreases. As $I_L$ is increased from zero in the positive bias (the spin injection geometry), the gradual change in $|V^{SR}(I_L, I_R)|$ along with $I_L$ is similar to that in the negative $I_L$ range. However, the $|V^{SR}(I_L, I_R)|$ value in the positive $I_L$ range is larger than that in the negative $I_L$ range when $|I_L|$ is the same value. The asymmetric $V^{SR}(I_L, I_R) -I_L$ shapes in Fig. 7(b) possibly reflect the change in $P_{\text{inj}}(I_L)$ with $I_L$, since the $V_{\text{JL}}-I_L$ curve (see Sec. S1 in S.M. [44]) has a nearly symmetric shape with respect to the origin. In Fig. 7(b), $V^{SR}(I_L, I_R)$ is not plotted in the $I_L$ range from −5 to 5 mA due to significantly lower signal-to-noise ratios.

Figure 7(c) shows $\Delta V_L-H$ characteristics measured with various $I_L$ values (= −0 - 20 mA) and $I_R = 20$ mA in Setup-B, where blue, green, orange, and red curves represent the signals obtained with $I_L = 5, -1, -5,$ and −20 mA, respectively. Figure 7(d) summarizes $V^{SL}(I_L, I_R)$ plotted as a function of $I_L$, in which $V^{SL}(I_L, I_R)$ values were estimated from these spin-valve signals using the same procedure as that in Setup-A cases. The $V^{SL}(I_L, I_R)$ feature was found to be complicated compared with $V^{SR}(I_L, I_R)$ in Fig. 7(b). As $|I_L|$ is increased from zero in the negative bias (the spin extraction geometry), $V^{SL}(I_L, I_R)$ steeply increases, then it shows a maximum at around −1 mA, and finally it decreases. On the other hand, as $I_L$ is increased from zero in the positive bias (the spin injection geometry), $V^{SL}(I_L, I_R)$ steeply decreases and it shows zero above 2 mA.

Figures 8(a) and (b) show $P_{\text{inj}}P_{\text{det}}-V_{\text{JL}}$ characteristics estimated in Setup-A and Setup-B, respectively, where (a) orange and green open dots denote the values with $I_R = 0$ and −1.5 mA, respectively, and (b) blue open dots denote the values with $I_R = 20$ mA. The shapes of the curves in Fig. 8(a) are different from those in Fig. 7(b), whereas the shape of the curve in Fig. 8(b) is similar to that in Fig. 7(d). This is because $V^{SR}(I_L, I_R)$ is proportional to $P_{\text{inj}}(I_L)$ in Setup-A, while $V^{SL}(I_L, I_R)$ is proportional to $P_{\text{det}}(I_L)$ in Setup-B (see Eqs. (11) and (12)). The gradual decrease of the slope in higher $|I_\text{L}|$ range in Fig. 7(b) reflects the gradual decrease of $P_\text{inj}$ in the higher $|V_\text{JL}|$ range in Fig. 8(a), because $P_\text{det}(I_\text{R})$ can be regarded as a constant value. Hence, it indicates that $P_\text{inj}$ decreases with increasing $|V_\text{JL}|$. On the other hand, since $V^\text{SL}(I_\text{L}, I_\text{R})$ is proportional to $P_\text{det}(I_\text{L})$, the $V^\text{SL}(I_\text{L}, I_\text{R})$−$I_\text{L}$ curve in Fig. 7(d) and the $P_\text{inj}P_\text{det}$−$I_\text{L}$ curve Fig. 8(b) are similar. Detailed analyses on the physical origins will be discussed in Sec. V.

**IV Analyses of the experimental results with our theoretical model**

Figures 8(a) and (b) fundamentally reflect $P_\text{inj}$ and $P_\text{det}$, respectively, based on the experimental setups, however, $P_\text{inj}$ and $P_\text{det}$ cannot be decomposed only by such experiments as pointed out previously. The purpose of this section is to separately estimate $P_\text{inj}$ and $P_\text{det}$ using our theoretical model described in Sec. II-C. First, the validity of the model junction structure in Figs. 2(a-d) is verified by fitting Eq. (10) to an experimental $V_\text{JL}$−$I_\text{L}$ curve. Next, the bias dependencies of $P_\text{inj}$ and $P_\text{det}$ are numerically calculated for our tunnel junction structure. Then, we show that our model can explain the experimental results in Figs. 8(a) and (b). From these results, we present how $P_\text{inj}$ and $P_\text{det}$ are determined, focusing on the role of the tunnel barrier, and show how $P_\text{inj}$ and $P_\text{det}$ are affected by SAS that is schematically illustrated in Figs. 4(a) and (b).

In the V-F model, the suppression of $P_\text{inj}$ and $P_\text{det}$ called "the conductivity mismatch problem" occurs in FM/SC junctions: Since the interface condition Eq. (4) is satisfied, increasing $r_\text{sr}$ results in the decrease of $J_\text{S}$ for the same $\Delta\mu$ value, which is associated with the bidirectional spin diffusion arising from the condition that $\mu_+$ and $\mu_-$ are *continuous* across the FM/SC interface. The same problem also occurs at both FM/I and I/SC interfaces in FM/I/SC junctions because the I layer serves as a low-conductance metal under the linear response approximation. To enhance $J_\text{S}$ by excluding this phenomenon, $\gamma > 0$ for the I layer is the final solution, because it can selectively filter out minority spins. On the other hand, in our theoretical model for FM/I/SC junctions, $\mu_+$ and $\mu_-$ are *discontinuous* at the I/SC or FM/I interface when the junction bias voltage is reasonably high because it is in the nonlinear response regime, as shown in Figs. 3(a) and (b). Hence, the system under this condition can inherently exclude the conductivity mismatch problem caused by the bidirectional spin diffusion in the V-F model even when $\gamma$ for the I layer is absent. In this section, however, we find that suppression of $P_\text{inj}$ and $P_\text{det}$, which is referred to as SAS, occurs due to another mechanism when $\Delta\mu$ is quite large.

**IV-A Validation of our theoretical model**

In Sec. II-C, we constructed the theoretical model based on the two assumptions: (i) A simplified MgO single layer replacing the series MgO/SiO$_x$ layers, and (ii) the same effective barrier height ($\Phi_0 = 1.15$ eV) with respect to the Fermi levels of Fe and $n^+$-Si at zero bias. Here, the validity of our model is verified as follows. First, the experimental $V_\text{JL}$−$I_\text{L}$ characteristics are well reproduced by fitting our numerical calculations using Eq. (10) (see Sec. S1 in S.M. [44]), which reveals that the assumptions and parameters used in our model are applicable. Second, the barrier height $\Phi_0 = 1.15$ eV used in the theoretical model is consistent with the realistic barrier height of our present junction structure Fe/MgO/SiO$_x$(0.2 nm)/$n^+$-Si, which is experimentally estimated to be ~1.1 eV and is higher than a single MgO barrier (~0.3 eV) probably due to the compensation of MgO oxygen deficiencies during the SiO$_x$ layer formation by RF oxygen plasma [14]. Therefore, our model constructed in Sec. II-C is useful and effective to analyze the experimental features of the spin polarization. In the following sections, we focus on the similarity in shape between experimental and theoretical $P_\text{inj}$−$V_\text{JL}$ (or $P_\text{det}$−$V_\text{JL}$) curves.

**IV-B Spin injection polarization and spin accumulation saturation**

In the experiments with Setup-A, $\Delta\mu_\text{L}(V_\text{JL})$ beneath the L electrode is composed of the spin injection form the L electrode $\Delta\mu_\text{L0}(V_\text{JL})$ and the spin transport from the R electrode. Here, the latter is neglected, i.e., $\Delta\mu_\text{L}(V_\text{JL}) = \Delta\mu_\text{L0}(V_\text{JL})$, since it is sufficiently small when $I_\text{R} = 0$ or −1.5 mA (see Sec. S1 in S.M. [44]). Under this approximation, the theoretical $P_\text{inj}(V_\text{JL})$ is calculated, and then it is directly compared with the experimental result in Fig. 8(a). To numerically calculate $P_\text{inj}(V_\text{JL})$, $J_\text{S}$ in Eq. (4) is expressed with $J_+(V_\text{JL}, \Delta\mu_\text{L}(V_\text{JL}))$ and $J_-(V_\text{JL}, \Delta\mu_\text{L}(V_\text{JL}))$ given by Eq. (10). Hence, $J_+(V_\text{JL}, \Delta\mu_\text{L}(V_\text{JL}))$, $J_-(V_\text{JL}, \Delta\mu_\text{L}(V_\text{JL}))$, and $\Delta\mu_\text{L}(V_\text{JL})$ are calculated using Eqs. (4) and (10) in a self-consistent manner. First, $V_\text{JL}$ is set at a target value, while an initial $\Delta\mu_\text{L}(V_\text{JL})$ value is set at zero. Subsequently, $J_+(V_\text{JL}, \Delta\mu_\text{L}(V_\text{JL}))$ and $J_-(V_\text{JL}, \Delta\mu_\text{L}(V_\text{JL}))$ are calculated using Eq. (10). Then, $\Delta\mu_\text{L}(V_\text{JL})$ is calculated using Eq. (4). This sequence is iteratively performed until $\Delta\mu_\text{L}(V_\text{JL})$ at the previous step differs little from that at the present step. After estimations with various $V_\text{JL}$ values in the same manner, finally $P_\text{inj}$−$V_\text{JL}$, $\Delta\mu_\text{L}$−$V_\text{JL}$, and $J_\text{JL}$−$V_\text{JL}$ curves are obtained. In this section, to investigate how the spin accumulation $\Delta\mu_\text{L}$ affects $P_\text{inj}$, two cases were calculated for $P_\text{inj}$ with $r_\text{sr} = 0$ and $r_\text{sr} = (\rho\lambda_\text{sf})\lambda_\text{sf}/t_\text{SOI} = 3.4\times10^{-10}$ Ω·m$^2$, respectively, where the correction term $\lambda_\text{sf}/t_\text{SOI}$ is added to the bulk spin resistance expression $\rho\lambda_\text{sf}$ because the Si channel thickness $t_\text{SOI} = 25$ nm is much smaller than the spin diffusion length $\lambda_\text{sf} = 1$ μm in our device [12,13,15]. The former case corresponds to $\lambda_{sf} = 0$ in the Si and it excludes the effect of SAS on $P_{inj}$ due to $\Delta\mu_L = 0$, as explained in Sec. II-C, and the latter case corresponds to the realistic Si spin resistance in our device and suppression of $P_{inj}$ through SAS occurs due to $\Delta\mu_L \neq 0$. On the other hand, Si band bending is considered in the calculation of $P_{inj}$, but our numerical calculations found that its influence on $P_{inj}$ is little (see Sec. S4 in S.M.).

Figure 9(a) shows $P_{inj}$ plotted as a function of $V_{JL}$ with $t_{MgO} = 1$ nm when $I_R = 0$ mA, where yellow and blue curves denote the values calculated with $r_{sr} = 0$ and $r_{sr} = 3.4\times10^{-10}$ Ω·m$^2$, respectively. The two curves reasonably reproduce the shape of the experimental results in Fig. 8(a). Moreover, $P_{inj} = P_{Fe}(\bar{\mu}^{Fe})$ (= 0.24) at $V_{JL} = 0$ is consistent with the V-F model with $\gamma = P_{Fe}(\bar{\mu}^{Fe})$, where $P_{Fe}(\bar{\mu}^{Fe})$ represents the spin polarization of the Fe band at $E = \bar{\mu}^{Fe}$. These two results validate our calculation method. One main feature of the $P_{inj}$–$V_{JL}$ curve is that $P_{inj}$ in the positive $V_{JL}$ range is larger than that in the negative $V_{JL}$ range. Figure 9(b) shows $J_{JL}$–$V_{JL}$ and $\Delta\mu_L$–$V_{JL}$ curves, where the former is the fitting to the experimental $J_{JL}$–$V_{JL}$ curve (see Sec. S1 in S.M. [44]) and the latter curve was calculated with $\Delta\mu_L = 2J_S r_{sr}$ ($r_{sr} = 3.4\times10^{-10}$ Ω·m$^2$). It is worth noting that the $\Delta\mu_L$–$V_{JL}$ curve is asymmetric with respect to $V_{JL} = 0$ although the $J_{JL}$–$V_{JL}$ curve is nearly symmetric. The saturation trend of $\Delta\mu_L$ in the higher negative $V_{JL}$ range leads to the difference between $r_{sr} = 0$ and $r_{sr} = 3.4\times10^{-10}$ Ω·m$^2$ cases in Fig. 9(a), which indicates SAS.

Here, the physics of the $P_{inj}$–$V_{JL}$ curve is explained by the band diagram, in which a constant junction voltage $V_{JL}$ is assumed. Figure 10(a) shows the band diagram of the Fe/MgO/$n^+$-Si junction in the positive $V_{JL}$ range (the spin injection geometry) with a low $V_{JL}$ value, where a dashed line in $n^+$-Si is $\bar{\mu}^{Si}$ and $\Delta\mu_L = \mu_+(V_{JL}) - \mu_-(V_{JL})$ is present in $n^+$-Si. Figure 10(b) shows the spin polarization $P_{Fe}(E)$ of electrons at energy $E$ in Fe and schematic distribution of normalized tunneling probability $T_{nor}(E)$ (=$T(E)/T(\bar{\mu}^{Fe})$) at the same low $V_{JL}$ value to facilitate understanding the situation of Fig. 10(a), where $\bar{\mu}^{Fe}$ is set at the zero potential ($\bar{\mu}^{Fe} = 0$). In a similar manner, the band diagram, $P_{Fe}(E)$, and $T_{nor}(E)$ under a high $V_{JL}$ value are shown in Figs. 10(c) and (d). The remarkable feature is that $P_{Fe}(E)$ monotonically decreases (increases) as $E$ becomes higher (lower) than $\bar{\mu}^{Fe}$, in other words, the capability of the spin selective filtering by Fe becomes lower/higher with increasing/decreasing $V_{JL}$. In the positive $V_{JL}$ range (the spin injection geometry), electrons (filled electron states) in Fe play important roles. Under very low $V_{JL}$ range ($V_{JL} \sim 0$ V), the conduction mechanism is nearly the same as that in the V-F model, where the electrons at $\bar{\mu}^{Fe}$ dominate the conduction, i.e., $P_{inj}(V_{JL}) = P_{Fe}(\bar{\mu}^{Fe})$. As $V_{JL}$ is increased, electrons with lower energies contribute to the tunneling current, as shown in Figs. 10(b) and (d), since the area in the $T(E)$ curve within $\bar{\mu}^{Fe} - V_{JL} < E < \bar{\mu}^{Fe}$ becomes larger. With this change, $P_{inj}$ becomes larger, since $P_{Fe}(E)$ becomes higher with lowering energy $E$, as the shown in Figs. 10(b) and (d), and the integration of $P_{Fe}$ from $\bar{\mu}^{Fe} - V_{JL}$ to $\bar{\mu}^{Fe}$ becomes larger. Due to this mechanism, $P_{inj}$ further increases to the maximum $P_{inj}(V_{JL}) = 0.27$ at around $V_{JL} = 0.8$ V in Fig. 9(a), because the energy range $\bar{\mu}^{Fe} - V_{JL} < E < \bar{\mu}^{Fe}$ becomes wider with a high $V_{JL}$ bias, as shown in Fig. 10(d). This phenomenon is not seen in the V-F model. The numerical $P_{inj}$–$V_{JL}$ curve in Fig. 9(a) shows that $P_{inj}$ slightly decreases as $V_{JL}$ is further increased from 0.8 V, which is caused by a sharper distribution of $T_{nor}(E)$ at around $\bar{\mu}^{Fe}$, and the details are explained in Sec. S5 in S.M. [44]. Unlike the experimental results in Fig. 8(a), where $P_{inj}$ has a steep decrease as $V_{JL}$ is increased, $P_{inj}$ has a slight decrease in our calculation. We attribute such discrepancy to the neglected spin-flip events during the energy relaxation in our model, as the assumption (III) mentioned in Sec. II-C. To explain further, after the direct tunneling, electrons with high energies above $\bar{\mu}^{Si}$ (the so-called "hot electrons") can undergo significant spin flips and reduce $P_{inj}$ due to the magnetic excitation at the I/SC interface [45]. The reduction in magnetoresistance induced by the hot-electron states has been frequently reported in tunneling magnetoresistance (TMR) devices at high biases [46-48].

Figures 11(a) and (b) show the band diagrams of the Fe/MgO/$n^+$-Si junction in the negative $V_{JL}$ range (the spin extraction geometry). Here, empty electron states in Fe play important roles. First, $\Delta\mu_L = 0$ is considered for a simple explanation. As $V_{JL}$ is increased from zero in the negative bias, the Si band bends toward downward and interface electrons with lower energies contribute to the tunneling current. The energy range for such electrons increases with negatively increasing $V_{JL}$. With such change, $P_{inj}$ becomes lower, since $P_{Fe}(E)$ becomes lower with increasing $E$, as shown in Figs. 10(b) and (d). Like the case of the positive $V_{JL}$ range, as $V_{JL}$ is increased in the negative bias, $T(E)$ has a sharp peak located at around $\bar{\mu}^{Si}$, which can be understood by replacing $\bar{\mu}^{Fe}$ with $\bar{\mu}^{Si}$ in Fig. 10(d). Thus, $J_S$ is almost determined by $P_{Fe}(\bar{\mu}^{Si})$ and it monotonically decreases with negatively increasing $V_{JL}$, as shown in Fig. 9(a).

Hereafter, the effect of $\Delta\mu_L = 2J_S r_{sr}$ on $P_{inj}(V_{JL})$ is considered. As indicated previously, SAS occurs in the higher negative $V_{JL}$ range, which can be seen as the difference between $r_{sr} = 0$ and $r_{sr} = 3.4\times10^{-10}$ Ω·m$^2$ cases in Fig. 9(a). Here, we first explain how the increase of $\Delta\mu_L$ can decrease $J_S$ using the schematic band diagrams shown in Figs. 12(a) and (b), where the spin extraction bias $V_{JL}$(< 0) is identical with each other, but (a) small and (b) large $\Delta\mu_L$ values are assumed, respectively. Under a larger $\Delta\mu_L$, the effective barrier height $\Phi_-$ for $\mu_-$ is lower and $\Phi_+$ for $\mu_+$ is higher, which leads to the decrease of $J_+$ and increase of $J_-$. This results in the decreases in $J_S$ (= $J_+ - J_-$) and $P_{inj}$. Next, we qualitatively explain how SAS occurs in FM/I/SC junctions in the spin injection and extraction geometries, respectively. When $V_{JL}$ is significantly high in the positive bias range (the spin injection geometry), as shown Fig. 10(c), electrons at $\bar{\mu}^{Si}$ in Fe have a quite small tunneling probability because the effective barrier height for these electrons is significantly high. Thus, the change in $\Delta\mu_L$ at $\bar{\mu}^{Si}$ in Si have little influence on $J_S$, namely, SAS does not appear in the positive $V_{JL}$ range. This is responsible for the identical curves at $V_{JL} > 0$ for both $\Delta\mu_L$ cases in Fig. 9(a). On the other hand, when $|V_{JL}|$ is significantly high in the negative range (the spin extraction geometry), as shown in Figs. 11(b) and 12(b), and $\Delta\mu_L$ is reasonably large, the effective barrier height $\Phi_-$ for $\mu_-$ is lower than $\Phi_+$ for $\mu_+$ and it considerably decreases as $V_{JL}$ is increased in the negative bias. From the $P_{Fe}(E)$ curve in Figs. 10(b) and (d), the empty states for up-spin electrons (denoted by red color and +) are still larger than those for up-spin electrons (denoted by blue color and −) even under this situation (Fig. 12(b)). Nonetheless, considering that $J_+$ and $J_-$ are exponential functions with respect to $\Phi_+$ and $\Phi_-$ (Eq. 10), respectively, and $T(\mu_+)$ and $T(\mu_-)$ are considerably larger than $T(E < \mu_+)$ and $T(E < \mu_-)$, respectively, and $J_-$ can become comparable to $J_+$ as $\Delta\mu_L$ is largely increased, i.e., $J_S$ decreases with increasing $|V_{JL}|$. This is the mechanism of SAS in FM/I/SC junctions, where the spin-selectivity of the empty electron states in Fe is less effective under higher $V_{JL}$ conditions in the spin extraction geometry.

For further study, the relationship between $P_{inj}|_{\Delta\mu_L \neq 0}(V_{JL})$ and $P_{inj}|_{\Delta\mu_L = 0}(V_{JL})$ is derived under the first order approximation of $\Delta\mu_L$ (see Sections S2 and S5 in S.M.):

$$P_{inj}|_{\Delta\mu_L \neq 0}(V_{JL}) = \frac{P_{inj}|_{\Delta\mu_L = 0}(V_{JL})}{1 + \left|\frac{\partial J(V_{JL})|_{\Delta\mu_L = 0}}{\partial \bar{\mu}^{Si}}\right| r_{sr}} = \frac{P_{inj}|_{\Delta\mu_L = 0}(V_{JL})}{1 + \frac{r_{sr}}{r_b^*}} = \frac{P_{inj}|_{\Delta\mu_L = 0}(V_{JL}) \, r_b^*}{r_b^* + r_{sr}}, \quad (14)$$

where $P_{inj}|_{\Delta\mu_L \neq 0}(V_{JL})$ and $P_{inj}|_{\Delta\mu_L = 0}(V_{JL})$ represent the spin injection polarizations with $r_{sr} \neq 0$ and $r_{sr} = 0$, respectively, and $1/r_b^* = \left|\frac{\partial J(V_{JL})|_{\Delta\mu_L = 0}}{\partial \bar{\mu}^{Si}}\right|$ expresses how the tunneling current changes with a small change in $\bar{\mu}^{Si}$ at a specific $V_{JL}$ value under $\Delta\mu_L = 0$. The functionality of $P_{inj}|_{\Delta\mu_L = 0}(V_{JL})$ in Eq. (14) is very similar to $\gamma$ in Eq. (6), in the sense that both parameters correspond to the junction capability of spin current generation when there is no spin accumulation. Thus, Eq. (14) can be regarded as a comprehensive model applicable to tunnel junctions with linear and nonlinear junction properties, while it converges to the V-F model under the linear response regime with a constant $r_b^*$ value. Equation (14) also indicates that the condition $r_b^* \gg r_{sr}$ is required for maximizing $P_{inj}(V_{JL})$, while the condition $r_b \gg r_{sr}$ predicted by the V-F model has been widely used in semiconductor-based spintronic devices. We show a large difference between these matching conditions ($r_b^* \gg r_{sr}$ and $r_b \gg r_{sr}$): Numerical analyses revealed that $P_{inj}|_{\Delta\mu_L \neq 0}(V_{JL})$ can be significantly reduced from $P_{inj}|_{\Delta\mu_L = 0}(V_{JL})$ under a quite large $\Delta\mu$ in nonlinear cases where $r_b^*$ decreases with decreasing $V_{JL}$ (see Sec.S3 in S.M. [44]).

**IV-C Spin detection and spin accumulation saturation**

To calculate $P_{det}(V_{JL})$ in Setup-B, we use Eq. (7) that is expressed with $\Delta\mu_L(V_{JL})$ and $\Delta V_{det}(V_{JL})$. Here, $\Delta\mu_L(V_{JL}, V_{JR})$ beneath the L electrode is composed of the spin injection from the L electrode $\Delta\mu_{L0}(V_{JL})$ and the spin transport from the R electrode $\Delta\mu_L(V_{JR}) = \Delta\mu_R(V_{JR}) \exp(-L_{ch}/\lambda_{sf})$, i.e., $\Delta\mu_L(V_{JL}, V_{JR}) = \Delta\mu_{L0}(V_{JL}) + \Delta\mu_L(V_{JR})$, where $\Delta\mu_R(V_{JR})$ is the spin accumulation beneath the R electrode. First, $\Delta\mu_R(V_{JR})$ with $I_R = 20$ mA ($J_R = 5.6\times10^7$ A/m$^2$) is estimated using the $\Delta\mu$–$V_{JL}$ and $J_{JL}$–$V_{JL}$ curves in Fig. 9(b). Then, $\Delta\mu_L(V_{JR})$ is calculated with $L_{ch} = 0.7$ μm and $\lambda_{sf} = 1$ μm [12,13,37−40]. Specific values are as follows: $V_{JR} = 0.49$ V, $\Delta\mu_R(V_{JR}) = 2.2$ mV, and $\Delta\mu_L(V_{JR}) = 1.1$ mV. Then, $\Delta\mu_{L0}(V_{JL})$ for a target $J_{JL}$ is estimated from the $\Delta\mu$–$V_{JL}$ and $J_{JL}$–$V_{JL}$ curves in Fig. 9(b). Next, $\Delta V_{det}(V_{JL})$ caused by $\Delta\mu_L(V_{JL}, V_{JR})$ is numerically calculated using a constant $J_{JL}$ condition for the P and AP magnetization states:

$$J_{JL}^P(V_{JL} + V_{det}(V_{JL}), \Delta\mu_L(V_{JL}, V_{JR})) = J_{JL}^{AP}(V_{JL} - V_{det}(V_{JL}), -\Delta\mu_L(V_{JL}, V_{JR})) = J_{JL}, \quad (15)$$

where $J_{JL}^P$ and $J_{JL}^{AP}$ represent the current densities of the L electrode in P and AP magnetization states, respectively. After calculations with various $V_{JL}$ values in the same manner, finally a $P_{det}$–$V_{JL}$ curve is obtained using Eq. (7) in the $V_{JL}$ range of −1.5 - 1.5 V (Figs. 13(a) and (b)). The properties of $P_{det}$ are analyzed under three cases with two parameters: $\Delta\mu_L$ (= 0 or $2J_S r_{sr}$ with $r_{sr} = 3.4\times10^{-10}$ Ω·m$^2$) and the Si band bending (present or absent). Hereafter, three cases are named: Case-1 ($\Delta\mu_L = 0$, with the Si band bending), Case-2 ($\Delta\mu_L = 0$, without the Si band bending), and Case-3 ($\Delta\mu_L = 2J_S r_{sr}$, with the Si band bending). Comparison of Case-1 with Case-2 can evaluate the effect of a semiconductor property (band bending) on $P_{det}$, while comparison of Case-1 with Case-3 can evaluate the effect of SAS on $P_{det}$. In Case-1 and Case-3, $V_{JL} = V_{MgO} + \Phi_{Si}$, as defined in Figs. 2(a) and (b), and the boundary condition for the Poisson equation is the continuity of the electric flux density at the MgO/$n^+$-Si interface. In Case-2, $V_{JL} = V_{MgO}$ and the boundary condition for the Poisson equation is neglected.

Figures 13(a) and (b) show $P_{det}$ plotted as functions of $V_{JL}$ with $t_{MgO} = 1$ nm, where yellow, gray, and blue curves represent Case-1, Case-2, and Case-3, respectively. Our method is validated by the fact that these curves reasonably reproduce the shape of the experimental result in Fig. 8(b). In addition, $P_{\text{det}} = P_{\text{Fe}}(\bar{\mu}^{Fe})$ (= 0.24) at $V_{\text{JL}} = 0$ in all the cases is consistent with the V-F model prediction assuming $V_{\text{JL}} \sim 0$. In the positive $V_{\text{JL}}$ range (the spin injection geometry), as $V_{\text{JL}}$ is increased from 0, $P_{\text{det}}$ steeply decreases from 0.24 to 0 and the yellow, blue, and gray curves are identical with each other. This phenomenon can be simply explained as follows. From Eq. (7), $P_{\text{det}}$ is the conversion factor between $\Delta V$ and $\Delta \mu$ in Si under a constant $J$ value. When $V_{\text{JL}}$ is increased, $\Phi_{\text{eff}}(\bar{\mu}^{Si})$ increases and the change in $\Delta \mu$ at $\bar{\mu}^{Si}$ have less impact on $\Delta V$ to keep a constant $J$ value, as shown in Fig. 10(c). Hence, SAS does not appear, and the Si band bending does not affect $P_{\text{det}}$. In the negative $V_{\text{JL}}$ range (the spin extraction geometry), the yellow, blue, and gray curves have a similar trend: As $|V_{\text{JL}}|$ is increased from 0, $P_{\text{det}}$ exhibits a steep increase, has a positive peak at $V_{\text{peak}} = -0.2$ V (yellow and blue) or $V_{\text{peak}} = -0.1$ V (gray), and then exhibits a gradual decrease. The peak position corresponds to the condition $|V_{\text{MgO}}| = |V_{\text{peak}}| - |\Phi_{\text{Si}}| = E_2$ where $\bar{\mu}^{Fe}$ and $E_C$ are matched, as shown in Fig. 13(c). It is noteworthy that the gray curve has $P_{\text{det}}$ larger than 1 in the $V_{\text{JL}}$ range of −0.1 to −0.3 V, namely, $P_{\text{det}}$ is significantly enhanced when the Si band bending is absent. Thus, this means that $P_{\text{det}}$ becomes larger as the Si band bending becomes smaller, which is possibly realized by a Si with a low doping concentration. This is unique for semiconductor-based tunnel junctions and likely used to enhance $P_{\text{det}}$. The peak height originates from multiple phenomena closely related with each other and thus in-depth analyses are given in Sec. S5 in S.M. [44]. In the higher negative $V_{\text{JL}}$ range in Fig. 13(a), the main difference between Case-1 and Case-3 is the rate of decrease with increasing $|V_{\text{JL}}|$ in the negative bias direction. The origin is similar to that for Fig. 9(a) in Sec. IV-B, i.e., SAS reduces $P_{\text{det}}$.

To facilitate understanding of the $P_{\text{det}}-V_{\text{JL}}$ curves, here we introduce an alternative form of Eq. (7) under the first-order approximation of $\Delta \mu_L$ (see Sec. S4 in S.M.):

$$P_{\text{det}}(V_{\text{JL}}) = 2\frac{\Delta V_{\text{det}}}{\Delta \mu_L} \sim \frac{\frac{\partial J_S|_{\Delta \mu_L=0}}{\partial \bar{\mu}^{Si}}}{\frac{\partial J|_{\Delta \mu_L=0}}{\partial V_{\text{JL}}}}, \qquad (16)$$

where $\Delta V_{\text{det}}$ is the spin detection voltage under a spin accumulation $\Delta \mu_L$, $\bar{\mu}^{Si}$ defined in Figs. 2(a) and (b) is the averaged energy level in the bulk Si region, the denominator $\frac{\partial J|_{\Delta \mu_L=0}}{\partial V_{JL}}$ is the junction differential conductance under $\Delta \mu_L = 0$, and the numerator $\frac{\partial J_s|_{\Delta \mu_L=0}}{\partial \bar{\mu}^{Si}}$ is a unique term derived from our model, which reflects how a small perturbation of $\bar{\mu}^{Si}$ affects the magnitude of $J_S$ under $\Delta \mu_L = 0$. One advantageous feature of Eq. (16) is that the two independent terms $\frac{\partial J_s|_{\Delta \mu_L=0}}{\partial \bar{\mu}^{Si}}$ and $\frac{\partial J|_{\Delta \mu_L=0}}{\partial V_{JL}}$ are separated, and $P_{\text{det}}(V_{\text{JL}})$ can be understood qualitatively by analyzing which term is dominant at various $V_{\text{JL}}$ values. As $V_{\text{JL}}$ is positively increased from 0 (the spin injection geometry), the numerator steeply decreases to zero and it becomes the dominant term in determining $P_{\text{det}}$. This is because $\Phi_{\text{eff}}(\bar{\mu}^{Si})$ increases and the change in $\Delta \mu$ at $\bar{\mu}^{Si}$ has less impact on $J_S$, as in the case shown in Fig. 10(c). As $V_{\text{JL}}$ is negatively increased from 0 (the spin extraction geometry), the increase of $P_{\text{det}}$ is attributed to the decrease of the denominator. Figure 14(a) shows a schematic band diagram in the range of $V_{\text{peak}} < V_{\text{JL}} < 0$, where $\frac{\partial J|_{\Delta \mu_L=0}}{\partial V_{JL}}$ is proportional to $D_{Si}(\bar{\mu}^{Fe})$ that is the Si density of states at $E = \bar{\mu}^{Fe}$. As $V_{\text{JL}}$ is negatively increased, $D_{Si}(\bar{\mu}^{Fe})$ decreases due to the parabolic distribution of $D_{Si}(E)$. Hence, $\frac{\partial J|_{\Delta \mu_L=0}}{\partial V_{JL}}$ proportionally decreases, which results in the increase of $P_{\text{det}}$. Such feature maintains until $V_{\text{JL}} = V_{\text{peak}}$ that corresponds to the condition $|V_{\text{MgO}}| = |V_{\text{peak}}| - |\Phi_{\text{Si}}| = E_2$, where $\bar{\mu}^{Fe}$ and $E_C$ are matched, as shown in Fig. 13(c). This is because $\bar{\mu}^{Fe}$ falls into the band gap of Si when $V_{\text{JL}} < V_{\text{peak}}$ and $\frac{\partial J|_{\Delta \mu_L=0}}{\partial V_{JL}}$ does not decrease because of $D_{Si}(\bar{\mu}^{Fe}) = 0$, as shown in Fig. 14(b). As $V_{\text{JL}}$ is further negatively increased from $V_{\text{peak}}$, like the case in $P_{\text{inj}}(V_{\text{JL}})$, the reduction in $P_{\text{det}}$ is attributed mainly to SAS and partially to the lower $P_{\text{Fe}}(E)$ values within $\bar{\mu}^{Si} - V_{\text{JL}} < E < \bar{\mu}^{Si}$ (see the detailed analysis in Sec. S6 in S.M. [44]).

## V. Discussion: How nonlinear junction properties influence the magnetoresistance ratio

In this section, we numerically analyze the magnetoresistance (MR) ratio of a two-terminal Si-based device with Fe/I/Si tunnel junctions for the source and drain contacts, and we find that the bias dependences of $P_{\text{inj}}$ and $P_{\text{det}}$ in the junctions considerably change the MR ratio from that expected from the V-F model with a constant $P_{\text{inj}}$ and $P_{\text{det}}$ value (= $\gamma$). In particular, SAS in $P_{\text{det}}$ dominantly suppresses the MR ratio in a higher junction bias range, which is a crucial issue for the device design to obtain high MR ratios.

Figure 15(a) shows a schematic device structure and its energy band diagram, where JCT1 (Fe1/I/Si) and JCT2 (Si/I/Fe2) are the source and drain contacts, respectively, JCT1 and JCT2 are in the P magnetization state, the length of the $n^+$-Si channel is 25 nm, $V_1^P$, $V_2^P$, and $V_{ch}$ are the voltage drops in JCT1, JCT2, and the $n^+$-Si channel, respectively, $V_{total}^P = V_1^P + V_2^P + V_{ch}$ is the total voltage drop of the device, $\Delta\mu$ is the spin accumulation in the $n^+$-Si channel, and $V_{total} > 0$ in any cases. For the AP magnetization state between JCT1 and JCT2, the relation $V_{total}^{AP} = V_1^{AP} + V_2^{AP} + V_{ch}$ is satisfied. The junctions have the same properties as those described in Sec. II-C. The $n^+$-Si channel length (25 nm) allows us to use the experimental $J$–$V$ curves in Sec. III and IV because it is longer than the typical depletion length ~3 nm, and also to use a constant $\Delta\mu$ in the channel because it is sufficiently shorter than the spin diffusion length of ~1 μm. The electron current $J$ and the spin current $J_S$ flow from JCT1 to JCT2, and vice versa. Thus, JCT1 has the blue $P_{inj}$ curve in the positive $V_{JL}$ range in Fig. 9(a), whereas JCT2 has the blue $P_{det}$ curve in the negative $V_{JL}$ range in Fig. 13(a). Here, we use the following formula to numerically calculate the MR ratio under a constant $J$ value:

$$MR\ ratio = \frac{\Delta R}{R_{total}^P} = \frac{V_{total}^{AP} - V_{total}^P}{V_{total}^P}, \quad (17)$$

where $V_{total}^P$ and $V_{total}^{AP}$ are the total voltage drops in the P and AP magnetization states, respectively, $\Delta R$ is the magnetoresistance, and $R_{total}^P$ is the total resistance-area product of the device in the P magnetization state. The values of $V_1^{P(AP)}$ and $V_2^{P(AP)}$ are obtained by a self-consistent calculation with the following continuous equation and a constant $J$ value:

$$J_1^{P(AP)}(V_1^{P(AP)}, \Delta\mu) = J_2^{P(AP)}(V_2^{P(AP)}, \Delta\mu) = J, \quad (18)$$

where $J_1$ and $J_2$ are the current densities in JCT1 and JCT2, respectively. In the following, $J_1$ and $J_2$ are calculated by Eq. (10) with the same parameters as those in Sec. II-C to express the same junction properties. Finally, $J$ is transformed to $V_{total}^P$ to plot the MR ratio as a function of $V_{total}^P$. Note that the calculation revealed $V_1$, $V_2 \gg V_{ch}$ and $V_1 \simeq V_2$. The detailed calculation method is described in Sec. S6 in the S.M. [44]. In addition, the MR ratio is also calculated using the V-F model, in which the $J$–$V$ curves of JCT1 and JCT2 are the same as that in our model calculation and both $P_{inj}$ and $P_{det}$ are $\gamma$ at any bias values.

Figure 15(b) shows the MR ratios as functions of $V_{total}^P$ with the insulator thickness $t_I = 1$ nm, where a blue curve and brown open dots represent those calculated by the V-F model and our model, respectively. The blue curve by the V-F model monotonically increases to ~4 % at $V_{total}^P = 3$ V, as $V_{total}^P$ is increased from 0. This is because $\Delta R$ is a constant value proportional to $4\gamma^2$, whereas $R_{total}^P$ decreases due to the decreases in the JCT1 and JCT2 resistances. As $V_{total}^P$ is further increased from 3 V, the MR ratio reaches a peak and then decreases (not shown). This is the well-known conductance matching curve that has been widely used to study semiconductor-based spintronic devices. As described below, when such devices have ferromagnetic tunnel junctions for the source and drain contacts and a large bias is applied between these contacts to obtain a large output current, the V-F model is *not* suitable for the guideline to maximize the MR ratio.

On the other hand, the brown open dots by our model does not monotonically increases with increasing $V_{total}^P$. This is clear evidence that the V-F model is not applicable to the analyses and design of devices in a higher bias range. The inset of Fig. 15(b) shows a magnified view of the brown open dots, indicating that the MR ratio has a peak of ~0.15 % at around 2 V and significantly decreases in the higher $V_{total}^P$ range as $V_{total}^P$ is increased. Note that the blue curve and brown open dots have the same values at around zero bias where the junction properties are under the linear approximation regime.

To clarify the detailed mechanism behind the features shown in the inset of Fig. 15(b), the MR ratio is calculated by the following equation (see Sec. S7 in S.M. [44]):

$$\begin{aligned}
MR\ ratio &= r_{sr} \frac{\left(P_{det}^{AP}(V_1^{AP}) + P_{det}^{AP}(V_2^{AP})\right)\left(P_{inj}^{AP}(V_1^{AP}) + P_{inj}^{AP}(V_2^{AP})\right) - \left(P_{det}^P(V_1^P) + P_{det}^P(V_2^P)\right)\left(P_{inj}^P(V_1^P) - P_{inj}^P(V_2^P)\right)}{R_{total}^P} \\
&= \frac{\Delta R}{R_{total}^P}, \quad (19)
\end{aligned}$$

where $r_{sr} = 3.4 \times 10^{-10}$ Ω·m², the superscripts P and AP denote the P and AP magnetization states, respectively, $P_{inj}(V_1)$ and $P_{inj}(V_2)$ ($P_{inj}(V_1)$ and $P_{inj}(V_2)$) are the spin injection (detection) polarizations of JCT1 and JCT2, respectively, $\Delta R$ is the magnetoresistance, and $R_{total}^P$ is the total resistance-area product of the device in the P magnetization state. The eight terms in Eq. (19), such as $P_{det}^{AP}(V_1^{AP})$ and $P_{det}^{AP}(V_2^{AP})$, are calculated, which can be done using $V_1$, $V_2$, and $\Delta\mu$ estimated from Eq. (18). The detailed observation of the parameters in Sec. S6 in S.M. revealed that $P_{det}^P(V_1^P)$ and $P_{det}^{AP}(V_1^{AP})$ are significantly smaller than the others and $V_1 \simeq V_2 \simeq V_{total}/2$. To facilitate the following analysis, Eq. (19) is approximately rewritten as

$$MR\ ratio = r_{sr} \frac{P_{det}^{AP}(V_2^{AP})\left(P_{inj}^{AP}(V_1^{AP})+P_{inj}^{AP}(V_2^{AP})\right) - P_{det}^P(V_2^P)\left(P_{inj}^P(V_1^P)-P_{inj}^P(V_2^P)\right)}{R_{total}^P}. \quad (20)$$

Figure 15(c) shows each term in Eq. (20) plotted as a function of $V_{total}^P$ with $t_I = 1$ nm, where green and purple open dots denote $P_{det}^{AP}(V_2^{AP})\left(P_{inj}^{AP}(V_1^{AP}) + P_{inj}^{AP}(V_2^{AP})\right)$ (the first term in the numerator) and $P_{det}^P(V_2^P)\left(P_{inj}^P(V_1^P) - P_{inj}^P(V_2^P)\right)$ (the second term in the numerator), respectively, and black open dots denote $r_{sr}/R_{total}^P$. The second term has smaller values than the first term and it is bias insensitive. Thus, the features in the inset of Fig. 15(b) are mostly dominated by the first term and $r_{sr}/R_{total}^P$. To gain a deeper insight into the bias dependence of the first term, the two $P_{inj}$ terms in the parentheses are excluded since $P_{inj}$ in the positive bias range is less sensitive to $V_{JL}$ as shown in Fig. 9(a). On the other hand, since $r_{sr}/R_{total}^P$ changes little in the $V_{total}^P$ range lower than 1 V, the peak of the brown open dots in Fig. 15(b) originates from $P_{det}^{AP}(V_2^{AP})$ that basically has the feature of the blue curve in Fig. 13(a). In fact, $V_2^P$ is ~0.2 V ($V_{total}^P$ = ~0.4 V) at the peak position in Fig. 15(c), which is consistent with the peak position $V_{JL}$ = ~0.2 V in Fig. 13(a). In addition, the drastic decrease of the first term in the higher bias range (green open dots in Fig. 15(c)) is also due to the feature of $P_{det}^{AP}(V_2^{AP})$ through SAS, as seen in the higher negative $V_{JL}$ range in Fig. 13(a). Our analyses indicate that $P_{det}^{AP}(V_2^{AP})$ is the dominant factor among many terms in Eq. (20) and its engineering is a key to obtain high MR ratios.

Our results can provide some other strategies to realize higher MR ratios. One approach is to reduce $R_{total}^P$. This can be achieved by a thinner I layer while $P_{inj}$ and $P_{det}$ remain unchanged. Figure 16(a) shows a calculated two-dimensional (2-D) contour map of the MR ratio plotted against $V_{total}^P$ and $t_I$, where a bar on the right-hand side is the correspondence between the color and MR ratio (0 − 5%). (The MR ratio *vs.* $V_{total}^P$ plot for $t_I = 0.4 – 1.0$ nm is shown in Fig. S10 in Sec. S8 in S.M. for easy viewing). The MR ratio has the following characteristics. As $V_{total}^P$ is increased at a specific $t_I$ value larger than 0.6 nm, the MR ratio increases at first and then decreases. This is due to the decrease in $P_{det}^{AP}(V_2^{AP})$ through SAS, like the green open dots in Fig. 15(c). On the other hand, as $V_{total}^P$ is increased at a specific $t_I$ value smaller than 0.6 nm, the MR ratio monotonically decreases since the decrease in $P_{det}^{AP}(V_2^{AP})$ through SAS becomes significant from $V_{total}^P = 0$. Note that thinner $t_I$ with lower voltage drop leads to higher voltage difference between $\bar{\mu}^{Fe}$ and $\bar{\mu}^{Si}$ under a specific $V_{total}^P$ value. As $t_I$ is increased at a specific $V_{total}^P$ value smaller than 0.6 V, the MR ratio monotonically decreases, owing to the exponential increase in the junction resistances added to $R_{total}^P$. According to such mechanism, the high MR ratio of ~5% is achieved in the region of $V_{total}^P < 0.5$ V and $t_I <$ 0.5 nm. This validates that reducing $R_{total}^P$ by thinner $t_I$ is an effective approach to enhance the MR ratio.

Another approach is to utilize spin-filter I tunnel barriers to suppress the decrease in $P_{det}^{AP}(V_2^{AP})$ caused by SAS in a higher $V_{total}^P$ range. Figures 17(a) and (b) show the schematic band diagrams in the spin extraction geometry ($V <$ 0) with a normal I layer and a spin-filter I layer, respectively, where $V$ and $\Delta\mu$ are identical with each other, the widths of the red and blue arrows in the I layer represent the magnitudes of $J_+$ and $J_−$, respectively, and $\Delta_{SF}$ in the spin-filter I layer is the spin-split energy to form a lower up-spin barrier height $\Phi_+$ (= $\Phi_{eff} − \Delta_{SF}/2$) and a higher down-spin barrier height $\Phi_−$ (= $\Phi_{eff} + \Delta_{SF}/2$). In Fig. 17(a), $J_S = J_+ − J_−$ is very small due to SAS and both $P_{inj}$ and $P_{det}$ are very small. Using the spin-filter I layer as shown in Fig. 17(b), however, $J_S$ is largely enhanced, which corresponds to the increase of $\Delta\mu_0$ in Figs. 4(a) and (b), and both $P_{inj}$ and $P_{det}$ are enhanced accordingly. In addition, the spin-filter I layer can also enhance $J_S$ in a very low $V_{total}^P$ range, since it basically plays the same role as $\gamma$ in the V-F model. Thus, it is expected that $P_{det}$ is improved in a wide $V_{total}^P$ range. To numerically examine the effect of a spin-filter layer, the spin-split energy $\Delta_{SF} = 0.2$ eV that is comparable to the experimental value in EuS [49] is introduced into the tunneling probabilities:

$$T^{+/-}(E) = \exp\left(-2t_{ox}\sqrt{\frac{2m_t(\Phi_{eff} \mp \Delta_{SF}/2 - E)}{\hbar^2}}\right), \quad (21)$$

and then Eq. (10) is calculated while other junction properties are identical with those described in Sec. II-C. Figure 17(c) shows schematic device structures when the spin-filter I layers are introduced in JCT1 and JCT2, where the upper and lower figures are the P and AP magnetization states, respectively. Figure 17(d) shows a schematic one-dimensional electron energy band diagram of a two-terminal spin transport device structure with the spin-filter I layers in the P magnetization state, where red and blue lines in the spin-filter I layers represent the conduction band minimum of the I layer for up- and down- spin electrons, respectively. Here the other device properties are the same as that described in Fig. 15(a). Figure 16(b) shows a 2-D contour map of the MR ratio plotted against $V_{total}^P$ and $t_I$, where a bar on the right-hand side is the correspondence between the color and MR ratio (0 − 16%). The MR ratio has higher values in the region of $V_{total}^P < 0.5$ V and $t_I < 0.6$ nm, which is a similar trend in Fig. 16(a). Furthermore, the MR ratio is much enhanced in the most region compared with that in Fig. 16(a), particularly at $t_I < 0.5$ nm, and its maximum value reaches 16%. The improvement of the MR ratio is more clearly verified when MR ratio vs. $V_{total}^P$ is plotted for $t_I = 1$ nm, as shown in Fig. 18(a), where brown open dots and red open rectangles are extracted from Figs. 16(a) and (b), respectively. Figure 18(a) reveals that the red open rectangles are considerably higher than the brown open dots in the entire $V_{total}^P$ range, whereas the peak positions for both cases are ~2.2 V.

To confirm the scenario of Fig. 17(b), the first term $P_{det}^{AP}(V_2^{AP})\left(P_{inj}^{AP}(V_1^{AP}) + P_{inj}^{AP}(V_2^{AP})\right)$ in Eq. (20) is plotted as a function of $V_{total}^P$ in Fig. 18(b), where pink open rectangles represent the values for a two-terminal device of Fig. 17(c) with spin-filter I layers and green open dots are the same as those in Fig. 15(c) with normal I layers. At $V_{total}^P = 0$, the first term for the spin-filter I layers is larger than that for the normal I layers, as expected from the V-F model. Hereafter, the features for the spin-filter I layers are discussed otherwise noted. As $V_{total}^P$ is increased, the first term steeply increases to the maximum of ~1.5 at $V_{total}^P = $ ~0.4 V that corresponds to the peak position of $P_{det}$ at $V_{JL} = −0.2$ V in Fig. 13(a) and the band diagram in Fig. 13(c). Thus, the series connection of the spin-filter I layer and Fe strengthens the spin selectivity of the junction, which leads to such significant increase. As $V_{total}^P$ is increased further, the first term has the maximum value until ~1.2 V, and then decreases. Suppressing SAS in the higher $V_{total}^P$ range is directly supported by the fact that the $V_{total}^P$ position where the first term starts to decrease is higher than that for the normal I layer. On the other hand, the MR ratio for the spin-filter I layer has the peak at the $V_{total}^P$ position similar to that for the normal I layer in Fig. 18(a), although the first terms in both cases have different characteristics in Fig. 18(b). The MR ratio is mainly determined by the product of the first term and $r_{sr}/R_{total}^P$ in Eq. (20) and the $r_{sr}/R_{total}^P$ vs. $V_{total}^P$ relationship shown by the black open dots in Fig. 15(c) is almost unchanged by introducing $\Delta_{SF} = 0.2$ eV since $\Phi_0 = 1.15$ eV is significantly higher than 0.2 eV and it dominantly determines $r_{sr}/R_{total}^P$. Hence, the $r_{sr}/R_{total}^P$ vs. $V_{total}^P$ relationship is the main factor for the $V_{total}^P$ position of the peak MR ratio under the condition examined here. The validity of this interpretation is verified in Fig. 16(b): The $V_{total}^P$ position of the peak MR ratio shifts to lower value as $t_I$ is decreased ($r_{sr}/R_{total}^P$ rises more sharply against $V_{total}^P$). Furthermore, according to such $t_I$ change, the peak MR ratio is enhanced particularly at $t_I < 0.6$ nm, since the maximum $P_{det}$ value at $V_{total}^P \sim 0.4$ V greatly contributes to the MR ratio.

The effect of spin-filter I layers on $P_{det}$ was further studied when $\Delta_{SF}$ is varied as well as when the Fe layer is changed to a nonmagnetic layer by setting $\Delta = 0$ (Sec. S6 in S.M.). Indeed, a spin-filter I layer without Fe layer enhances $P_{det}$ and suppresses SAS in the higher $V_{total}^P$ range, and the combination of a spin-filter I layer with an Fe layer is quite effective to achieve superior characteristics. Therefore, Si-based tunnel junctions with a spin-filter I layer and a ferromagnetic electrode for both source and drain contacts are very promising to realize spin transport devices with high MR ratios in a broad bias range.

## VI. Conclusion

We experimentally and theoretically studied the spin injection and detection in a ferromagnetic Fe/amorphous-MgO/SiO$_2$/$n^+$-Si tunnel junction, particularly clarified how spin injection/detection polarizations are achieved as well as how they change in a higher bias range where the nonlinear electrical property appears. Our theoretical model based on the band diagram of the junction and the direct tunneling formula can explain the experimentally-estimated spin injection/detection polarizations in a wide bias range, and it is consistent with the widely-used model proposed by Valet and Fert in the linear bias range.

Our other findings are summarized as follows: First, the bias-dependent features of the spin polarizations originate from the band diagrams of the Fe layer and $n^+$-Si in the junction, which means that a spin-selective insulator layer is not always necessary. Second, as the junction bias voltage is negatively increased (the spin extraction geometry), the spin accumulation becomes saturated and the spin current and spin polarizations are significantly reduced. This phenomenon named "spin accumulation saturation (SAS)" is inevitable in semiconductor-based ferromagnetic tunnel junctions. Third, when the same junction characteristics extracted from the experiments are used, the numerical study in the two-terminal device structure demonstrates that the MR ratio increases in the lower range and decreases in the higher range as the total bias voltage is increased. The latter feature mainly comes from SAS in the drain junction that has the spin extraction geometry, and it can be significant when the tunnel barrier is extremely thin. The further numerical study demonstrates that using spin-selective insulating layers with spin-split barrier heights for up- and down-spin electrons, instead of normal tunnel barriers like amorphous MgO, at the source and drain contacts considerably enhances the MR ratio through the improvement of the drain junction characteristics.

Semiconductor-based spin transport devices with ferromagnetic tunnel junctions require the application of reasonably high biases. Thus, our model can be a versatile tool to analyze as well as to optimize the spin-transport properties of such devices aiming at practical applications with high MR ratios.

**Acknowledgments**
This work was partly supported by Grants-in-Aid for Scientific Research (20H05650, 23K17324), CREST Program (JPMJCR1777) of Japan Science and Technology Agency, and Spintronics Research Network of Japan (Spin-RNJ). B.Y. thanks the financial support from the MERIT-WINGS Program at the University of Tokyo.

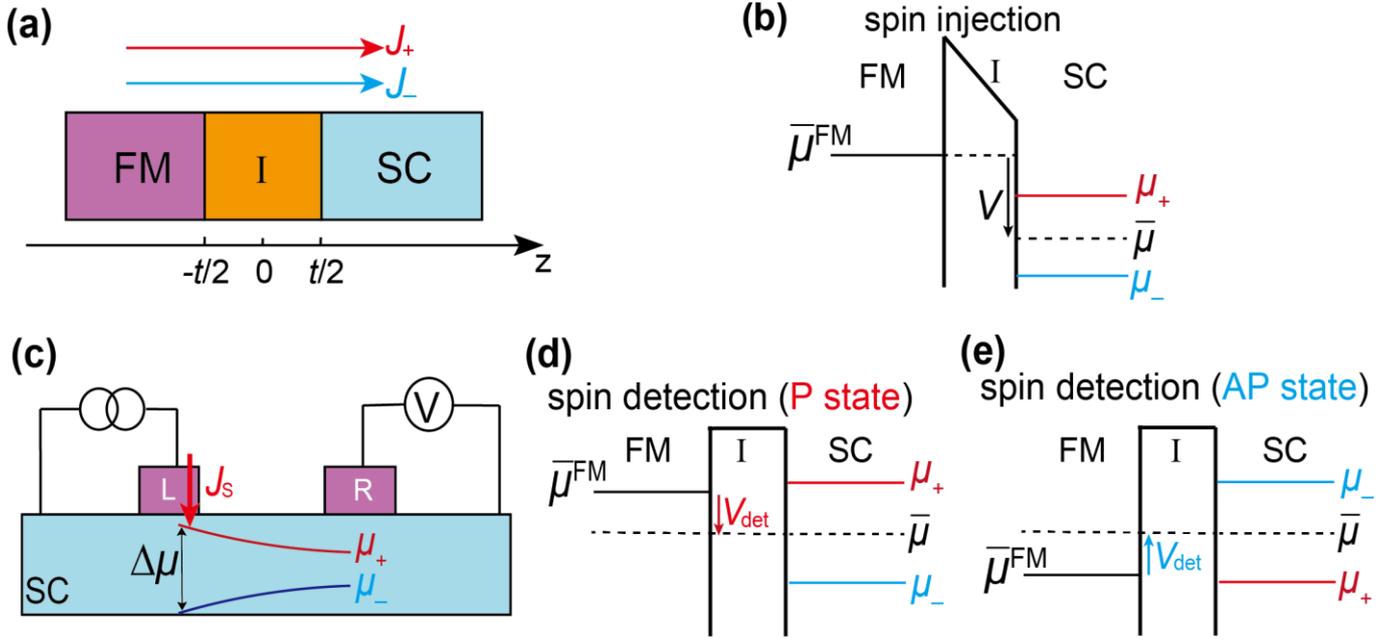

Fig. 1. (a) Semi-infinite one-dimensional FM/I/SC structure, where FM is a ferromagnetic metal, SC is a semiconductor and I is an insulator with a thickness of $t$. The origin of the $z$ axis is located at the center of the I layer. The red and blue arrows represent up-spin current ($J_+$) and down-spin current ($J_-$), respectively. (b) Schematic band diagram of the FM/I/SC under a spin injection bias $V > 0$, where black solid line in the FM represents the Fermi level ($\bar{\mu}^{FM}$), black dashed line represents the averaged Fermi level of the SC ($\bar{\mu}$), and red and blue solid lines represent the up-spin and down-spin electrochemical potentials ($\mu_+$ and $\mu_-$), respectively. (c) Schematic illustration of a spin transport geometry through a SC channel, where the L and R electrodes are ferromagnetic tunnel junctions for spin injection and detection, respectively, and red and blue solid curves represent the up-spin and down-spin electrochemical potentials ($\mu_+$ and $\mu_-$), respectively. The spin accumulation ($\Delta\mu = \mu_+ - \mu_-$) is induced beneath the L electrode by the injection of a constant spin current $J_S$ ($= J_+ - J_-$), and it is transported through the SC and detected at the R electrode by a voltage meter. (d)(e) Schematic band diagrams of the FM/I/SC junction (R junction) during spin detection in the parallel (P) and anti-parallel (AP) magnetization states, respectively, where the junction is under a zero-current-bias condition, $V_{det}$ is the spin detection voltage drop, $\bar{\mu}^{FM}$, $\bar{\mu}$, $\mu_+$, and $\mu_-$ are the same symbols as those in (b).

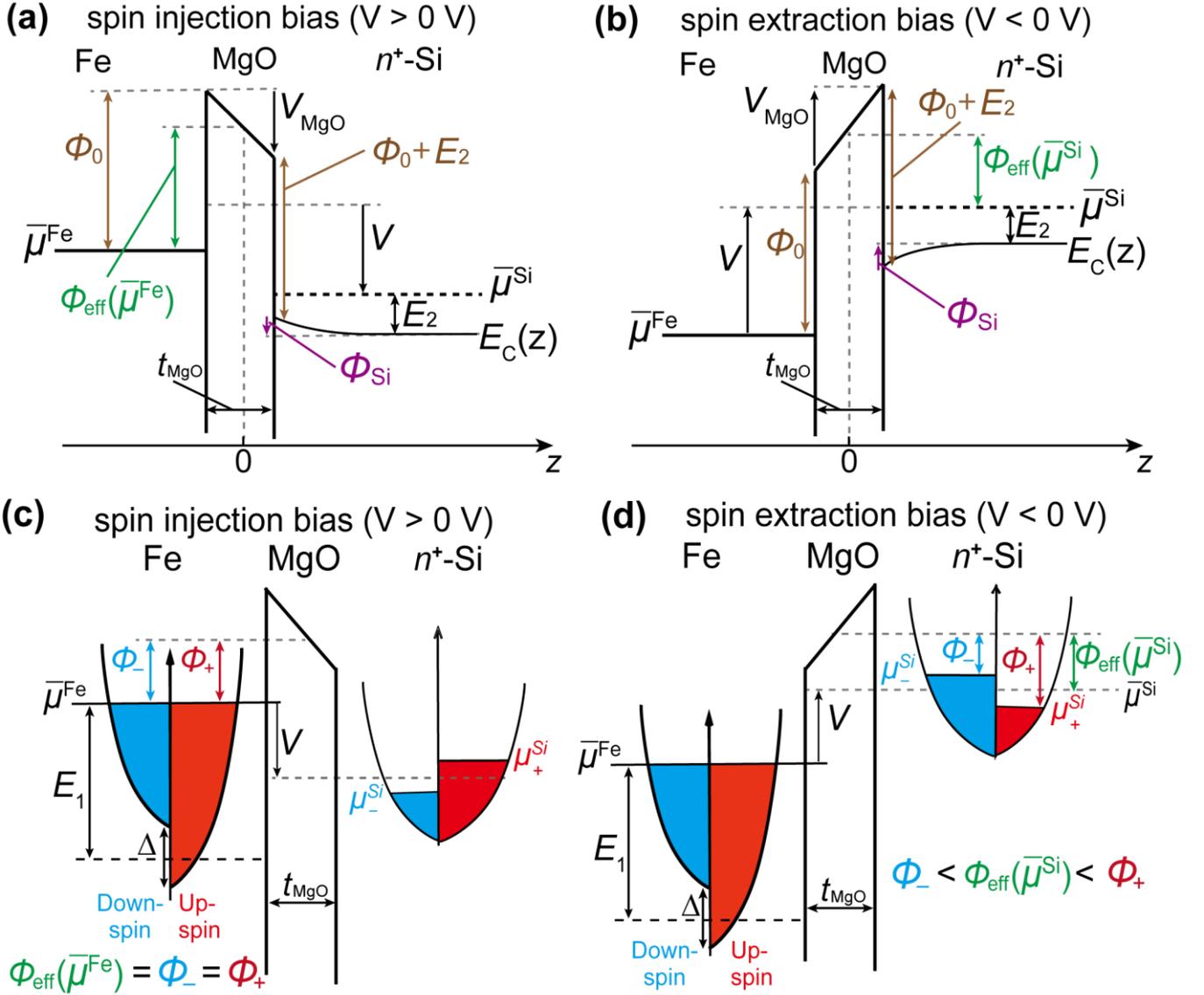

Fig. 2 (a)(b) Semi-infinite one-dimensional electron energy band diagrams of a Fe/MgO/$n^+$-Si junction along the $z$-axis in the spin injection ($V > 0$) and extraction ($V < 0$) geometries, respectively, where the origin of the $z$ axis is located at the center of the I layer, $\bar{\mu}^{Fe}$ and $\bar{\mu}^{Si}$ are the Fermi levels of the Fe and $n^+$-Si, respectively, $E_C(z)$ is the Si conduction band minimum, $E_2$ is defined by the energy difference between $\bar{\mu}^{Si}$ and $E_C(z=+\infty)$, $V$ is the applied junction voltage drop that equals to the difference between $\bar{\mu}^{Fe}$ and $\bar{\mu}^{Si}$, $V_{MgO}$ is the voltage drop of the MgO layer, $\Phi_{Si}$ is the band bending at the Si surface. In this paper, the downward and upward arrows represent the positive and negative polarities of the voltage, respectively. The MgO has a barrier height of 1.15 eV (flat band voltage is set at $V = 0$). The green double-headed arrows represent the effective barrier heights ($\Phi_{eff}$) for the electrons at $\bar{\mu}^{Fe}$ in (a) and at $\bar{\mu}^{Si}$ in (b), respectively. (c)(d) Schematic illustrations of the electron distributions and density of states in the spin injection ($V > 0$) and extraction ($V < 0$) geometries, respectively, where $\mu_+^{Si}$ and $\mu_-^{Si}$ are the Si electrochemical potentials for up-spin and down-spin electrons, respectively, $\bar{\mu}^{Fe}$ is the Fermi level of the Fe, $\Delta$ is the spin split energy of Fe, and $E_1$ is defined by the difference between $\bar{\mu}^{Fe}$ and the energy level in the middle of $\Delta$. The red and blue colored regions represent the occupied states for up-spin and down-spin electrons, respectively. The red and blue double-headed arrows represent the effective barrier heights for up-spin ($\Phi_+$) and down-spin ($\Phi_-$) electrons, respectively.

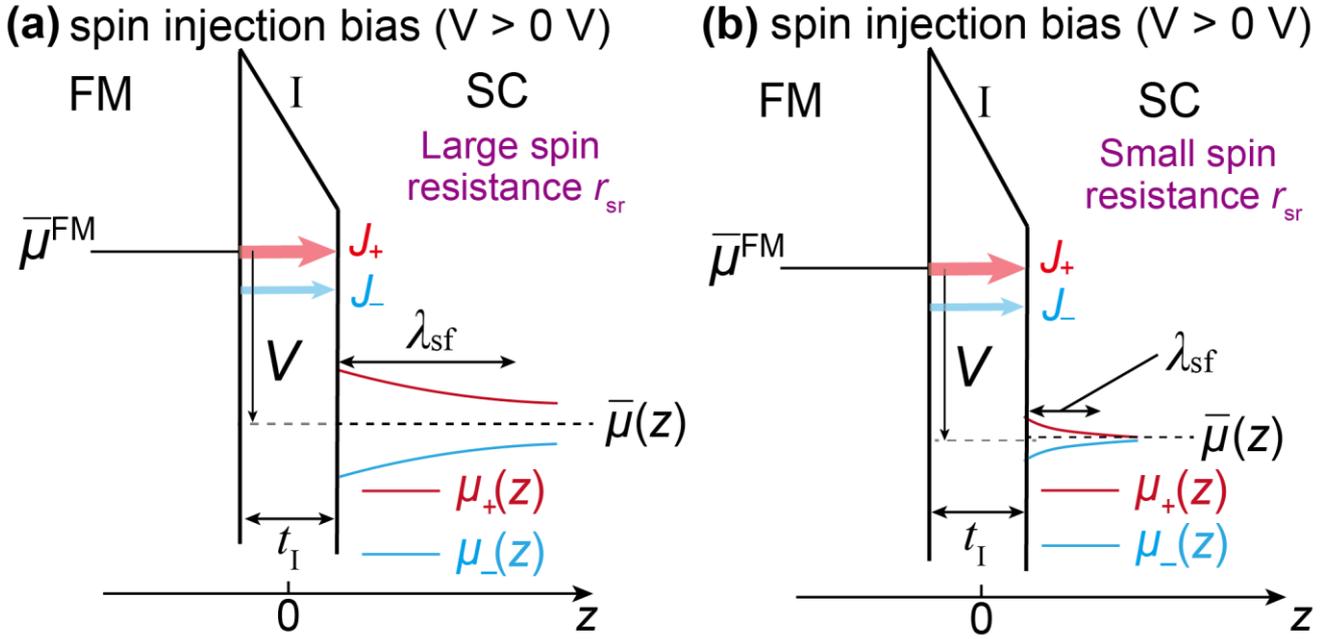

Fig.3 (a)(b) Schematic band diagrams in the spin injection geometry ($V > 0$) with (a) large and (b) small $r_{sr}$ values, where $V$, $J_+$, and $J_-$ are identical with each other, $t_I$ is the thickness of the I layer, $\lambda_{sf}$ is the spin diffusion length in SC, black solid line in FM represents the Fermi level ($\bar{\mu}^{FM}$), black dashed line represents the averaged Fermi level in the SC ($\bar{\mu}$), and red and blue solid lines represent the up-spin and down-spin electrochemical potentials ($\mu_+$ and $\mu_-$), respectively.

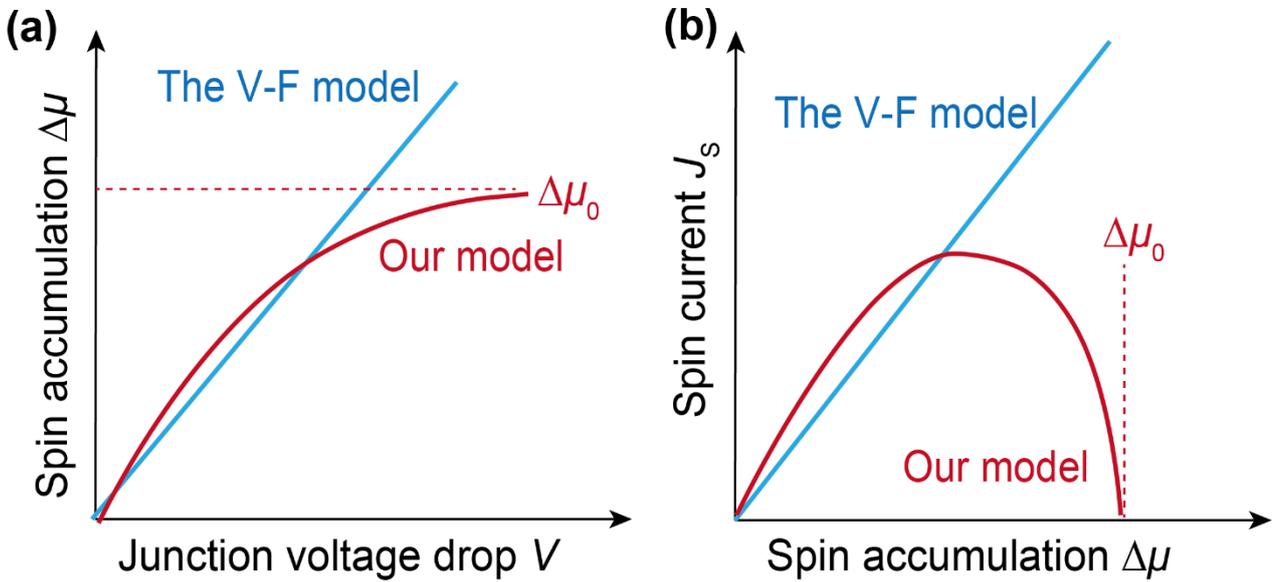

Fig.4 Schematic illustrations of (a) $\Delta\mu$–$V$ and (b) $J_S$–$\Delta\mu$ relationships in the spin extraction geometry, where red and blue curves are the relationships derived from our model and the V-F model, respectively, and $\Delta\mu_0$ represents a specific saturated value of $\Delta\mu$.

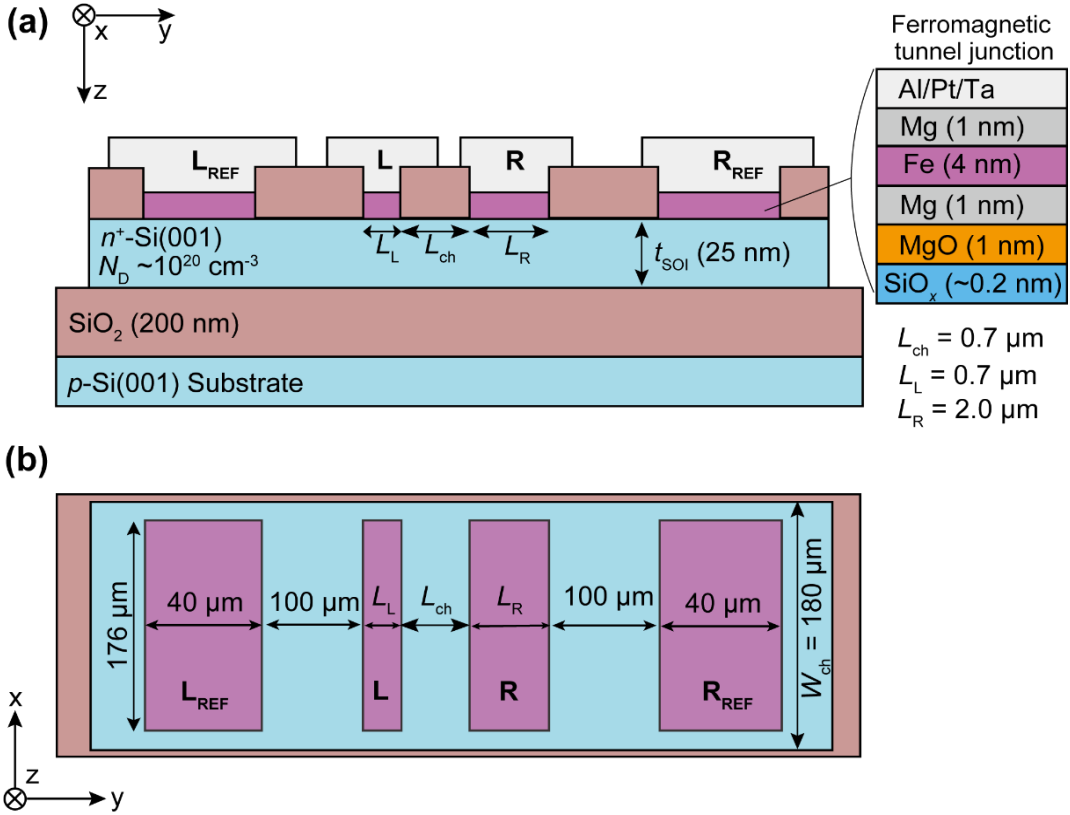

Fig. 5 (a) Cross-sectional and (b) top views of the four-terminal spin transport device structure having a 25-nm-thick $n^+$-Si channel and Fe(4 nm)/Mg(1 nm)/MgO(1 nm)/SiO$_x$(~0.2 nm)/$n^+$-Si junctions prepared on a silicon-on-insulator (SOI) substrate, where the phosphorus doping concentration of the $n^+$-Si channel is ~1×10$^{20}$ cm$^{-3}$, and the thickness of the buried oxide (BOX) SiO$_2$ layer is 200 nm. The Cartesian coordinate system is defined as follows: $x$ and $y$ are parallel to the longitudinal and transverse directions of the L/R electrode, respectively, and $z$ is normal to the substrate plane. The channel length between the L and R electrodes along the $y$ direction is $L_{ch}$ = 0.7 μm, the channel width along the $x$ direction is $W_{ch}$ = 180 μm, and the short-side lengths along the $y$ direction of the L and R electrodes are $L_L$ = 0.7 μm and $L_R$ = 2.0 μm, respectively. The L$_{REF}$ and R$_{REF}$ electrodes are located at ~100 μm away from the L and R electrodes, respectively, and their short-side lengths along the y direction is 40 μm.

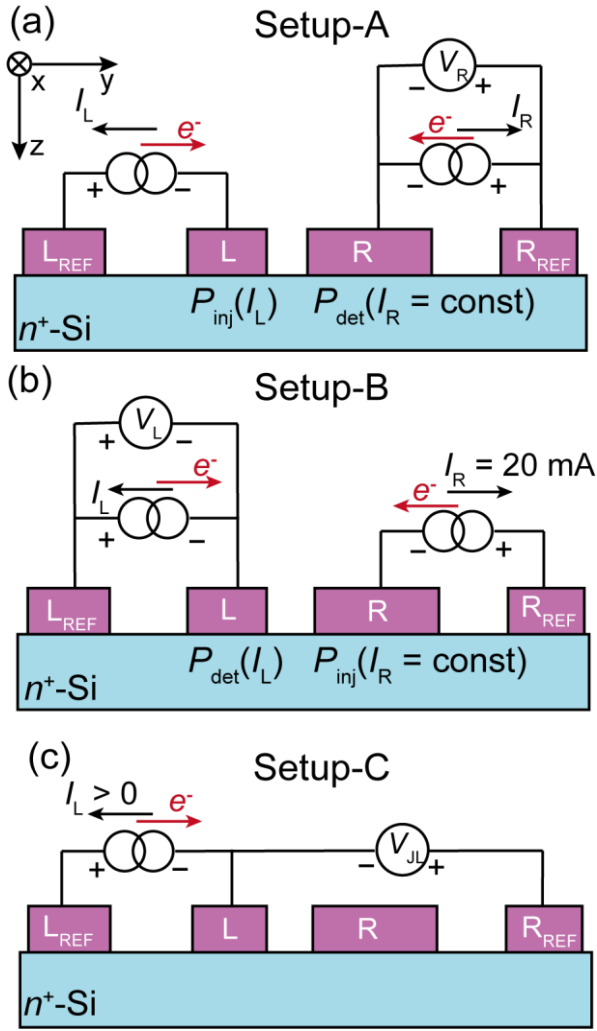

Fig. 6 (a) Spin-valve measurement Setup-A, where the L and R electrodes are the spin injector and detector, respectively, and the polarities of the applied currents ($I_L$ and $I_R$) and voltage ($V_R$) are defined. The voltage change $\Delta V_R$ is measured at 4 K under a constant $I_R$ (0 or −1.5 mA) and various $I_L$ (from −20 to 20 mA) while an external magnetic field (−200-200 Oe) is applied parallel to the $x$ axis. (b) Spin-valve measurement Setup-B, where the L and R electrodes are spin detector and injector, respectively, the voltage change $\Delta V_L$ is measured at 4 K under a constant $I_R$ (20 mA) and various $I_L$ (from −20 to 20 mA) while an external magnetic field (−200-200 Oe) is applied parallel to the $x$ axis. (c) Voltage − current ($V_{JL}$–$I_L$) characteristics measurement Setup-C, where the L junction voltage drop $V_{JL}$ is measured at 4 K while applying various $I_L$ (from −20 to 20 mA).

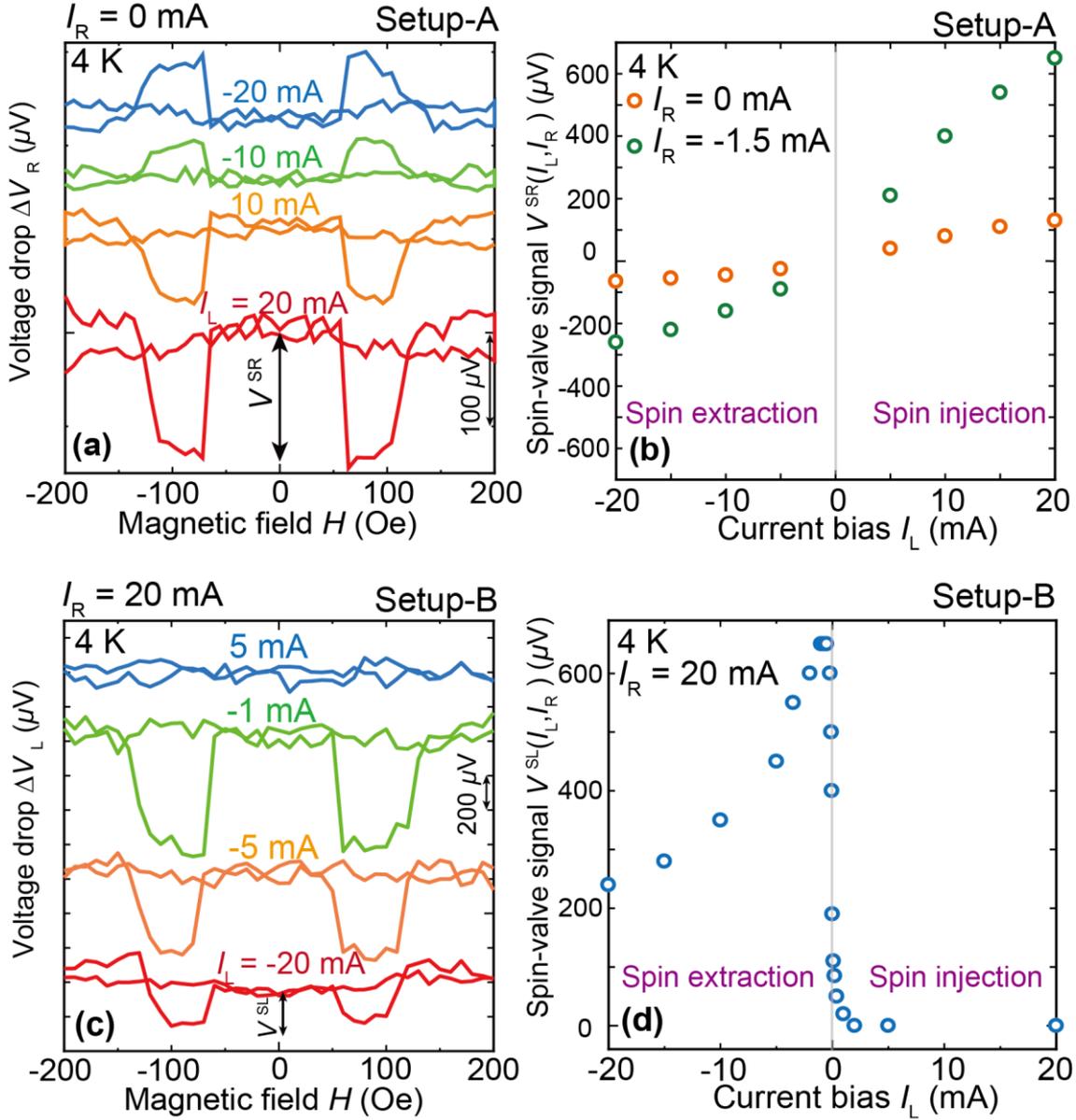

Fig. 7 (a) Voltage change $\Delta V_R$ measured at 4 K with $I_R = 0$ mA and various $I_L$ values in Setup-A, where blue, green, orange, and red curves are spin-valve signals measured with $I_L = -20, -10, 10$ and $20$ mA, respectively. The magnitude of the spin-valve signal $V^{SR}(I_L, I_R)$ is defined as the maximum voltage change of $\Delta V_R$ between antiparallel and parallel magnetization states. (b) The magnitudes of the spin-valve signals $V^{SR}(I_L, I_R)$ at 4 K plotted as a function of $I_L$ in Setup-A, where orange and green open dots represent the measurement conditions with $I_R = 0$ and $-1.5$ mA, respectively. (c) Voltage change $\Delta V_L$ measured at 4 K with $I_R = 20$ mA and various $I_L$ in Setup-B, where blue, green, orange, and red curves are spin-valve signals measured with $I_L = 5, -1, -5$, and $-20$ mA, respectively. The magnitude of the spin-valve signal $V^{SL}(I_L, I_R)$ is defined as the maximum voltage change of $\Delta V_L$ between antiparallel and parallel magnetization states. (d) The magnitudes of the spin-valve signals $V^{SL}(I_L, I_R)$ at 4 K plotted as a function of $I_L$ in Setup-B with $I_R = 20$ mA.

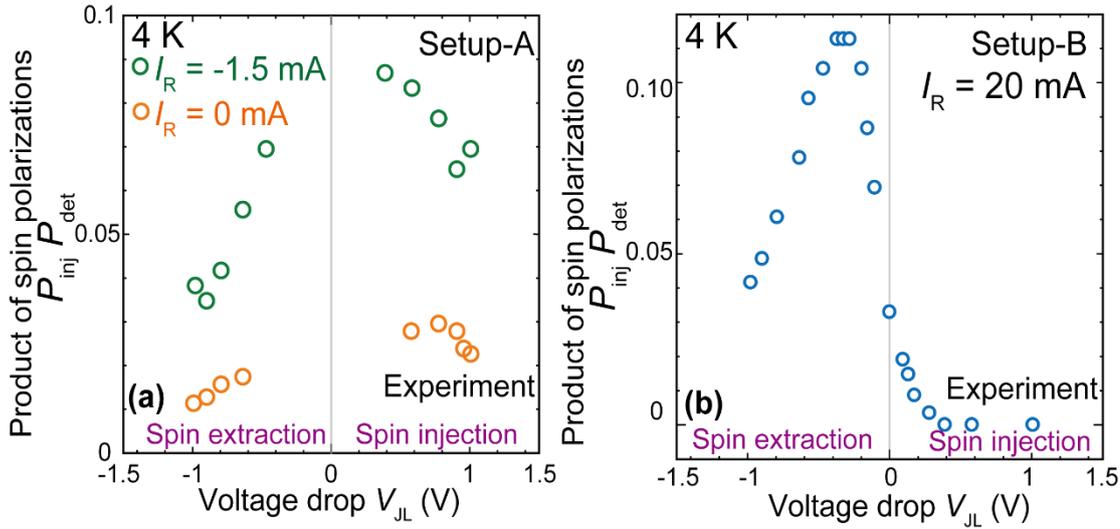

Fig. 8 (a) Product of the spin polarizations $P_{inj}P_{det}$ at 4 K plotted as the function of the L junction voltage drop $V_{JL}$ in Setup-A, where $P_{inj}P_{det}$ was estimated from $V^{SR}(I_L, I_R)$ using Eqs. (11) and (13), and $V_{JL}$ was transformed from $I_L$ by the $V_{JL}-I_L$ characteristics measured in Setup-C. The orange and green open dots represent the measurement conditions with $I_R = 0$ and $-1.5$ mA, respectively. (b) Product of the spin polarizations $P_{inj}P_{det}$ at 4 K plotted as a function of $V_{JL}$ in Setup-B, where $P_{inj}P_{det}$ was estimated from $V^{SL}(I_L, I_R)$ using Eqs. (12) and (13), and $V_{JL}$ was transformed from $I_L$ by the $V_{JL}-I_L$ characteristics measured in Setup-C.

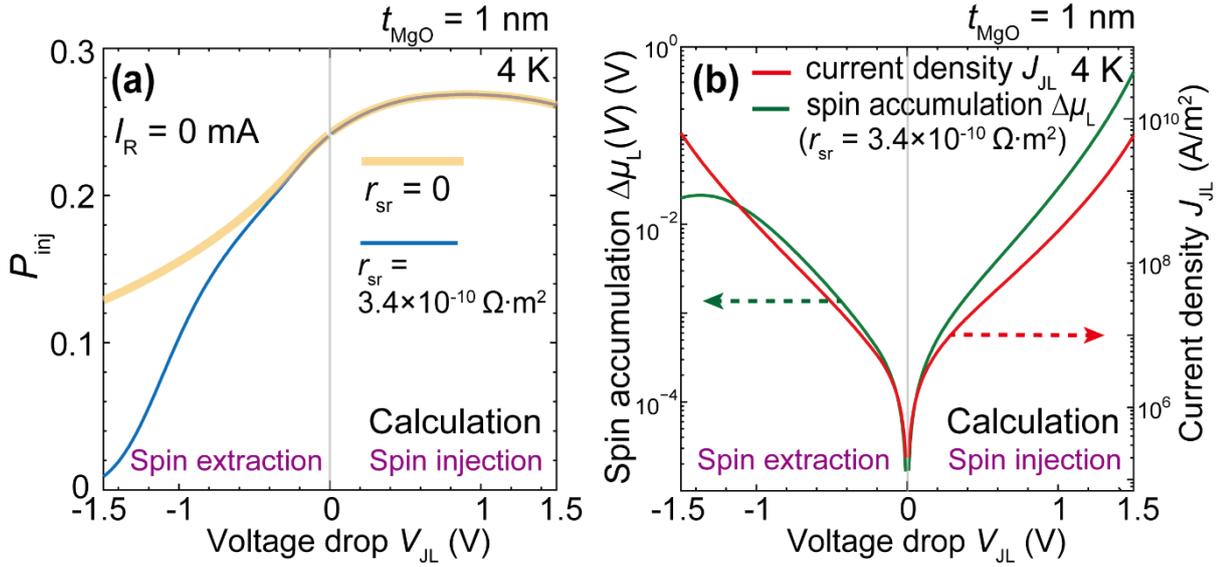

Fig. 9 (a) Calculated spin injection polarization $P_{inj}$ at 4 K plotted as a function of $V_{JL}$ with $I_R = 0$ mA, where yellow and blue curves are the results calculated with $t_{MgO} = 1$ nm under the Si spin resistances $r_{sr} = 0$ Ω·m² ($\Delta\mu_L = 0$) and $r_{sr} = 3.4 \times 10^{-10}$ Ω·m² ($\Delta\mu_L \neq 0$), respectively. (b) Calculated spin accumulation $\Delta\mu_L$ (left axis) beneath the L junction and calculated L junction current density $J_{JL}$ (right axis) plotted as a function of $V_{JL}$ with $t_{MgO} = 1$ nm when $I_R = 0$ mA at 4 K.

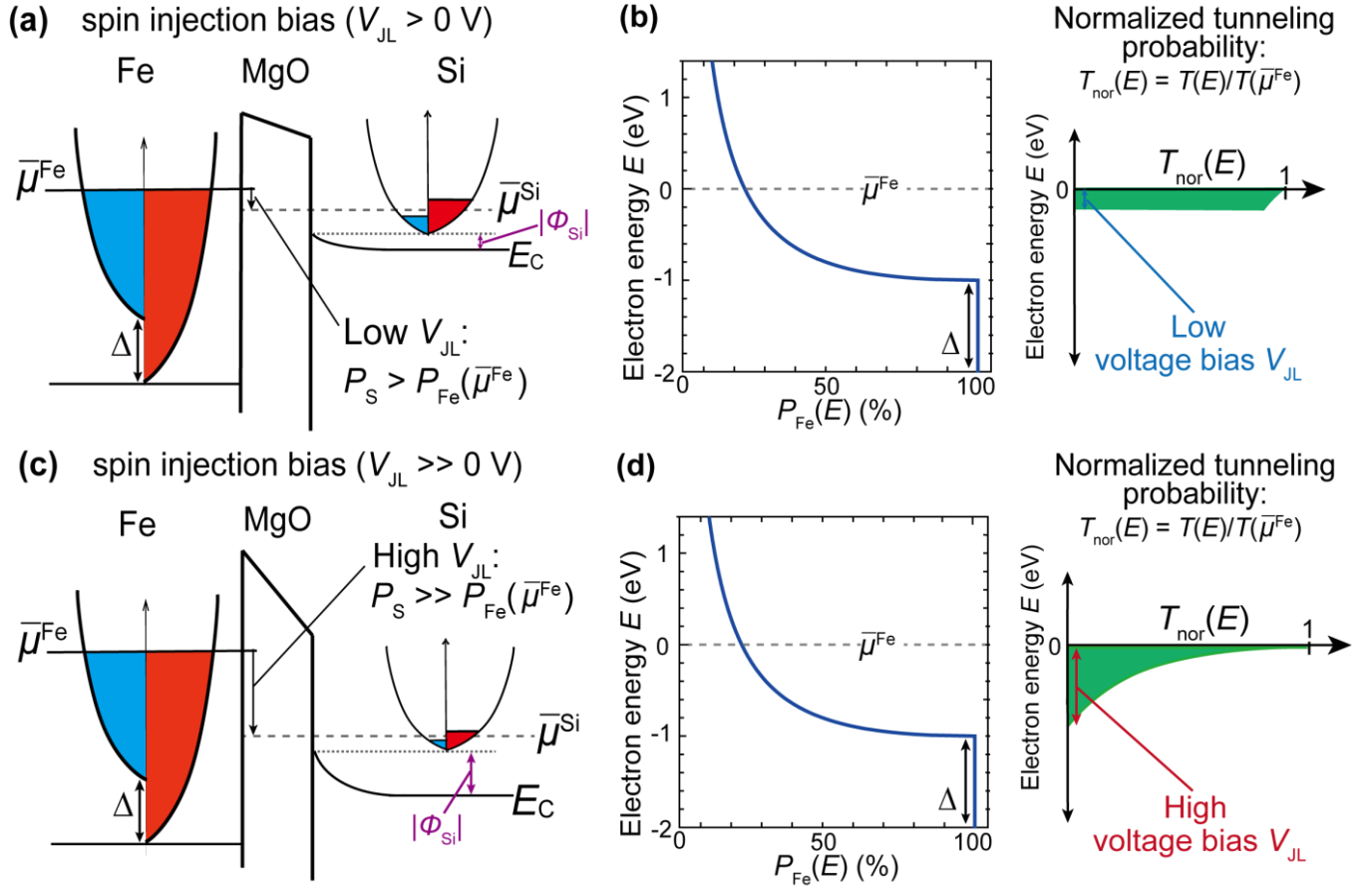

Fig. 10 (a)(c) One-dimensional energy band diagrams of a Fe/MgO/$n^+$-Si junction in the spin injection geometry with low ($V_{JL} > 0$) and high ($V_{JL} \gg 0$) junction voltage drops, respectively where $\bar{\mu}^{Fe}$ and $\bar{\mu}^{Si}$ are the Fermi levels of the Fe and $n^+$-Si, respectively, $E_C$ is the Si conduction band minimum, and red and blue regions represent the occupied up-spin and down-spin electrons, respectively. (b)(d) Left side: Calculated Fe band spin polarization $P_{Fe}(E)$ plotted as a function of electron energy $E$, where $\bar{\mu}^{Fe}$ is the Fermi level of Fe that is defined as the zero potential. Right side:

Normalized tunneling probability $T_{\text{nor}}(E) = T(E) / T(\bar{\mu}^{Fe})$ schematically plotted as a function of $E$ at the same voltage bias as (a) and (c), respectively.

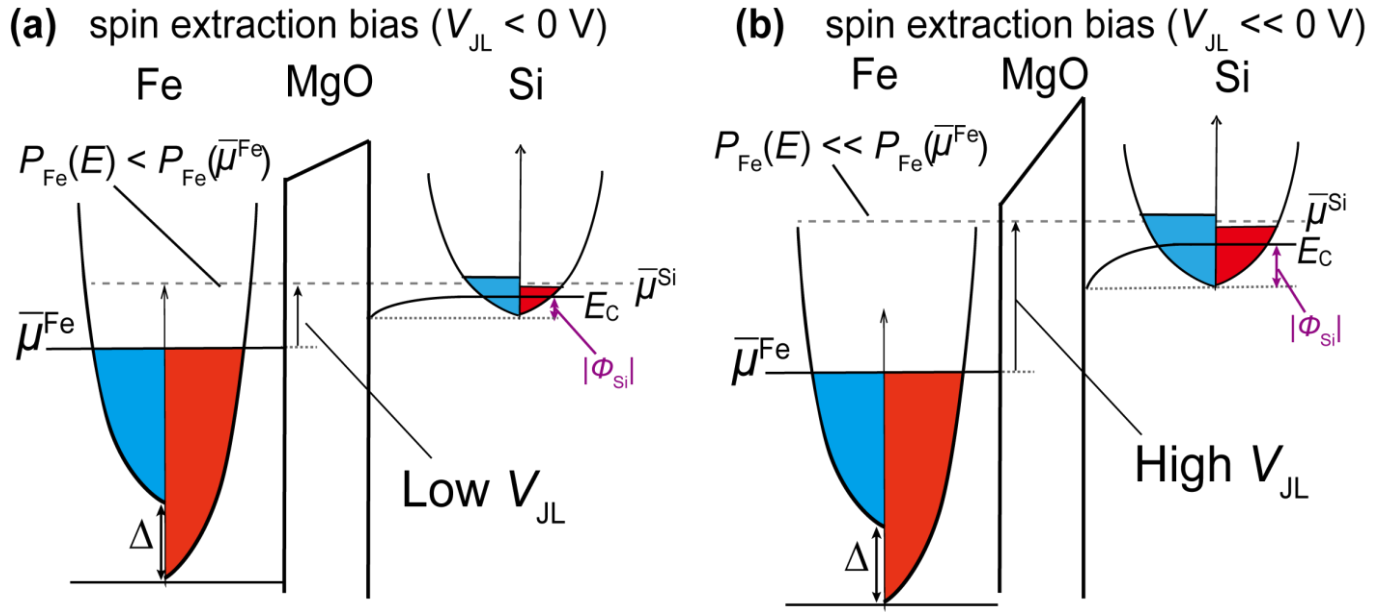

Fig. 11(a)(b) One-dimensional energy band diagrams of a Fe/MgO/$n^+$-Si junction in the spin extraction geometry with a low ($V_{JL} < 0$) and a high ($V_{JL} \ll 0$) junction voltage drop, respectively.

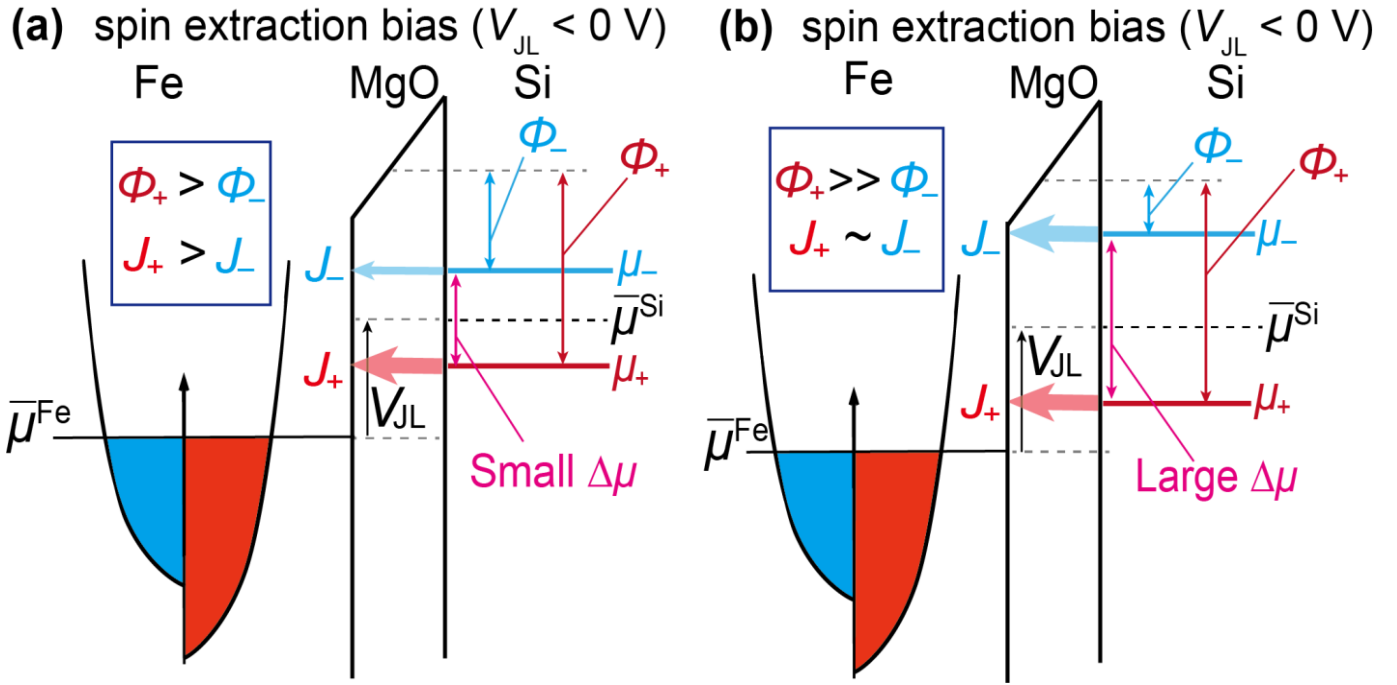

Fig.12 (a)(b) Schematic band diagrams in the spin extraction geometry ($V_{JL} < 0$) with (a) small and (b) large $\Delta\mu$, where the black solid line in Fe represents the Fermi level ($\bar{\mu}^{Fe}$), black dashed line represents the averaged Fermi level of Si ($\bar{\mu}^{Si}$), $V_{JL}$ is identical with each other, $\Phi_+$ and $\Phi_-$ are the effective barrier heights for up-spin and down-spin electrons, respectively, and $J_+$ and $J_-$ are the up-spin and down-spin electron current densities, respectively. The widths of the light red and light blue arrows in the MgO layer express the magnitudes of $J_+$ and $J_-$, respectively.

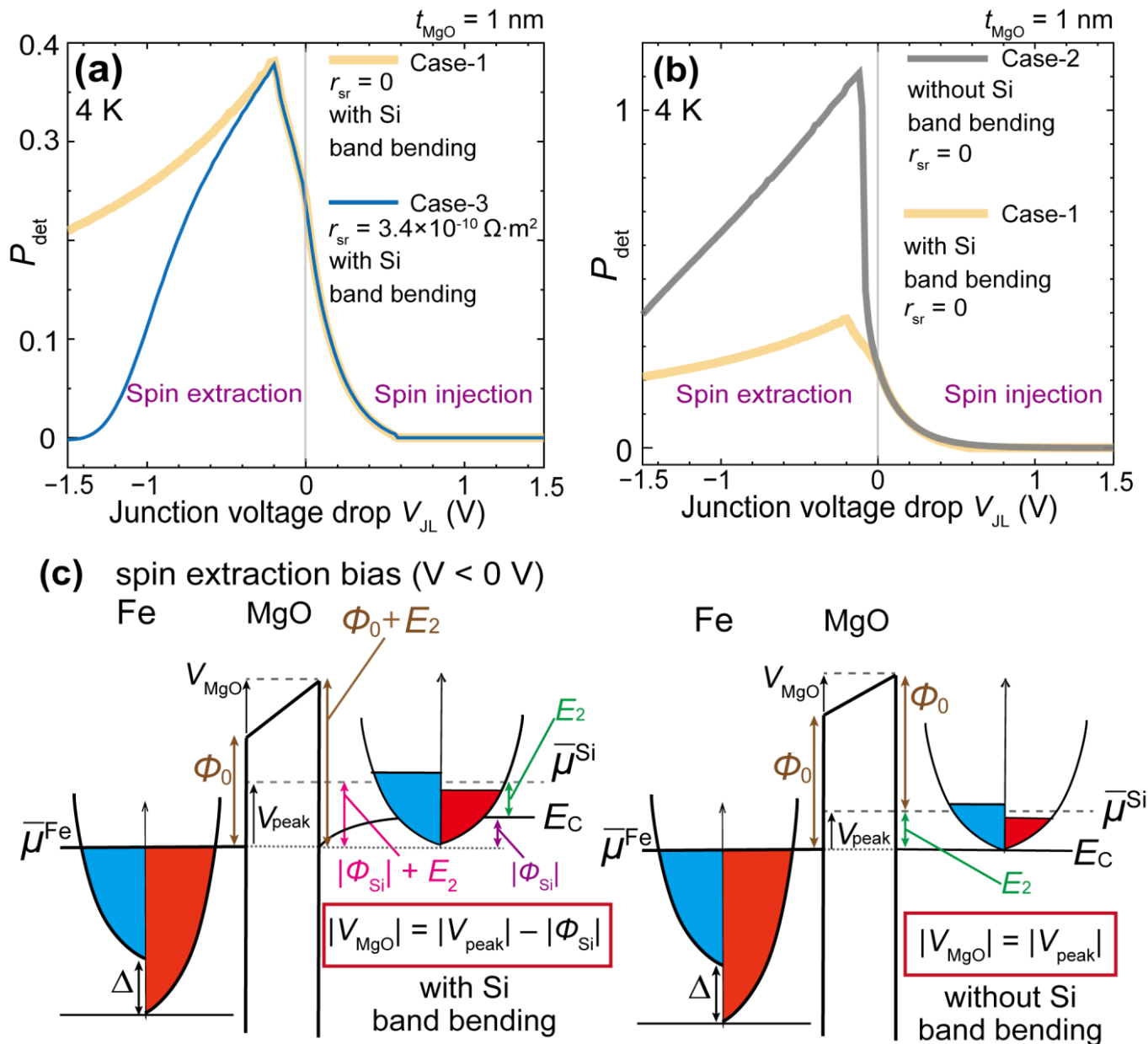

Fig. 13 (a) Calculated $P_{det}$ at 4 K plotted as functions of the junction voltage drop $V_{JL}$ with the Si band bending, where yellow and blue lines are calculated with $t_{MgO}$ = 1nm under the Si spin resistances $r_{sr}$ = 0 and $r_{sr}$ = 3.4×10$^{-10}$ Ω·m$^2$, respectively. (b) Calculated $P_{det}$ at 4 K plotted as functions of $V_{JL}$ with $t_{MgO}$ = 1nm when $r_{sr}$ = 0, where yellow and grey lines are the results calculated with and without the Si bend bending, respectively. (c) Schematic band diagrams of a Fe/MgO/$n^+$-Si junction in the spin extraction geometry when $P_{det}$ reaches its maximum at $V = V_{peak}$ for the cases (left) with considering the Si band bending and (right) without considering the Si band bending, where the Fermi level of Fe $\bar{\mu}^{Fe}$ is matched with the Si conduction band minimum $E_C$.

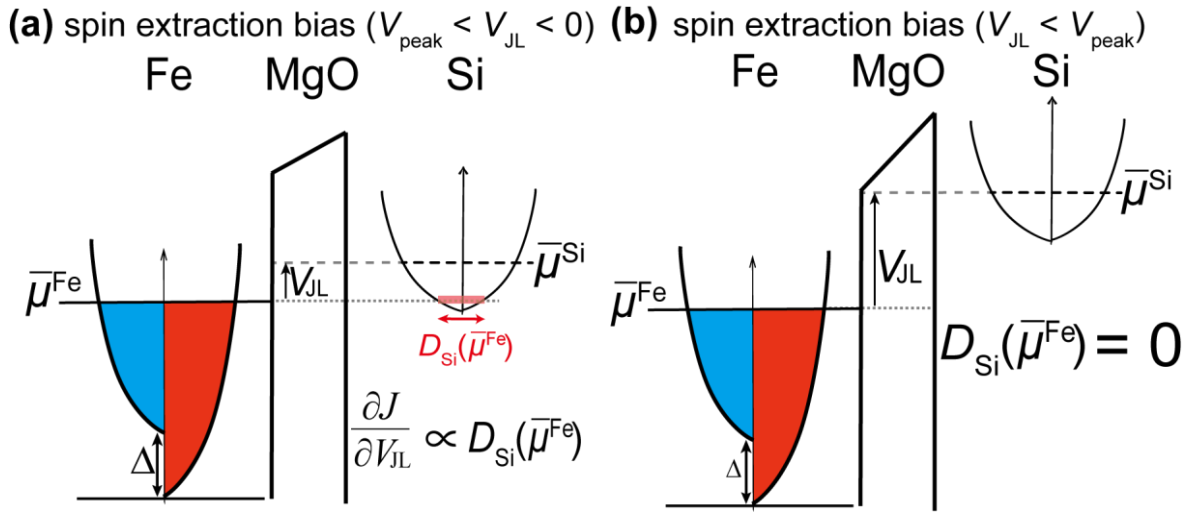

Fig. 14 (a) One-dimensional energy band diagram of a Fe/MgO/$n^+$-Si junction in the spin extraction geometry in $V_{\text{peak}} < V_{\text{JL}} < 0$, where the length of the red double arrow represents the magnitude of $D_{\text{Si}}(\bar{\mu}^{Fe})$ that is the Si density of states at $\bar{\mu}^{Fe}$, and $\frac{\partial J|_{\Delta\mu_L=0}}{\partial V_{JL}}$ is proportional to $D_{\text{Si}}(\bar{\mu}^{Fe})$ in this bias range. (b) One-dimensional energy band diagram of a Fe/MgO/$n^+$-Si junction in the spin extraction geometry in $V_{\text{JL}} < V_{\text{peak}}$, where $D_{\text{Si}}(\bar{\mu}^{Fe}) = 0$ holds in this bias range.

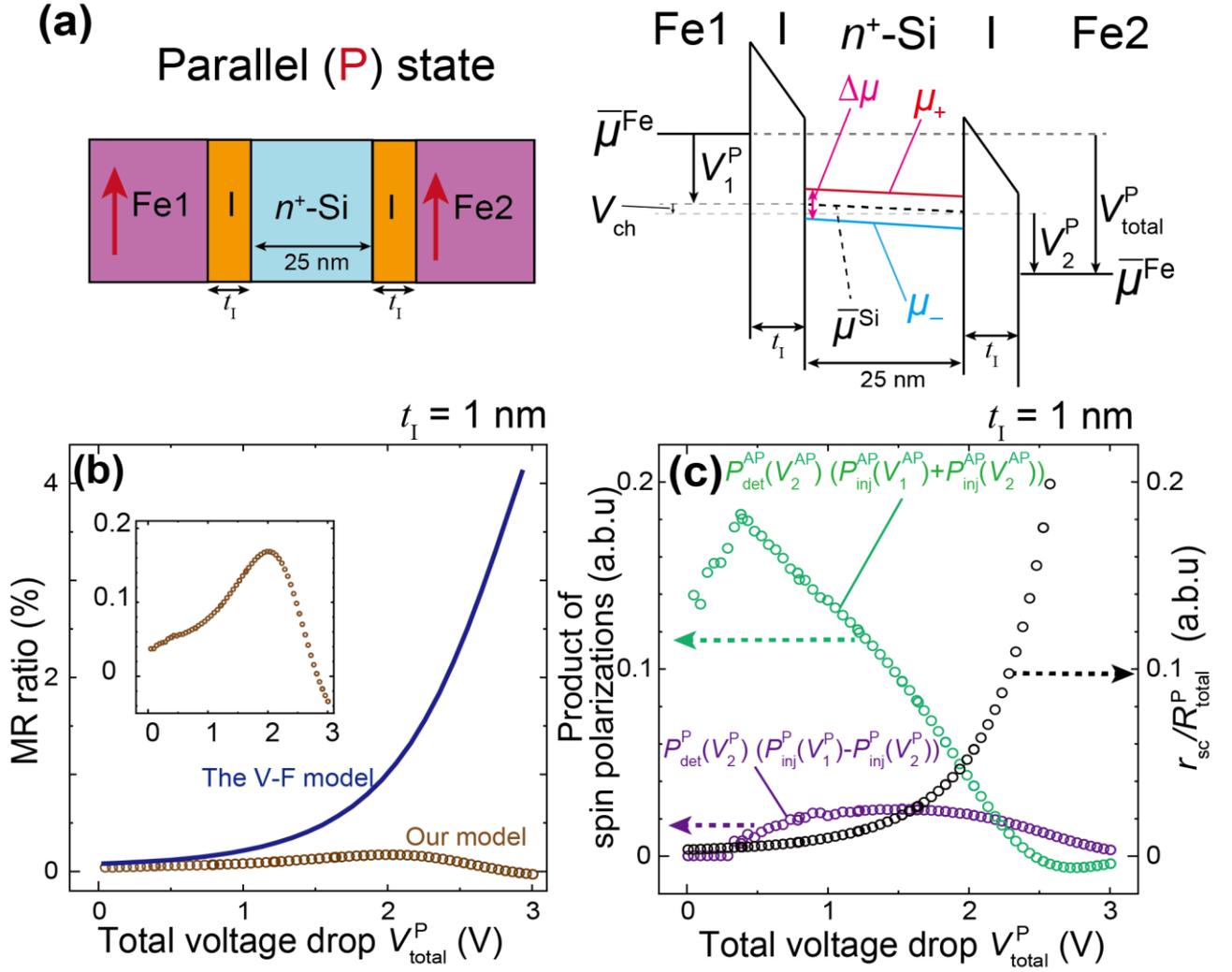

Fig.15 (a) Schematic cross-sectional view of a two-terminal spin transport device structure (left) and its one-dimensional electron energy band diagram (right) in the parallel (P) magnetization state, where a $n^+$-Si channel has the length of 25 nm, Fe1/I/Si tunnel junction (JCT1) and Fe2/I/Si tunnel junction (JCT2) have the same properties as those described in Sec. II-C, $\bar{\mu}^{Fe}$ and $\bar{\mu}^{Si}$ are the Fermi levels of the Fe and $n^+$-Si, respectively, $\mu_+^{Si}$ and $\mu_-^{Si}$ are the electrochemical potentials in Si for up-spin and down-spin electrons, respectively, $V_{total}^P$ is the total voltage drop of the device, $V_1^P$, $V_2^P$, and $V_{ch}$ are the voltage drops in JCT1, JCT2, and the Si channel, respectively. (b) Numerically calculated MR ratios plotted as functions of the total device voltage drop $V_{total}^P$ with $t_I = 1$ nm at 4 K, where brown open dots and blue curve are calculated by our model and the V-F model, respectively. The inset is a magnified view of the MR ratio curve calculated by our model. (c) Numerically calculated $P_{det}^{AP}(V_2^{AP})\left(P_{inj}^{AP}(V_1^{AP}) + P_{inj}^{AP}(V_2^{AP})\right)$ (green open dots), $P_{det}^P(V_2^P)\left(P_{inj}^P(V_1^P) - P_{inj}^P(V_2^P)\right)$ (purple open dots), and $r_{sr}/R_{total}^P$ (black open dots) are plotted as functions of $V_{total}^P$ with $t_I = 1$ nm at 4 K.

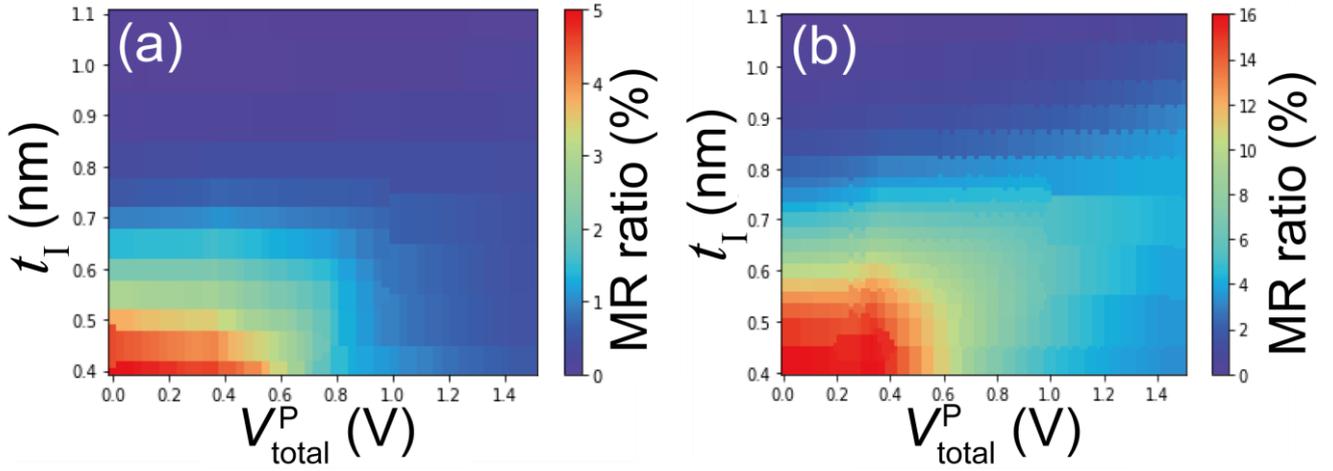

Fig. 16 (a)(b) Numerically calculated 2-D contour MR ratio map of a Fe1/I/$n^+$-Si/I/Fe2 structure plotted against the total device voltage drop $V_{total}^P$ and the I layer thickness $t_I$, where Fe1/I/Si tunnel junction (JCT1) and Fe2/I/Si tunnel junction (JCT2) have the same properties as those in Sec. II-C. The I layers are (a) normal I layer and (b) spin filter I layers with the spin-split energy $\Delta_{SF} = 0.2$ eV.

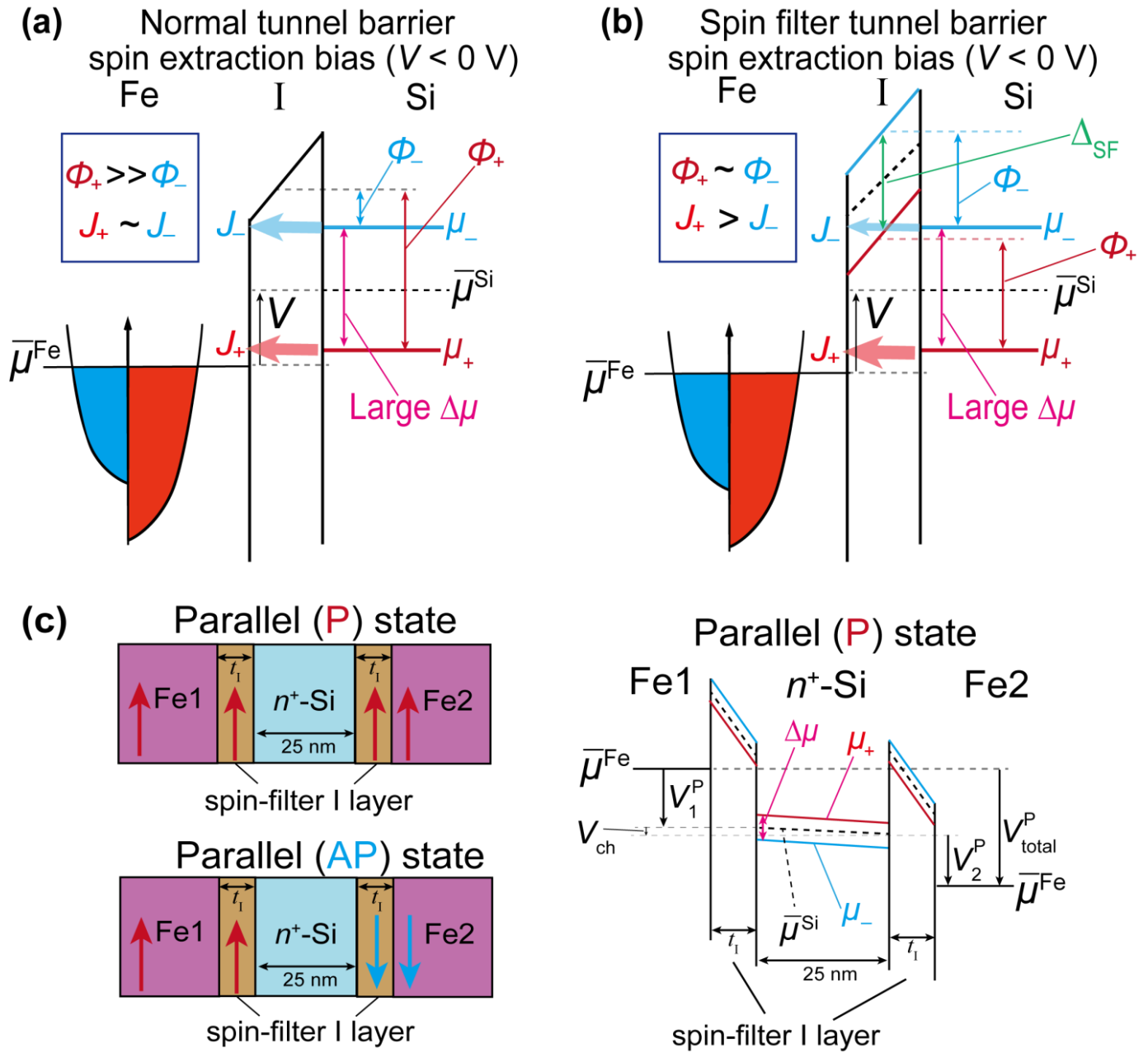

Fig.17 Schematic band diagrams in the spin extraction geometry ($V_{JL} < 0$) with (a) a normal I layer and (b) a spin-filter I layer, where $\Delta_{SF}$ is the spin split energy of the barrier with respect to the dotted line, $\Delta\mu$ and $V_{JL}$ are identical with each other, $\Phi_+$ and $\Phi_-$ are the effective barrier heights for up-spin and down-spin electrons, respectively, and $J_+$ and $J_-$ are the up-spin and down-spin electron current densities, respectively. The widths of the light red and light blue arrows in the I layer express the magnitudes of $J_+$ and $J_-$, respectively. (c) Schematic cross-sectional views of a two-terminal spin transport device structure with spin-filter I layers in the parallel (the upper side) and anti-parallel (the lower side) magnetization states. (d) Schematic one-dimensional electron energy band diagram of a two-terminal spin transport device structure with spin-filter I layers in the parallel (P) magnetization state.

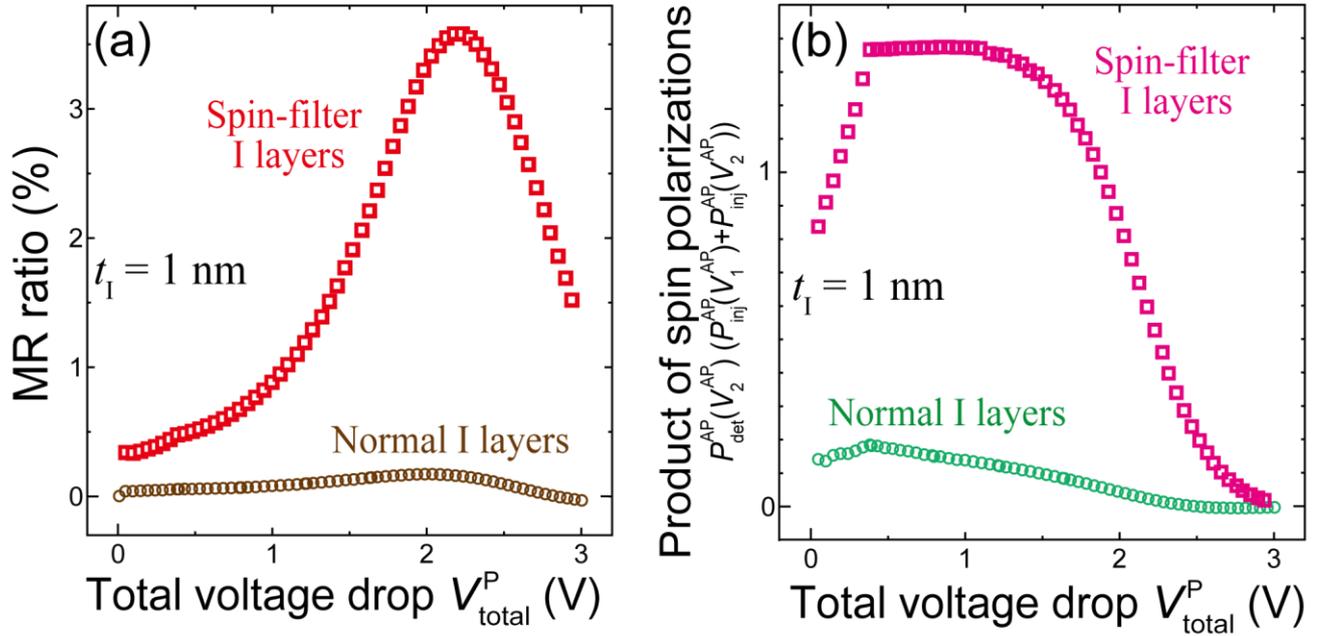

Fig. 18 (a) Numerically calculated MR ratios of two-terminal devices plotted as functions of $V_{\text{total}}^{\text{P}}$ when $t_{\text{I}} = 1$ nm, where red open rectangles and brown open dots are calculated with spin-filter I layers (see Fig.17 (c)) and normal I layers (see Fig. 15(a)), respectively. (b) Numerically calculated $P_{det}^{AP}(V_2^{AP})\left(P_{inj}^{AP}(V_1^{AP}) + P_{inj}^{AP}(V_2^{AP})\right)$ plotted as functions of $V_{\text{total}}^{\text{P}}$ when $t_{\text{I}} = 1$ nm, where pink open rectangles and green open dots are calculated with spin-filter I layers (see Fig.17 (c)) and anormal I layers (see Fig. 15(a)), respectively.

# Spin injection and detection in a Si-based ferromagnetic tunnel junction: A theoretical model based on the band diagram and experimental demonstration

Baisen Yu[1], Shoichi Sato[1, 2], Masaaki Tanaka[1, 2, 3], and Ryosho Nakane[1, 3, 4]

[1]*Deptartment of Electrical Engineering and Information Systems, The University of Tokyo,*
*7-3-1 Hongo, Bunkyo-ku, Tokyo 113-8656, Japan*
[2]*Center for Spintronics Research Network (CSRN), The University of Tokyo,*
*7-3-1 Hongo, Bunkyo-ku, Tokyo 113-8656, Japan*
[3]*Institute for Nano Quantum Information Electronics, The University of Tokyo,*
*4-6-1 Meguro-ku, Tokyo 153-8505, Japan*
[4]*System Design Lab (d.lab), The University of Tokyo,*
*7-3-1 Hongo, Bunkyo-ku, Tokyo 113-8656, Japan*

## S1. Experimental $J-V$ characteristics in the L junction and its fitting curve

Figure. S1(a) shows the experimental $J_L - V_{JL}$ characteristics (blue curve) measured in Setup-C at 4 K and its fitting curve (red curve) with Eq. (10) under $\Delta\mu_L = 0$, from which A is estimated to be $7\times10^{11}$ A/m$^2$/(eV)$^2$.

Figure. S1(b) shows the experimental R junction voltage drop $V_R$ plotted as a function of the L junction current bias $I_L$ (from −20 to 20 mA) in Setup-A, when $I_R = 20$ mA is set. It is confirmed that $V_R$ has negligible changes with $I_L$, which indicates that the current bias applied between the L and L$_{REF}$ electrodes has little influence on the voltage drop measured between the R and R$_{REF}$ electrodes, namely, the spin-valve signal $V^{SR}(I_L, I_R)$ is only determined by the changes of $P_{inj}(I_L)$. In a similar manner, Fig. S1(c) shows the experimental voltage drop $V_L$ (blue dots) plotted as a function of the R junction current bias $I_R$ (from −20 to 20 mA) in Setup-B, when $I_L = 20$ mA is set (Noted that here $I_R$ is not a fixed constant value). It is confirmed that $V_L$ has negligible changes with $I_R$, which indicates that the current bias applied between the R and R$_{REF}$ electrodes has little influence on the voltage drop measured between the L and L$_{REF}$ electrodes, namely, the spin-valve signal $V^{SL}(I_L, I_R)$ is only determined by the changes of $P_{det}(I_L)$.

Fig. S1(a) Experimental $J_L$-$V_{JL}$ characteristics (blue curve) measured in Setup-C at 4 K and its fitting curve (red line)

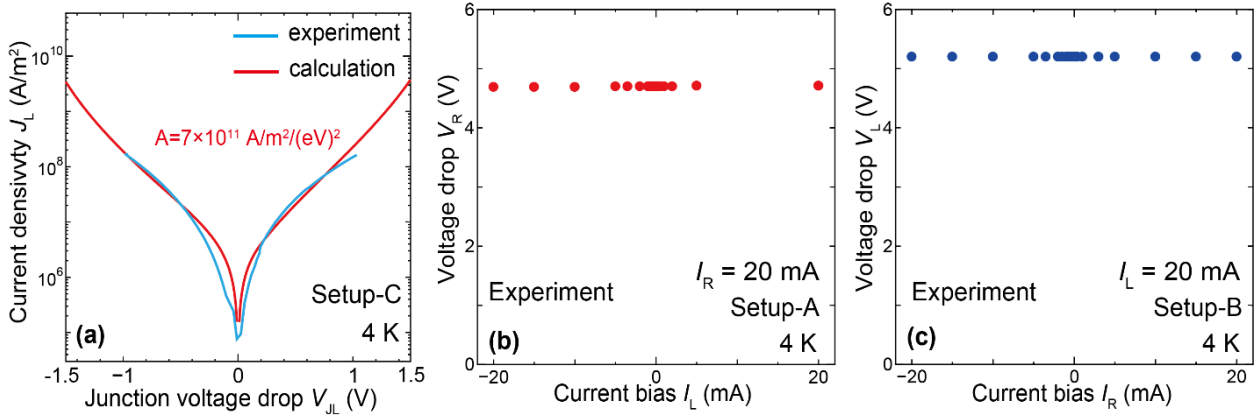

calculated by Eq. (10), where A =$7\times10^{11}$ A/m$^2$/(eV)$^2$. (b) Experimentally measured $V_R$ plotted as a function of the L junction current bias $I_L$ in Setup-A at 4 K, where $I_R = 20$ mA. (c) Experimentally measured $V_L$ plotted as a function of the R junction current bias $I_R$ in Setup-B at 4 K, where $I_L = 20$ mA.

## S2. Analytical expressions of $P_{inj}$ and $P_{det}$

In the main manuscript, the features of $P_{inj}(V_{JL})$ and $P_{det}(V_{JL})$ are discussed based on the band diagrams and direct tunneling mechanism. To obtain a deeper physical insight, here we analytically derive the approximated formulas of $P_{inj}$ and $P_{det}$ by performing Taylor expansion on Eq. (10).

### S2-A Approximate expressions of $P_{inj}$

Here, the expression of $P_{inj}$ is derived in Setup-A with $I_R = 0$ by retaining the first-order term of spin accumulation $\Delta\mu_L$. The L junction currents for up-spin ($J_+$) and down-spin ($J_-$) electrons are expressed as follows:

$$J_+ = J_+\left(V_{JL}, \bar{\mu}^{Si} + \frac{\Delta\mu_L}{2}\right) = J_+(V_{JL}, \bar{\mu}^{Si}) + \frac{\partial J_+(V_{JL}, \bar{\mu}^{Si})}{\partial \bar{\mu}^{Si}} \frac{\Delta\mu_L}{2}, \quad (S1)$$

$$J_- = J_-\left(V_{JL}, \bar{\mu}^{Si} - \frac{\Delta\mu_L}{2}\right) = J_-(V_{JL}, \bar{\mu}^{Si}) - \frac{\partial J_-(V_{JL}, \bar{\mu}^{Si})}{\partial \bar{\mu}^{Si}} \frac{\Delta\mu_L}{2}, \quad (S2)$$

where $V_{JL}$ is the L junction voltage drop, $\bar{\mu}^{Si}$ is the average Si Fermi level, $\Delta\mu_L = 2J_S r_{sr}$ is the spin accumulation induced by the spin injection at the L junction, and $r_{sr} = (r_{Si} \lambda_s)\lambda_s/t_{SOI}$ is the effective spin resistance of the Si channel. Using Eqs. (S1) and (S2), the electron current ($J = J_+ + J_-$) and the spin current ($J_S = J_+ - J_-$) can be expressed as:

$$J = J_0(V_{JL}, \bar{\mu}^{Si}) + \frac{\partial J_{S0}(V_{JL}, \bar{\mu}^{Si})}{\partial \bar{\mu}^{Si}} \frac{\Delta\mu_L}{2}, \quad (S3)$$

$$J_S = J_{S0}(V_{JL}, \bar{\mu}^{Si}) + \frac{\partial J_0(V_{JL}, \bar{\mu}^{Si})}{\partial \bar{\mu}^{Si}} \frac{\Delta\mu_L}{2}$$

$$= P_{inj}|_{\Delta\mu_L=0} J_0(V_{JL}, \bar{\mu}^{Si}) + \frac{\partial J_0(V_{JL}, \bar{\mu}^{Si})}{\partial \bar{\mu}^{Si}} \frac{\Delta\mu_L}{2}, \quad (S4)$$

where $J_0(V_{JL}, \bar{\mu}^{Si}) = J_+(V_{JL}, \bar{\mu}^{Si}) + J_-(V_{JL}, \bar{\mu}^{Si})$ and $J_{S0}(V_{JL}, \bar{\mu}^{Si}) = J_+(V_{JL}, \bar{\mu}^{Si}) - J_-(V_{JL}, \bar{\mu}^{Si})$ are the electron and spin currents under $\Delta\mu_L = 0$, respectively. For simplification, we refer to $J_0(V_{JL}, \bar{\mu}^{Si})$ and $J_{S0}(V_{JL}, \bar{\mu}^{Si})$ as $J_0$ and $J_{S0}$, respectively, in the following discussion. Besides, $P_{inj}|_{\Delta\mu_L=0} = J_{S0}/J_0$ is the magnitude of the spin injection polarization under $\Delta\mu_L = 0$. By taking the expression of $J_0$ in Eq. (S3) into Eq. (S4), the spin current $J_S$ can be expressed as:

$$J_S = P_{inj}|_{\Delta\mu_L=0}\left(J - \frac{\partial J_{S0}}{\partial \bar{\mu}^{Si}} \frac{\Delta\mu_L}{2}\right) + \frac{\partial J_0}{\partial \bar{\mu}^{Si}} \frac{\Delta\mu_L}{2}$$

$$= P_{inj}|_{\Delta\mu_L=0} J - \left(P_{inj}|_{\Delta\mu_L=0} \frac{\partial J_{S0}}{\partial \bar{\mu}^{Si}} - \frac{\partial J_0}{\partial \bar{\mu}^{Si}}\right) J_S r_{sr}, \quad (S5)$$

for which the equation $\Delta\mu_L = 2J_S r_{sr}$ is used. From the Eq. (S5), the spin injection polarization $P_{inj}$ with non-zero $\Delta\mu_L$ can be represented as follows:

$$P_{inj} = \frac{J_S}{J} = P_{inj}|_{\Delta\mu_L=0} \frac{1}{1 + \left(P_{inj}|_{\Delta\mu_L=0} \frac{\partial J_{S0}}{\partial \bar{\mu}^{Si}} - \frac{\partial J_0}{\partial \bar{\mu}^{Si}}\right) r_{sr}}, \quad (S6)$$

where the signs of $\frac{\partial J_0}{\partial \bar{\mu}^{Si}}$ and $\frac{\partial J_{S0}}{\partial \bar{\mu}^{Si}}$ are negative, since the electron current flowing from Fe to Si is defined as positive. The effect of SAS on $P_{inj}$ can be analyzed by Eq. (S6). In the spin injection geometry ($V_{JL} > 0$), as shown Fig. 10(c), electrons at $\bar{\mu}^{Si}$ in Si have a quite small tunneling probability because the effective barrier height for these electrons is significantly high. For this reason, the change in $\Delta\mu_L$ at $\bar{\mu}^{Si}$ in Si have little influence on $J_0$ and $J_{S0}$, while $\frac{\partial J_0}{\partial \bar{\mu}^{Si}}$ and $\frac{\partial J_{S0}}{\partial \bar{\mu}^{Si}}$ are nearly zero. Thus, SAS does not appear and $P_{inj} \sim P_{inj}|_{\Delta\mu_L=0}$. On the other hand, in the spin extraction geometry ($V_{JL} < 0$), SAS appears since electrons at $\bar{\mu}^{Si}$ in Si have the lower effective barrier height so that the change in $\Delta\mu_L$ at $\bar{\mu}^{Si}$ in Si have significant influence on $J_0$ and $J_{S0}$, while $\frac{\partial J_0}{\partial \bar{\mu}^{Si}}$ and $\frac{\partial J_{S0}}{\partial \bar{\mu}^{Si}}$ are non-zero. To explain further, we simplify Eq. (6) into a more concise form by neglecting the term $P_{inj}|_{\Delta\mu_L=0} \frac{\partial J_{S0}}{\partial \bar{\mu}^{Si}}$ in Eq. (S6):

$$P_{inj} = P_{inj}|_{\Delta\mu_L=0} \frac{1}{1 + \left|\frac{\partial J_0}{\partial \bar{\mu}^{Si}}\right| r_{sr}}, \quad (S7)$$

where we use $\left|\frac{\partial J_0}{\partial \bar{\mu}^{Si}}\right|$ to express the positive magnitude of $-\frac{\partial J_0}{\partial \bar{\mu}^{Si}}$. Equation (S7) is a reasonable approximation for the

following reason: Our numerical calculations confirm that $\frac{\partial J_0}{\partial \bar{\mu}^{Si}}$ is larger by ~1 order of magnitude than that of $\frac{\partial J_{S0}}{\partial \bar{\mu}^{Si}}$ and $P_{inj}|_{\Delta\mu_L=0} < 0.24$ is satisfied in the spin extraction geometry (see the yellow curve in Fig. 9(a)). Hence, the magnitude of $\frac{\partial J_0}{\partial \bar{\mu}^{Si}}$ is larger by ~100 times than that of $P_{inj}|_{\Delta\mu_L=0} \frac{\partial J_{S0}}{\partial \bar{\mu}^{Si}}$. In the spin extraction geometry, the effect of SAS on $P_{inj}$ is analyzed using Eq. (S7) as follows: As $V_{JL}$ is negatively increased from 0, $\left|\frac{\partial J_0}{\partial \bar{\mu}^{Si}}\right|$ steeply increases since the effective barrier height for the electrons at $\bar{\mu}^{Si}$ in Si becomes lower, the term $\frac{1}{1+\left|\frac{\partial J_0}{\partial \bar{\mu}^{Si}}\right|r_{sr}}$ becomes significantly smaller than 1, and thus $P_{inj}$ is largely reduced. Remarkably, in terms of the relation between $P_{inj}$ and $r_{sr}$, Eq. (S7) shows a mathematical form similar to that in the V-F model [S1,S2]:

$$P_{inj}^{VF} = \gamma \frac{1}{1+\frac{r_{sr}}{r_b}}. \qquad (S8)$$

The significant difference between our model and the V-F model is that the critical term for determining the spin injection polarization is the differential term $\left|\frac{\partial J_0}{\partial \bar{\mu}^{Si}}\right|$ in our model, instead of the linear conductance $1/r_b$ of a barrier in the V-F model.

**S2-B Approximate expressions of $P_{det}$**

The approximated form of $P_{det}(V_{JL})$ is derived in the measurement Setup-B, where the L and R electrodes are the spin detector and injector, respectively. Figures S2(a) and (b) show the schematic illustrations of the parallel (P) and anti-parallel (AP) magnetization states in Setup-B, respectively, where the L and R electrodes are biased by a constant $I_L$ value (−20 ~ 20 mA) and $I_R$ = 20 mA, respectively, arrows represent the magnetization directions of the electrodes, and red and blue lines represent electrochemical potentials for up-spin ($\mu_+$) and down-spin ($\mu_-$) electrons, respectively. As described in Sec. IV-C, the spin accumulation $\Delta\mu_L(V_{JL}, V_{JR})$ beneath the L electrode is composed of the spin injection from the L electrode $\Delta\mu_{L0}(V_{JL})$ and the spin transport from the R electrode $\Delta\mu_L(V_{JR}) = \Delta\mu_R(V_{JR}) \exp(-L_{ch}/\lambda_{sf})$, i.e., $\Delta\mu_L(V_{JL}, V_{JR}) = \Delta\mu_{L0}(V_{JL}) + \Delta\mu_L(V_{JR})$, where $\Delta\mu_R(V_{JR})$ is the spin accumulation beneath the R electrode. To simplify the derivation, here we denote $\Delta\mu_L = \Delta\mu_{L0}(V_{JL})$ and $\Delta\mu_R = \Delta\mu_L(V_{JR})$, namely, $\Delta\mu_L(V_{JL}, V_{JR}) = \Delta\mu_L + \Delta\mu_R$. In the P magnetization state, the up-spin current ($J_+$) and down-spin current ($J_-$) at the L electrode are expressed as:

$$J_{\pm}^P = J_{\pm}\left(V_{JL} - \Delta V_{det}, \bar{\mu}^{Si} \pm \frac{\Delta\mu_L}{2} \pm \frac{\Delta\mu_R}{2}\right)$$
$$= J_{\pm}(V_{JL}, \bar{\mu}^{Si}) - \frac{\partial J_{\pm}(V_{JL}, \bar{\mu}^{Si})}{\partial V_{JL}} \Delta V_{det} \pm \frac{\partial J_{\pm}(V_{JL}, \bar{\mu}^{Si})}{\partial \bar{\mu}^{Si}} \frac{\Delta\mu_L + \Delta\mu_R}{2}$$
$$+ \frac{1}{2} \frac{\partial^2 J_{\pm}(V_{JL}, \bar{\mu}^{Si})}{\partial \bar{\mu}^{Si2}} \left\{\left(\frac{\Delta\mu_L}{2}\right)^2 + \left(\frac{\Delta\mu_R}{2}\right)^2 + \frac{\Delta\mu_L\Delta\mu_R}{2}\right\}, \qquad (S9)$$

where $\Delta\mu_L$ and $\Delta\mu_R$ represent the spin accumulation induced by the L and R electrodes, respectively, the first- and second-order terms of $\Delta\mu_L$ and $\Delta\mu_R$ are retained in the Talyor expansion, the superscript P represents the P magnetization state between the L and R electrodes, $\Delta V_{det}$ is the amplitude of the spin detection voltage, and $V_{JL}$ is the applied voltage drop. In a similar manner, in the AP magnetization state, $J_+$ and $J_-$ at the L electrode are expressed as:

$$J_{\pm}^{AP} = J_{\pm}\left(V_{JL} + \Delta V_{det}, \bar{\mu}^{Si} \pm \frac{\Delta\mu_L}{2} \mp \frac{\Delta\mu_R}{2}\right)$$
$$= J_{\pm}(V_{JL}, \bar{\mu}^{Si}) + \frac{\partial J_{\pm}(V_{JL}, \bar{\mu}^{Si})}{\partial V_{JL}} \Delta V_{det} \pm \frac{\partial J_{\pm}(V_{JL}, \bar{\mu}^{Si})}{\partial \bar{\mu}^{Si}} \frac{\Delta\mu_L - \Delta\mu_R}{2}$$
$$+ \frac{1}{2} \frac{\partial^2 J_{\pm}(V_{JL}, \bar{\mu})}{\partial \bar{\mu}^{Si2}} \left\{\left(\frac{\Delta\mu_L}{2}\right)^2 + \left(\frac{\Delta\mu_R}{2}\right)^2 - \frac{\Delta\mu_L\Delta\mu_R}{2}\right\}. \qquad (S10)$$

The total currents flowing through the L junction in P and AP magnetization states are expressed as:
$$J^P = J_+^P + J_-^P$$
$$= J_0 - \frac{\partial J_0}{\partial V_{JL}} \Delta V_{det} + \frac{\partial J_{S0}}{\partial \bar{\mu}^{Si}} \frac{\Delta\mu_L + \Delta\mu_R}{2}$$

$$+\frac{1}{2}\frac{\partial^2 J_0}{\partial \bar{\mu}^{Si^2}}\left\{\left(\frac{\Delta\mu_L}{2}\right)^2+\left(\frac{\Delta\mu_R}{2}\right)^2+\frac{\Delta\mu_L\Delta\mu_R}{2}\right\} \quad (S11)$$

$$\begin{aligned}J^{AP}&=J_+^{AP}+J_-^{AP}\\&=J_0+\frac{\partial J_0}{\partial V_{JL}}\Delta V_{det}+\frac{\partial J_{S0}}{\partial \bar{\mu}^{Si}}\frac{\Delta\mu_L-\Delta\mu_R}{2}\\&+\frac{1}{2}\frac{\partial^2 J_0}{\partial \bar{\mu}^{Si^2}}\left\{\left(\frac{\Delta\mu_L}{2}\right)^2+\left(\frac{\Delta\mu_R}{2}\right)^2-\frac{\Delta\mu_L\Delta\mu_R}{2}\right\}.\end{aligned} \quad (S12)$$

Since the condition $J^P = J^{AP}$ is satisfied for any $I_L$ values, the following equation is obtained:

$$2\frac{\partial J_0}{\partial V_{JL}}\Delta V_{det}-\frac{\partial J_{S0}}{\partial \bar{\mu}^{Si}}\Delta\mu_R-\frac{\partial^2 J_0}{\partial^2\bar{\mu}^{Si}}\frac{\Delta\mu_L\Delta\mu_R}{2}=0. \quad (S13)$$

From Eq. (S13), we find that $\Delta\mu_R$ is the determining factor to generate a non-zero $\Delta V_{det}$, because $\Delta V_{det} = 0$ is the only solution when $\Delta\mu_R = 0$. This is reasonable, because the sign of $\Delta\mu_R$ is changed, while the sign of $\Delta\mu_L$ does not change, by switching the magnetization state from P to AP. Hence, the spin detection polarization ($P_{det}$) is calculated by taking the ratio between $\Delta V_{det}$ and $\Delta\mu_R$ through Eq. (S13):

$$P_{det}=2\frac{\Delta V_{det}}{\Delta\mu_R}=\frac{\frac{\partial J_{S0}}{\partial\bar{\mu}^{Si}}+\frac{\partial^2 J_0}{\partial\bar{\mu}^{Si^2}}\frac{\Delta\mu_L}{2}}{\frac{\partial J_0}{\partial V_{JL}}}, \quad (S14)$$

where the second term in the numerator indicates the effect of SAS on $P_{det}$. Our numerical calculations confirm that $\frac{\partial^2 J_0}{\partial\bar{\mu}^{Si^2}} < 0$ in the spin extraction geometry, namely, SAS reduces the spin detection efficiency. This fact reveals that the effect of SAS on $P_{det}$ originates from the second-order (and higher-order) terms of $\Delta\mu_L$. The numerator in Eq. (S14) can be transformed into the following form by taking the expression of $J_{S0}$ in Eq. (S4) into Eq. (S14)

$$\begin{aligned}\frac{\partial J_{S0}}{\partial\bar{\mu}^{Si}}+\frac{\partial^2 J_0}{\partial\bar{\mu}^{Si^2}}\frac{\Delta\mu_L}{2}&=\frac{\partial J_S}{\partial\bar{\mu}^{Si}}-\frac{\partial}{\partial\bar{\mu}^{Si}}\left(\frac{\partial J_0}{\partial\bar{\mu}^{Si}}\frac{\Delta\mu_L}{2}\right)+\frac{\partial^2 J_0}{\partial\bar{\mu}^{Si^2}}\frac{\Delta\mu_L}{2}\\&=\frac{\partial J_S}{\partial\bar{\mu}^{Si}}-\frac{\partial^2 J_0}{\partial\bar{\mu}^{Si^2}}\frac{\Delta\mu_L}{2}-\frac{\partial J_0}{\partial\bar{\mu}^{Si}}\frac{\partial}{\partial\bar{\mu}^{Si}}\left(\frac{\Delta\mu_L}{2}\right)+\frac{\partial^2 J_0}{\partial\bar{\mu}^{Si^2}}\frac{\Delta\mu_L}{2}\\&=\frac{\partial J_S}{\partial\bar{\mu}^{Si}}-\frac{\partial J_0}{\partial\bar{\mu}^{Si}}\frac{\partial}{\partial\bar{\mu}^{Si}}\left(\frac{\Delta\mu_L}{2}\right)\\&=P_{inj}\frac{\partial J}{\partial\bar{\mu}^{Si}}+J\frac{\partial P_{inj}}{\partial\bar{\mu}^{Si}}-\frac{\partial J_0}{\partial\bar{\mu}^{Si}}\frac{\partial}{\partial\bar{\mu}^{Si}}\left(\frac{\Delta\mu_L}{2}\right),\end{aligned} \quad (S15)$$

where the $J_S = P_{inj} J$ is used. When the second and third terms are neglected, the following expression is obtained,

$$P_{det}=P_{inj}\frac{\frac{\partial J}{\partial\bar{\mu}^{Si}}}{\frac{\partial J_0}{\partial V_{JL}}}. \quad (S16)$$

Equation (S16) shows that $P_{det}$ is proportional to $P_{inj}$. This can provide us an alternative way to understand the reduction of $P_{det}$ caused by SAS in the spin extraction geometry ($V_{JL} < 0$). In Fig. 9(a), as $V_{JL}$ is negatively increased from 0, the blue $P_{inj}$–$V_{JL}$ curve decreases faster than the yellow $P_{inj}$–$V_{JL}$ curve due to SAS. Such decrease of $P_{inj}$ accounts for the more rapid decrease of the blue $P_{det}$–$V_{JL}$ curve than that of the yellow $P_{det}$–$V_{JL}$ curve in Fig. 13(a), as $V_{JL}$ is negatively increased from 0. In other words, SAS affects $P_{det}$ through $P_{inj}$. Equation (S16) is useful to gain an insight into the relationship between $P_{det}$ and $P_{inj}$, whereas it is an approximated form because the neglected two terms in (S15) may not be always significantly small to be neglected.

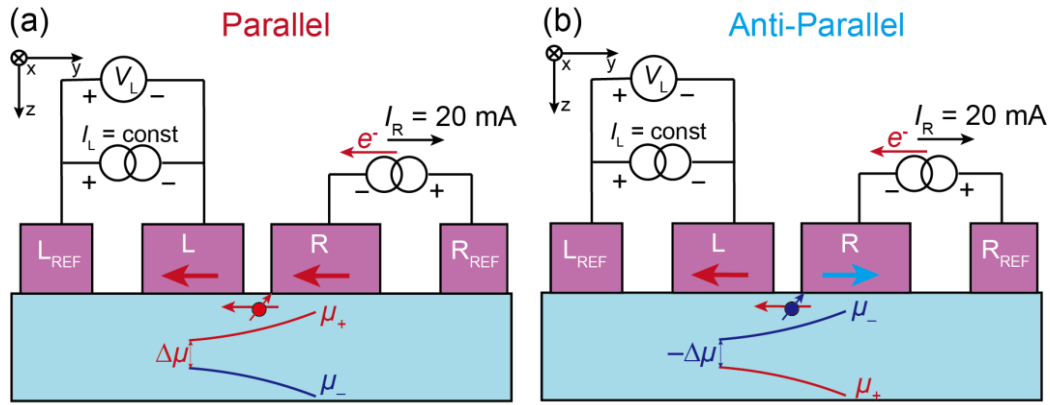

Fig. S2 (a)(b) Measurement Setup-B in the parallel (P) and anti-parallel (AP) magnetization states with $I_R$ = 20 mA and a constant $I_L$ value, respectively, where red and blue lines represent the Si electrochemical potentials of up-spin ($\mu_+$) and down-spin ($\mu_-$) electrons, respectively. Colored arrows represent the magnetization directions of the ferromagnetic electrodes.

## S3. Effect of SAS on $P_{inj}$ and the comparison with the V-F model

The analysis in Sec. IV reveals that $P_{inj}$ is largely suppressed in the high negative $V_{JL}$ range owing to "spin accumulation saturation (SAS)" that is analogous to the conductivity mismatch problem in the V-F model. In Sec. S2-A, we have derived Eq. (S7) that analytically describes how $P_{inj}$ is affected by SAS. In this section, we validate the effectiveness of Eq. (S7) by numerical calculations, and compare the different effects on $P_{inj}$ as the tunnel junction resistance (junction voltage bias $V_{JL}$) is changed between our model and the V-F model, where the junction properties extracted from the experimental device are used.

Here, we recall the differences between our model and the V-F model as follows: First, $P_{inj}$ is a function of the junction bias $V_{JL}$ in our model, whereas $P_{inj}$ is equal to a constant $\gamma$ for any $V_{JL}$ in the V-F model. Second, the critical term for determining $P_{inj}$ is the differential term $|\frac{\partial J_0}{\partial \bar{\mu}^{Si}}|$ in our model (see Eq. (S7)) and the linear conductance $1/r_b$ of a barrier in the V-F model (see Eq. (S8)), respectively. The following procedures were performed to utilize the fitting $J_L$–$V_{JL}$ characteristics in Fig. S1(a) and to compare the two models on how $P_{inj}$ is affected by the change of the junction resistance: First, the linear conductance $1/r_{bT} = I_L/V_{JL}$ defined in the tunnel junction is used to replace $1/r_b$ in the V-F model. It is noted that $1/r_{bT}$ is a function of $V_{JL}$ in our model, whereas $1/r_b$ keeps a constant at any $V_{JL}$ the V-F model. Second, the normalized $P_{inj}$ defined by the ratio $P_{inj}/P_{inj}|_{\Delta\mu=0}$ is calculated for both models to exclude the bias-dependent influence on $P_{inj}$, where $P_{inj}/P_{inj}|_{\Delta\mu=0}$ is numerically calculated using Eq. (10) in our model and Eq. (S8) in the V-F model.

Figure S3(a) shows the numerically-calculated normalized $P_{inj}$ in the spin injection geometry, where the red line corresponding to our model is plotted against $|\frac{\partial J_0}{\partial \bar{\mu}^{Si}}|$ and blue line corresponding to the V-F model is plotted against $r_{bT}$. The V-F model (blue line) shows that the normalized $P_{inj}$ decreases as $r_{bT}$ is decreased, which reflects the well-known conductance mismatching curve. On the other hand, our model (red line) shows that the normalized $P_{inj}$ keeps a constant (100%) as $|\frac{\partial J_0}{\partial \bar{\mu}^{Si}}|$ is decreased. Thus, SAS does not appear in the spin injection geometry, which is consistent with the conclusion in Sec. IV-A.

Figure S3(a) shows the numerically-calculated normalized $P_{inj}$ in the spin extraction geometry, where the red line corresponding to our model is plotted against $|\frac{\partial J_0}{\partial \bar{\mu}^{Si}}|$ and blue line corresponding to the V-F model is plotted against $r_{bT}$. Both blue and red curves show a similar decreasing trend as $|\frac{\partial J_0}{\partial \bar{\mu}^{Si}}|$ or $r_{bT}$ is decreased, while our model (the red curve) shows a deviation from the V-F model. This result indicates that the magnitude of $|\frac{\partial J_0}{\partial \bar{\mu}^{Si}}|$ in Eq. (S7) is smaller than $r_{bT}$ in Eq. (S8) at a specific $V_{JL}$ in the spin extraction geometry. The pink dashed line in Fig.S3 (b) is calculated using our analytical Eq. (S7). The good agreement with the numerical calculation (red line) validates that Eq. (S7) is an accurate approximation for estimating the reduction of $P_{inj}$ caused by SAS.

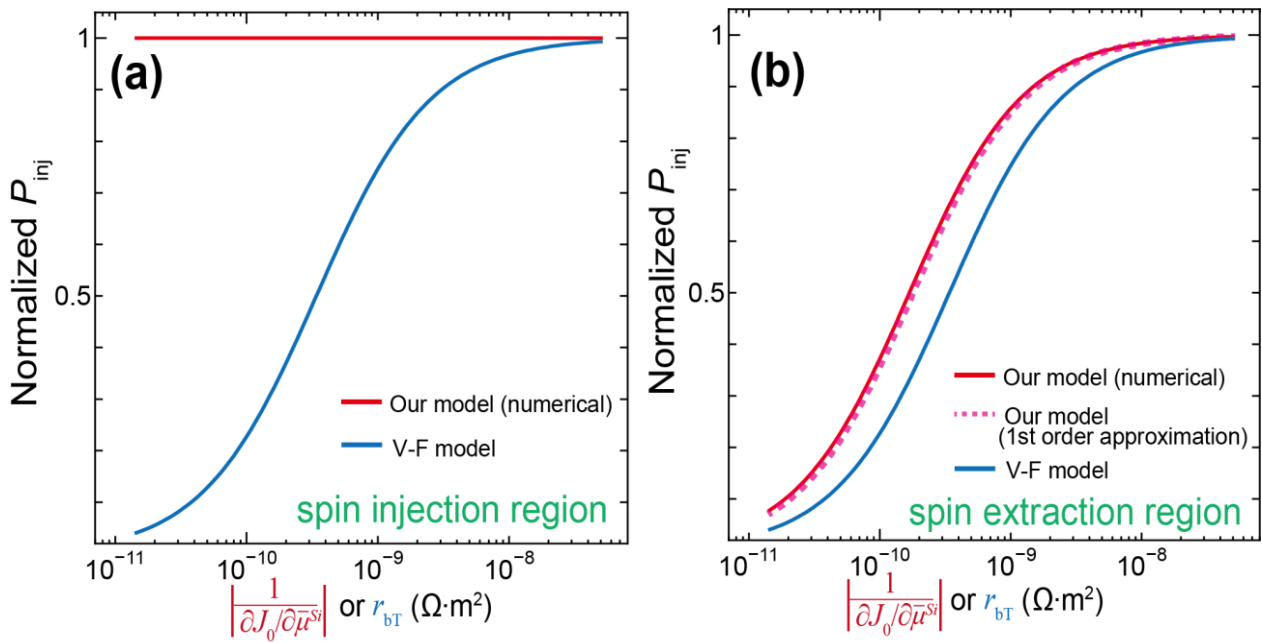

Fig. S3(a) In the spin injection geometry ($V_{JL} > 0$). The red line: The normalized $P_{inj}$ plotted as a function of differential resistance $1/|\frac{\partial J_0}{\partial \bar{\mu}^{Si}}|$ numerically calculated using our model. The blue line: The normalized $P_{inj}$ plotted as a function of the tunneling linear resistance $r_{bT}$ using the V-F model. The Si spin resistance $r_{sr} = 3.4 \times 10^{-10}$ Ω·m². (b) In the spin extraction geometry ($V_{JL} < 0$). The red line: The normalized $P_{inj}$ plotted as a function of differential resistance $1/|\frac{\partial J_0}{\partial \bar{\mu}^{Si}}|$ numerically calculated using our model. Pink broken line: The normalized $P_{inj}$ plotted as a function of differential resistance $1/|\frac{\partial J_0}{\partial \bar{\mu}^{Si}}|$ calculated by the analytical Eq. (S7). The blue line: The normalized $P_{inj}$ plotted as a function of the tunneling linear resistance $r_{bT}$ using the V-F model.

## S4. Effect of the Si band bending on $P_{\text{inj}}$

Figure S4(a) shows the calculated $P_{\text{inj}}$ plotted as a function of $V_{\text{JL}}$, where yellow and gray curves correspond to the cases with and without Si band bending (SBB), respectively. In the spin extraction geometry ($V_{\text{JL}} < 0$), the two curves are almost overlapped, which reveals that SBB has almost no effect on $P_{\text{inj}}$. In the spin injection geometry ($V_{\text{JL}} > 0$), there is a slight difference between the two curves, i.e., the yellow curve is lower than and gray curve in the high bias range of $V_{\text{JL}}$. This is explained as follows: In the high bias range of the spin injection geometry, $P_{\text{inj}}$ is dominantly determined by the electron tunneling at $E = \bar{\mu}^{Fe}$, as explained in Sec. IV-A and the schematic normalized tunneling possibility $T_{\text{nor}}(E)$ shown in Fig. 10(d). For a specific $V_{\text{JL}}$ value, when SBB is present, $D_{\text{Si}}(\bar{\mu}^{Si})$ decreases while $D_{\text{Fe}}(\bar{\mu}^{Fe})$ does not change, which means that $P_{\text{inj}}$ is slightly reduced by a factor of $D_{\text{Si}}(\bar{\mu}^{Si}) / D_{\text{Si}}(E - \Phi_{\text{Si}})$ ($> 1$). This leads to a slight decrease of the yellow curve relative to the gray curve. Therefore, to simplify the qualitative analysis, we do not discuss the effects of SBB on the calculated $P_{\text{inj}}$–$V_{\text{JL}}$ curves in Fig. 9(a).

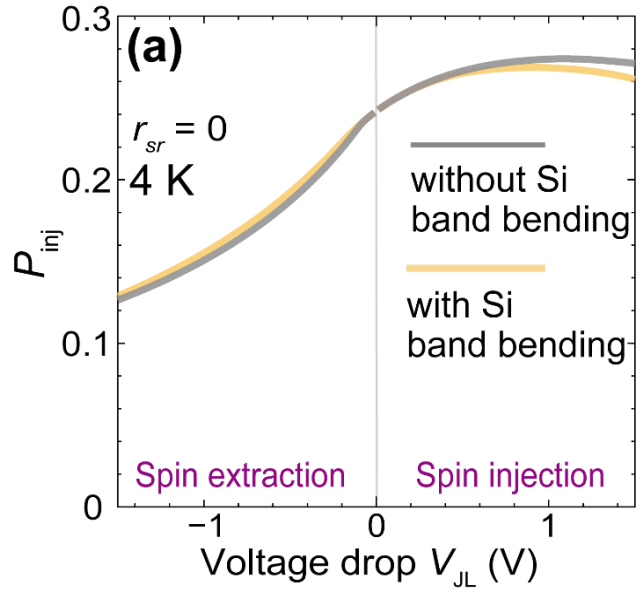

Fig. S4 Calculated spin injection polarization $P_{\text{inj}}$ at 4 K plotted as a function of $V_{\text{JL}}$ under $r_{\text{sr}} = 0$ with $I_R = 0$ mA in Setup-A, where yellow and gray curves are the results calculated with and without considering the Si band bending, respectively.

## S5. Reason for the slight reduction in $P_{inj}$ in the spin injection geometry

In Sec. IV-B, the increasing trend of $P_{inj}$ in Fig. 9(a) in the spin injection geometry has been explained using the band diagrams and normalized tunneling probability. In a similar manner, here we explain the reason why $P_{inj}$ has a slight decrease as $V_{JL}$ is further increased from 0.8 V. Figure S5(a) shows the band diagram of the Fe/MgO/$n^+$-Si junction in the positive $V_{JL}$ range (the spin injection geometry) with a very high $V_{JL}$ (> 0.8 V). Figure S5(b) shows the $P_{Fe}(E)$ curve and schematic normalized tunneling probability $T_{nor}(E)$ $(=T(E)/T(\bar{\mu}^{Fe}))$ at the same $V_{JL}$ value. An important feature in a very high $V_{JL}$ range is that the distribution of $T_{nor}(E)$ becomes more concentrated at $E = \bar{\mu}^{Fe}$, because $T_{nor}(E)$ exponentially increases as $V_{JL}$ is increased. Thus, as $V_{JL}$ is increased in a very high bias range, the current contributed by the electron tunneling at $E = \bar{\mu}^{Fe}$ becomes more dominant in the total tunneling current, which leads to the asymptotic convergence of $P_{inj}(V_{JL})$ to $P_{Fe}(\bar{\mu}^{Fe}) = 0.24$. This explains the slight reduction in $P_{inj}$ in the range of $V_{JL}$ from 0.8 V to 1.5 V.

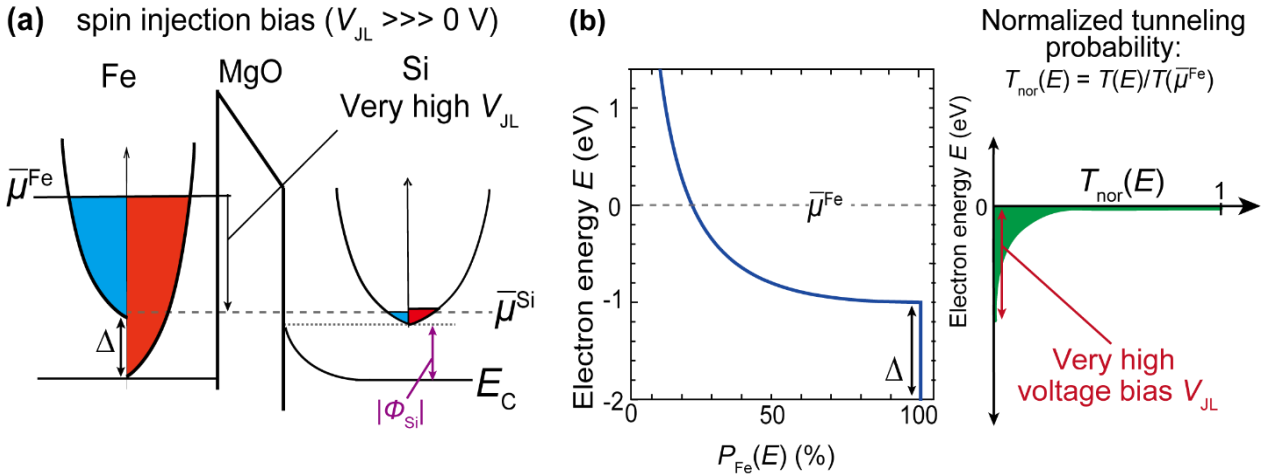

Fig. S5 (a) One-dimensional energy band diagram of a Fe/MgO/$n^+$-Si junction in the spin injection geometry with a very high ($V_{JL} \gg 0$) junction voltage drop, where $\bar{\mu}^{Fe}$ and $\bar{\mu}^{Si}$ are the Fermi levels of Fe and $n^+$-Si, respectively, $E_C$ is the Si conduction band minimum, and red and blue regions represent the occupied up-spin and down-spin electrons, respectively. (b) Left side: Calculated spin polarization $P_{Fe}(E)$ in the Fe band plotted as a function of electron energy $E$, where $\bar{\mu}^{Fe}$ is defined as the zero potential $E = 0$. Right side: Normalized tunneling probability $T_{nor}(E)$ schematically plotted as a function of $E$ at the same voltage bias in (a).

**S6. Numerical analysis on $P_{det}$**

In this section, we utilize analytical equations derived in Sec. S2 to show that the $P_{det}$–$V_{JL}$ curves in Fig. 13 (b) without SAS can be analyzed by two separated differential terms. The numerical calculations of those two terms can facilitate understanding the changing trends of $P_{det}$ in various $V_{JL}$ ranges. First, the first-order approximated expression of $P_{det}$ can be obtain as

$$P_{det} = \frac{\frac{\partial J_{S0}}{\partial \bar{\mu}^{Si}}}{\frac{\partial J_0}{\partial V_{JL}}}, \quad (S17)$$

by neglecting the second-order differential term $\frac{\partial^2 J_0}{\partial \bar{\mu}^{Si^2}} \frac{\Delta \mu_L}{2}$ in Eq. (14). Hereafter, both $\frac{\partial J_{S0}}{\partial \bar{\mu}^{Si}}$ and $\frac{\partial J_0}{\partial V_{JL}}$ are represented by the absolute values to focus on their magnitudes. It is numerically confirmed that Eq. (S17) is a good approximation of $P_{det}$ when SAS is not considered. Here we numerically calculate $\frac{\partial J_{S0}}{\partial \bar{\mu}^{Si}} - V_{JL}$ and $\frac{\partial J_0}{\partial V_{JL}} - V_{JL}$ curves to clarify how they affect the changing trends of the $P_{det}$–$V_{JL}$ curves in Figs. 13(a) and (b) in various $V_{JL}$ ranges. Figure S6(a) shows the numerically-calculated $\frac{\partial J_{S0}}{\partial \bar{\mu}^{Si}} - V_{JL}$ and $\frac{\partial J_0}{\partial V_{JL}} - V_{JL}$ curves using Eq. (10) in a semi-log plot, where green and purple lines are the $\frac{\partial J_0}{\partial V_{JL}} - V_{JL}$ and $\frac{\partial J_{S0}}{\partial \bar{\mu}^{Si}} - V_{JL}$ curves, respectively, and the solid and broken lines represent the cases with and without considering Si band bending (SBB), respectively.

**S6-A. Analysis of the $\frac{\partial J_0}{\partial V_{JL}} - V_{JL}$ curve when Si band bending (SBB) is neglected**

Here we analyze the $\frac{\partial J_0}{\partial V_{JL}} - V_{JL}$ curve without considering SBB (the broken green curve) in Fig. S6. As $V_{JL}$ is increased from 0 V to 1.5 V (in the spin injection geometry), the broken green curve in Fig. S6(a) exponentially increases. This is because the effective barrier heights $\Phi_{eff}(E)$ for the Fe electrons become lower as $V_{JL}$ is increased. Fig. S7(a) and (b) show the schematic band diagrams of the Fe/MgO/$n^+$-Si junction at low and high positive $V_{JL}$ biases with $\Delta \mu_L = 0$, respectively, where $\Phi_{eff}(\bar{\mu}^{Fe})$ becomes lower at a high positive bias. Such lowering of $\Phi_{eff}$ also occurs for the electrons with $E < \bar{\mu}^{Fe}$ in Fe.

As $V_{JL}$ is negatively increased from 0 to −0.1 V (in the spin extraction geometry), the broken green curve in Fig. S6(a) decreases and shows a minimum at $V_{JL}$ = −0.1 V. This originates from the gradual decrease of $D_{Si}(\bar{\mu}^{Fe})$ that is the Si density of states at $E = \bar{\mu}^{Fe}$, because the magnitude of $\frac{\partial J_0}{\partial V_{JL}}$ is proportional to $D_{Si}(\bar{\mu}^{Fe})$, as shown in the band diagram of Fig. 14(a). The minimum $\frac{\partial J_0}{\partial V_{JL}}$ condition corresponds to the situation that $\bar{\mu}^{Fe}$ matches the conduction band bottom of the Si when $V_{JL}$ = −0.1 V, as shown in the band diagram of Fig. 13(c). In the bias range of −0.1 V < $V_{JL}$ < 0 V, the term $\frac{\partial J_0}{\partial V_{JL}}$ determines the magnitude of $P_{det}$: For the gray $P_{det}$–$V_{JL}$ curve shown in Fig. 13(b), the decrease of $\frac{\partial J_0}{\partial V_{JL}}$ (the denominator in Eq. (S17)) accounts for the steep increase of $P_{det}$ as $V_{JL}$ is changed from 0 to −0.1 V.

As $V_{JL}$ is further negatively increased from −0.1 V (in the spin extraction geometry), the broken green curve in Fig. S6(a) exponentially increases, because the effective barrier height $\Phi_{eff}(E)$ for the electrons in Si becomes lower. Fig. S7(c) and (d) show the schematic band diagrams of the Fe/MgO/$n^+$-Si junction at low and high negative $V_{JL}$ biases with $\Delta \mu_L = 0$, respectively, where $\Phi_{eff}(\bar{\mu}^{Si})$ becomes lower at a high negative $V_{JL}$. Such lowering of $\Phi_{eff}$ also occurs for the electrons with $E < \bar{\mu}^{Si}$ in Si.

**S6-B. Effect of Si band bending (SBB) on the $\frac{\partial J_0}{\partial V_{JL}} - V_{JL}$ curve**

Here we analyze the effects of SBB on the $\frac{\partial J_0}{\partial V_{JL}} - V_{JL}$ curve (the solid green curve) in Fig. S6. The most significant difference between the solid and broken curves in Fig. S6 is in the range of −0.2 V < $V_{JL}$ < 0 V, where SBB has an important influence on $\frac{\partial J_0}{\partial V_{JL}}$.

As $V_{JL}$ is negatively increased from 0 to −0.2 V (in the spin extraction geometry), unlike the decreasing trend of the broken green curve in Fig. S6, the solid green curve slightly increases and shows a cusp at $V_{JL}$ = −0.2 V. Such slight increase comes from the electron accumulation caused by SBB in Si nearby the tunnel barrier(MgO) / Si interface: The number of electrons contributing to the tunneling current gradually increases with negatively increasing $V_{JL}$. This explains why the yellow $P_{det}$−$V_{JL}$ curve (with SBB) is much lower than the gray $P_{det}$−$V_{JL}$ curve (without SBB) in Fig. 13(b) in the spin extraction geometry: SBB-induced electron accumulation enhances $\frac{\partial J_0}{\partial V_{JL}}$, which corresponds to the increase of the denominator in Eq. (S17) and reduces the magnitude of $P_{det}$.

When $V_{JL}$ > 0 (in the spin injection geometry), the yellow and gray $P_{det}$−$V_{JL}$ curves in Fig. 13(b) are almost same, which indicates that SBB has little influence on $P_{det}$. Thus, we do not discuss the effects of SBB on $\frac{\partial J_0}{\partial V_{JL}}$ in the positive bias range.

### S6-C. Analysis of the $\frac{\partial J_{S0}}{\partial \bar{\mu}^{Si}} - V_{JL}$ curve when Si band bending (SBB) is neglected

We analyze the $\frac{\partial J_{S0}}{\partial \bar{\mu}^{Si}} - V_{JL}$ curve without considering SBB (the broken purple curve) in Fig. S6.

As $V_{JL}$ is negatively increased from 0 V to −1.5 V (in the spin extraction geometry), $\frac{\partial J_{S0}}{\partial \bar{\mu}^{Si}}$ monotonically increases. This is explained as follows. The magnitude of $\frac{\partial J_{S0}}{\partial \bar{\mu}^{Si}}$ reflects how a small perturbation of $\bar{\mu}^{Si}$ affects the change of $J_{S0}$ that exponentially increases with the decrease of the effective barrier height $\Phi_{eff}(\bar{\mu}^{Si})$. Figures. S7 (c) and (d) show the band diagrams of the Fe/MgO/$n^+$-Si junction at low and high negative $V_{JL}$ biases with $\Delta \mu_L = 0$, respectively, where the effective barrier height $\Phi_{eff}(\bar{\mu}^{Si})$ is lower when the magnitude of $V_{JL}$ is larger. Hence, $\frac{\partial J_{S0}}{\partial \bar{\mu}^{Si}}$ exponentially increases as $V_{JL}$ is negatively increased from 0.

As $V_{JL}$ is increased from 0 V to 1.5 V (in the spin injection geometry), $\frac{\partial J_{S0}}{\partial \bar{\mu}^{Si}}$ monotonically decreases. This is explained as follows. The magnitude of $\frac{\partial J_{S0}}{\partial \bar{\mu}^{Si}}$ reflects how a small perturbation of $\bar{\mu}^{Si}$ affects the change of $J_{S0}$ that exponentially increases with the decrease of the effective barrier height $\Phi_{eff}(\bar{\mu}^{Fe} - V_{JL})$. Figures. S7 (a) and (b) show the band diagrams of the Fe/MgO/$n^+$-Si junction at low and high positive $V_{JL}$ biases with $\Delta \mu = 0$, respectively, where the effective barrier height $\Phi_{eff}(\bar{\mu}^{Fe} - V_{JL})$, denoted by red characters, for the electrons in Fe is higher when the magnitude of $V_{JL}$ is larger. Hence, $\frac{\partial J_{S0}}{\partial \bar{\mu}^{Si}}$ exponentially decrease as $V_{JL}$ is positively increased from 0. It is noted that the small magnitude of $\frac{\partial J_{S0}}{\partial \bar{\mu}^{Si}}$ accounts for the nearly zero values of $P_{det}$ in Fig. 13(b) when $V_{JL}$ > 0.

### S6-D. Effect of Si band bending (SBB) on the $\frac{\partial J_{S0}}{\partial \bar{\mu}^{Si}} - V_{JL}$ curve

We analyze the effects of SBB on the $\frac{\partial J_{S0}}{\partial \bar{\mu}^{Si}} - V_{JL}$ curve (the solid purple curve) in Fig. S6.

Both solid and broken purple curves in Fig. S6 show a similar changing trend: As $V_{JL}$ is changed from −1.5 to 1.5 V, $\frac{\partial J_{S0}}{\partial \bar{\mu}^{Si}}$ monotonically decreases. Their slight differences are as follows:

In the spin extraction geometry ($V_{JL}$ < 0), as $V_{JL}$ is negatively increased, the solid purple curve increases more rapidly than the broken purple curve, because the accumulation region is formed in Si nearby the MgO/Si interface where the energy difference $\bar{\mu}^{Si} - E_C$ becomes larger and the electron density is enhanced, as shown in the band diagram of Fig. S7 (e). Since more electrons accumulated nearby the MgO/Si interface contribute to the tunneling current, and the magnitude of $\frac{\partial J_{S0}}{\partial \bar{\mu}^{Si}}$ is enhanced.

In the spin injection geometry ($V_{JL}$ > 0), as $V_{JL}$ is positively increased, the solid purple curve decreases much more rapidly than the broken purple curve, because the depletion region is formed in Si nearby the MgO/Si interface where $\bar{\mu}^{Si}$ becomes lower than the Si conduction band minimum $E_C$, as shown in the band diagram of Fig. S7(f). In such situation, a small change of $\bar{\mu}^{Si}$ has little influence on the change of $J_{S0}$, because there are less empty states in Si which can accept the Fe electrons. Nevertheless, the depletion at the MgO/Si interface does not make a difference in the values of $P_{det}$ in the spin injection geometry, as shown in the yellow and gray $P_{det}$−$V_{JL}$ curves of Fig. 13(b). This is because the increase of $\Phi_{eff}(\bar{\mu}^{Fe} - V_{JL})$ can significantly reduce the magnitude of $\frac{\partial J_{S0}}{\partial \bar{\mu}^{Si}}$, as $V_{JL}$ is positively increased from 0, even when the electron depletion is not considered.

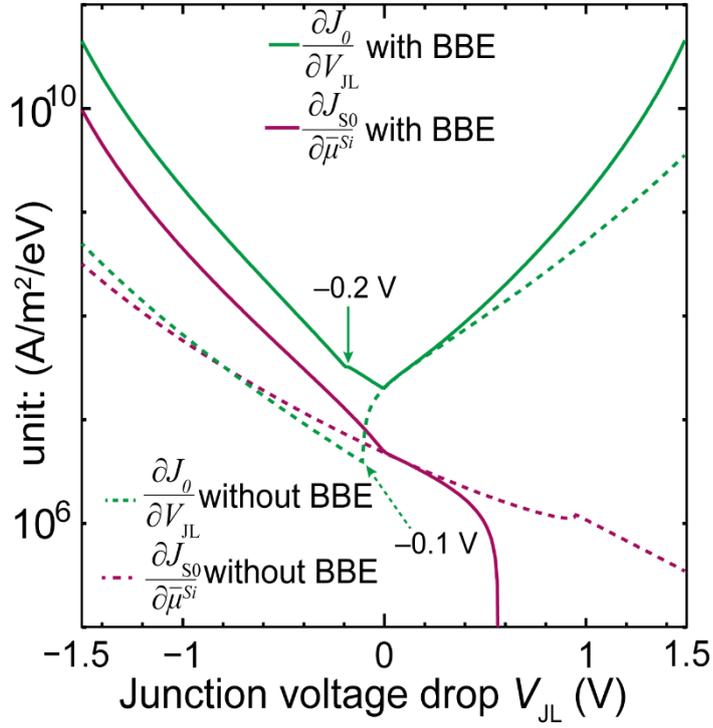

Fig. S6 Green and purple solid lines are calculated, $\frac{\partial J_0}{\partial V_{JL}} - V_{JL}$ and $\frac{\partial J_{S0}}{\partial \bar{\mu}^{Si}} - V_{JL}$ curves with considering Si band bending (SBB), respectively. Green and purple broken lines are calculated $\frac{\partial J_0}{\partial V_{JL}} - V_{JL}$ and $\frac{\partial J_{S0}}{\partial \bar{\mu}^{Si}} - V_{JL}$ curves without considering SBB, respectively. Both $\frac{\partial J_0}{\partial V_{JL}}$ and $\frac{\partial J_{S0}}{\partial \bar{\mu}^{Si}}$ represent their absolute values.

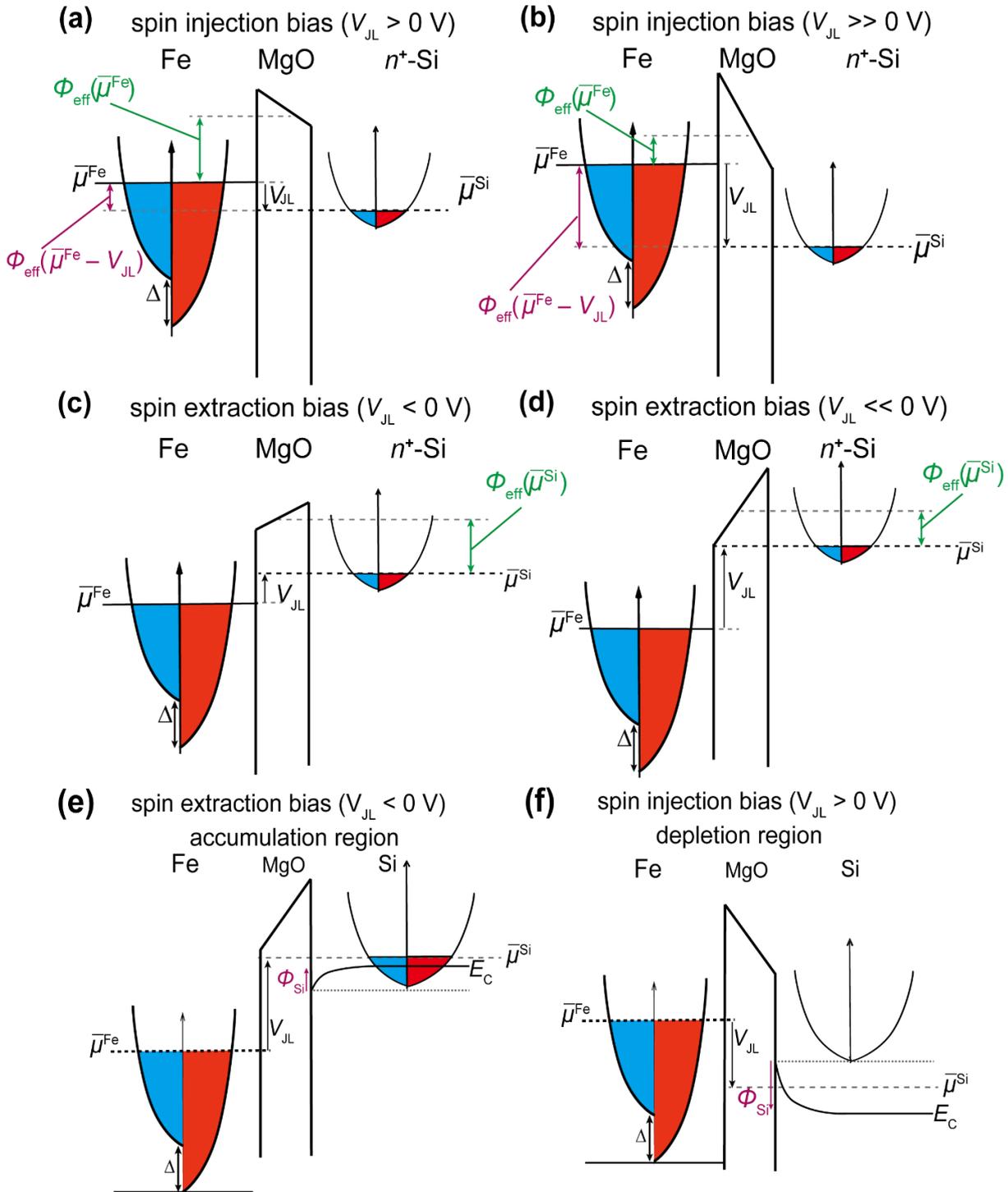

Fig. S7 (a)(b) One-dimensional energy band diagrams of a Fe/MgO/$n^+$-Si junction in the spin injection geometry with low ($V_{JL} > 0$) and high ($V_{JL} \gg 0$) junction voltage drops with $\Delta\mu_L = 0$. (c)(d) One-dimensional energy band diagrams of a Fe/MgO/$n^+$-Si junction in the spin extraction geometry with low ($V_{JL} < 0$) and high ($V_{JL} \ll 0$) junction voltage drops with $\Delta\mu_L = 0$. (e)(f) One-dimensional energy band diagrams of a Fe/MgO/$n^+$-Si junction (e) in the spin extraction geometry ($V_{JL} < 0$) when an electron accumulation region is formed in Si nearby the MgO/Si interface, and (f) in the spin injection geometry ($V_{JL} > 0$) when a depletion region is formed in Si nearby the MgO/Si interface.

## S7. Numerical calculations of the MR ratio in the two-terminal (2T) geometry

In this section, we first explain the details of the MR ratio calculations for the two-terminal (2T) Si-based device shown in Fig. 15(a). Then, to optimize the design parameters of the device, we calculated 2-D contour maps of the MR ratio plotted as a function of the device voltage drop and tunnel barrier thickness. Finally, we propose an device structure where spin filter tunnel barriers are combined with ferromagnetic metal electrodes, as shown in the schematic band diagram in Fig. 17(c). In the proposed device, a MR ratio up to 15% is theoretically predicted.

For simplicity, we assume that spin accumulation $\Delta\mu$ is uniform through the thin (25 nm) Si layer, namely, the decay of $\Delta\mu$ caused by spin transport is neglected. In our calculations, the parallel (P) and anti-parallel (AP) magnetization states are realized by setting the same and opposite sign of the spin-splitting term $\Delta$ in Eq. (10) for the two electrodes, respectively. For a given electron current $J$, first the JCT1 voltage drop $V_1^{P(AP)}$ and JCT2 voltage drop $V_2^{P(AP)}$ are self-consistently calculated with $\Delta\mu = 0$ while maintaining the current continuous condition

$$J_1^{P(AP)}(V_1^{P(AP)}, \Delta\mu^{P(AP)}) = J_2^{P(AP)}(V_2^{P(AP)}, \Delta\mu^{P(AP)}) = J, \tag{S18}$$

for the P and AP magnetization states, respectively. Then, a renewed $\Delta\mu$ is calculated by considering the injected spin currents as follows,

$$\frac{\Delta\mu^{P(AP)}}{2} = [J_{S1}^{P(AP)}(V_1^{P(AP)}, \Delta\mu^{P(AP)}) + J_{S2}^{P(AP)}(-V_2^{P(AP)}, \Delta\mu^{P(AP)})]r_{sr}, \tag{S19}$$

where $J_{S1}$ and $J_{S2}$ are the spin currents injected by the JCT1 and JCT2 junctions, respectively, and the Si spin resistance $r_{sr} = (r_{Si}\lambda_{sf})\lambda_{sf}/t_{Si} = 3.4\times10^{-10}$ $\Omega\cdot m^2$ is used. The minus sign of $V_2^{P(AP)}$ comes from the fact that the definition of $V$ and $J$ is opposite at JCT2. Next, $V_1^{P(AP)}$ and $V_2^{P(AP)}$ are renewed using the $\Delta\mu^{P(AP)}$ values obtained in the last step. The sequence is iteratively performed until $\Delta\mu^{P(AP)}$ converges to a sufficiently non-zero stable value. Finally, $\Delta\mu^{P(AP)}$, $V_1^{P(AP)}$, and $V_2^{P(AP)}$ are obtained. Here, the MR ratio and total device voltage drop ($V_{total}^P$) in the P state are calculated by the following equations,

$$MR\ ratio = \frac{V_{total}^{AP} - V_{total}^{P}}{V_{total}^{P}}, \tag{S20}$$

$$V_{total}^{P(AP)} = V_1^{P(AP)} + V_2^{P(AP)} + Jr_{ch}, \tag{S21}$$

where $r_{ch} = \rho t_{Si}$ is the Si channel resistance. Equation (S20) can be expressed by the spin polarizations of JCT1 and JCT2 as follows:

$$\begin{aligned}
MR\ ratio &= \frac{V_{total}^{AP} - V_{total}^{P}}{V_{total}^{P}} \\
&= r_{sc}\frac{\left(P_{det}^{AP}(V_1^{AP}) + P_{det}^{AP}(V_2^{AP})\right)\left(P_{inj}^{AP}(V_1^{AP}) + P_{inj}^{AP}(V_2^{AP})\right)}{R_{total}} \\
&\quad -r_{sc}\frac{\left(P_{det}^{P}(V_1^{P}) + P_{det}^{P}(V_2^{P})\right)\left(P_{inj}^{P}(V_1^{P}) - P_{inj}^{P}(V_2^{P})\right)}{R_{total}},
\end{aligned} \tag{S22}$$

where

$$\begin{aligned}
\Delta V_{total}^P &= \left(P_{det}^P(V_1^P) + P_{det}^P(V_2^P)\right)\Delta\mu^P \\
&= r_{sc}J\left(P_{det}^P(V_1^P) + P_{det}^P(V_2^P)\right)\left(P_{inj}^P(V_1^P) - P_{inj}^P(V_2^P)\right),
\end{aligned} \tag{S23}$$

$$\begin{aligned}
\Delta V_{total}^{AP} &= \left(P_{det}^{AP}(V_1^{AP}) + P_{det}^{AP}(V_2^{AP})\right)\Delta\mu^{AP} \\
&= r_{sc}J\left(P_{det}^{AP}(V_1^{AP}) + P_{det}^{AP}(V_2^{AP})\right)\left(P_{inj}^{AP}(V_1^{AP}) + P_{inj}^{AP}(V_2^{AP})\right),
\end{aligned} \tag{S24}$$

and

$$\Delta\mu^P = r_{sc}(J_{s1} - J_{s2}) = r_{sc}J\left(P_{inj}^P(V_1^P) - P_{inj}^P(V_2^P)\right), \tag{S25}$$

$$\Delta\mu^{AP} = r_{sc}(J_{s1} - J_{s2}) = r_{sc}J\left(P_{inj}^{AP}(V_1^{AP}) + P_{inj}^{AP}(V_2^{AP})\right). \tag{S26}$$

The eight spin polarizations are defined as follows:

$$P_{inj}^{P(AP)} = \frac{J_s^{P(AP)}(V_i^{P(AP)}, \Delta\mu^{P(AP)})}{J^{P(AP)}(V_i^{P(AP)}, \Delta\mu^{P(AP)})}, \tag{S27}$$

$$P_{\det(i)}^{P(AP)} = \frac{2\Delta V_i^{P(AP)}}{\Delta \mu^{P(AP)}}, \quad (S28)$$
$$i = 1 \text{ or } 2$$

where $P_{\text{inj}}^{P(AP)}$ and $P_{\det}^{P(AP)}$ are the spin injection and detection polarizations in the P(AP) magnetization state at JCT-$i$ ($i$ =1 or 2), they can be estimated by $J^{P(AP)}$, $J_S^{P(AP)}$, $V_1^{P(AP)}$, $V_2^{P(AP)}$, and $\Delta\mu^{P(AP)}$. Figures S8(a-d) show the numerically-calculated eight spin polarizations plotted as functions of $V_1$ or $V_2$ in the P and AP magnetization states. Their features are basically similar to $P_{\text{inj}}(V_{\text{JL}})$ and $P_{\det}(V_{\text{JL}})$ shown in Fig. 7(a) and Fig. 11(a). Figures. S8(a) and (c) correspond to the positive and negative bias ranges of Fig. 9(a), respectively, whereas Figs. S8(b) and (d) correspond to the positive and negative bias ranges of Fig. 13(a), respectively. It is noted that the bias-dependent features of $P_{\text{inj}}(V_2)$ and $P_{\det}(V_2)$ in Figs. S8(c) and (d) are not similar to that of the single junction in Fig. 13(a); Especially the dependence on the magnetization state (P or AP) is observed in a high bias range. This comes from more significant decrease (increase) of $\Delta\mu$ at the Si interface in the P (AP) magnetization state in the 2T device structure than that in a single junction structure.

In Sec. V, it has been demonstrated that the introduction of spin-filter I layers can effectively enhance the MR ratio in a wide range of $V_{\text{total}}^P$. To further clarify the effects of a spin-filter I layer on the MR ratio, here we show the numerical calculation results of several important terms in Eq. (19). Figure S9(a) shows the calculated $P_{det}^{AP}(V_2^{AP})\left(P_{inj}^{AP}(V_1^{AP}) + P_{inj}^{AP}(V_2^{AP})\right)$ plotted as functions of $V_{\text{total}}^P$, where brown, pink, and green dots are the results with considering both ferromagnetic electrodes and spin-filter I layers, only spin-filter I layers, and only ferromagnetic electrodes, respectively. The green dots are the same data as the green dots in Fig. 18(b). The pink dots are higher than the green dots in $V_{\text{total}}^P < 3$ V, which reveals that the spin-filter I layers can mitigate SAS in the high bias range and can also enhance the spin polarizations in the low bias range like the effect of $\gamma$ in the V-F model. It is remarkable that the brown dots are much higher than the sum of the pink and green dots, which validates the advantage of combining a spin-filter I layer with a ferromagnetic electrode. For the case with both a spin-filter I layer and a ferromagnetic electrode in each junction, Fig. S9(b) shows the calculated spin polarizations and spin accumulation in the AP magnetization state plotted as functions of $V_{\text{total}}^P$, where brown dots are the same as that in Fig. S9(a), red dots are $P_{det}^{AP}(V_2^{AP})$, blue dots in the upper side are $P_{inj}^{AP}(V_1^{AP})$, blue dots in the lower side are $P_{inj}^{AP}(V_2^{AP})$, and black dots are the spin accumulation $\Delta\mu$. An important feature is that a saturation trend of $\Delta\mu$ does not occur in the high bias range of $V_{\text{total}}^P$ and the magnitude of $\Delta\mu$ is largely enhanced compared with the results in Fig. 9(b). The red dots have a similar decreasing trend as the brown dots, which reflects that $P_{det}^{AP}(V_2^{AP})$ is the dominant term in $P_{det}^{AP}(V_2^{AP})\left(P_{inj}^{AP}(V_1^{AP}) + P_{inj}^{AP}(V_2^{AP})\right)$. As $V_{\text{total}}^P$ is increased, unlike $P_{\text{inj}}$ in the positive $V_{\text{JL}}$ range in Fig. 9(a), a slight increase of $P_{inj}^{AP}(V_1^{AP})$ is observed due to the difference of the barrier heights for up- and down-spin electrons induced by the spin-filter I layer. On the other hand, like $P_{\text{inj}}$ in the negative $V_{\text{JL}}$ range in Fig. 9(a), $P_{inj}^{AP}(V_2^{AP})$ decreases as $V_{\text{total}}^P$ is increased. Since SAS is mitigated by the spin-filter I layer, negative values of $P_{inj}^{AP}(V_2^{AP})$ are observed in $V_{\text{total}}^P > 2$V due to the large values of $\Delta\mu$ ($> 0.2$ V) in the Si channel.

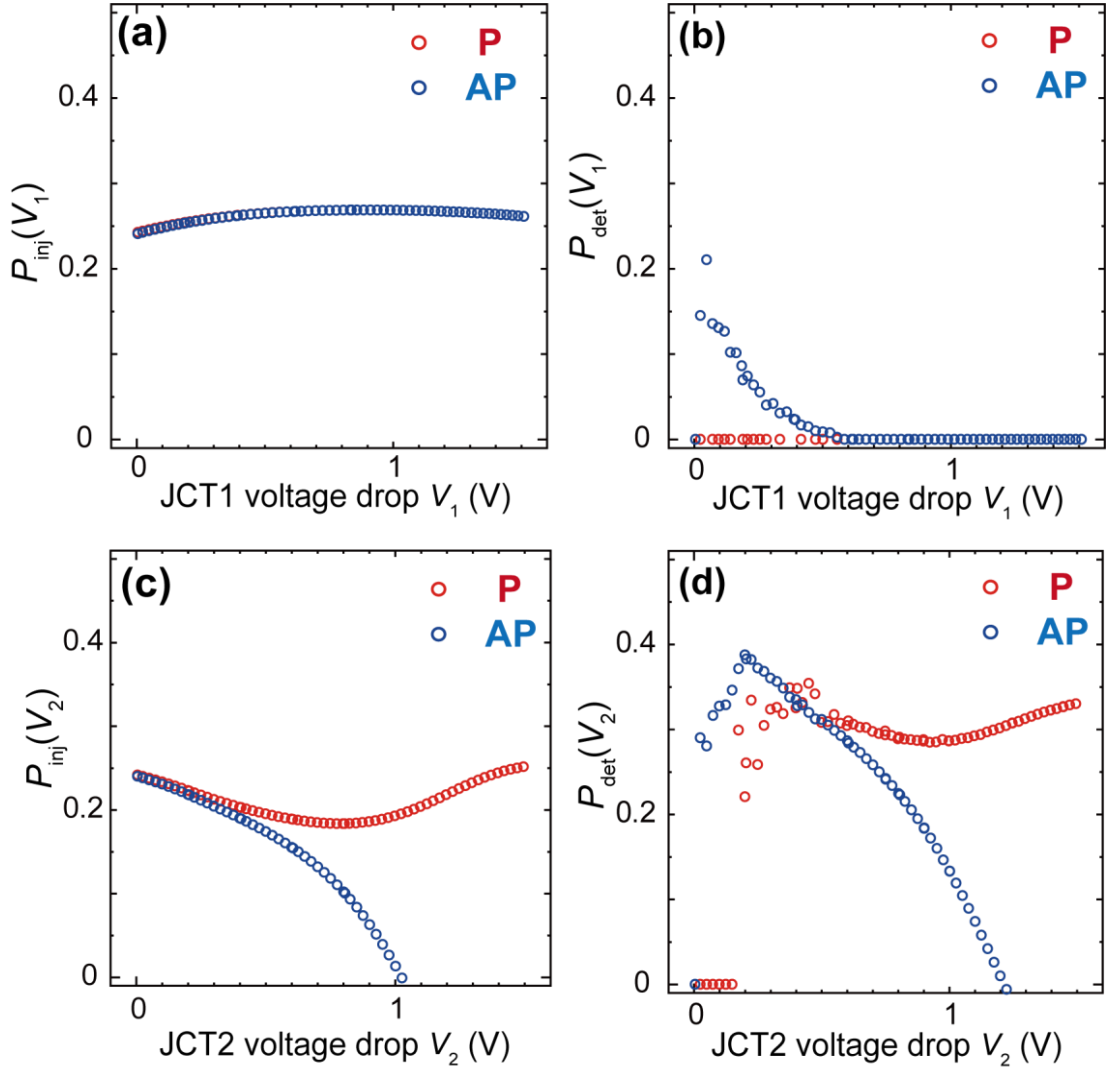

Fig. S8 (a) Spin injection polarization at JCT1 $P_{\mathrm{inj}}$ plotted as functions of JCT1 junction voltage drop $V_1$ in P (red open circle) and AP (blue open circle) states, respectively. (b) Spin detection polarization at JCT1 $P_{\mathrm{det}}$ plotted as functions of JCT1 junction voltage drop $V_1$ in P (red open circle) and AP (blue open circle) states, respectively. (c) Spin injection polarization at JCT2 $P_{\mathrm{inj}}$ plotted as functions of $V_2$ in P (red open circle) and AP (blue open circle) states, respectively. (d) Spin detection polarization at JCT2 $P_{\mathrm{det}}$ plotted as functions $V_2$ in P (red open circle) and AP (blue open circle) states, respectively.

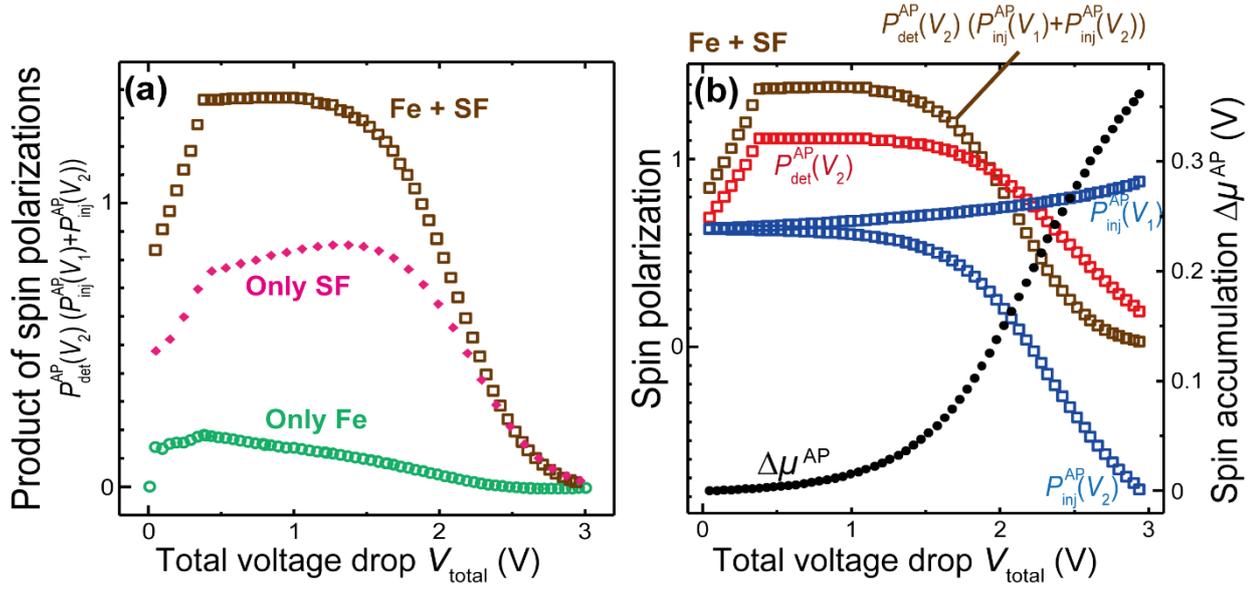

Fig. S9(a) Numerically calculated $P_{det}^{AP}(V_2^{AP})\left(P_{inj}^{AP}(V_1^{AP}) + P_{inj}^{AP}(V_2^{AP})\right)$ plotted as functions of $V_{total}^{P}$, where brown, pink and green dots are the results with considering both ferromagnetic electrodes and spin-filter I layers (Fe + SF), only spin-filter I layers (Only SF), and only ferromagnetic electrodes (Only Fe), respectively. (b) Numerically calculated spin polarizations and spin accumulation in the AP state plotted as functions of $V_{total}^{P}$ with considering both ferromagnetic electrodes and spin-filter I layers, where brown dots are the same as that in (a), red dots are $P_{det}^{AP}(V_2^{AP})$, blue dots in the upper side are $P_{inj}^{AP}(V_1^{AP})$, blue dots in the lower side are $P_{inj}^{AP}(V_2^{AP})$, and black dots are the spin accumulation $\Delta\mu$.

## S8. MR ratio *vs.* $V_{total}^P$ plot with various insulator thicknesses ($t_I$)

In Sec. V, we present a two-dimensional (2-D) contour map of the MR ratio calculated for the 2T device of Fig. 15(a) plotted against $V_{total}^P$ and $t_I$ in Fig. 16(a) when normal insulator tunnel barriers and ferromagnetic electrodes are considered. To observe the detailed MR ratio *vs.* $V_{total}^P$ relationship, we show the MR ratio *vs.* $V_{total}^P$ with various $t_I$ values in Fig. S10, where the inset is a magnified view of the MR ratio curves for $t_I$ = 0.8, 0.9 and 1.0 nm, and the brown open dots are the same data as that in Fig. 15(b) with $t_I$ = 1.0 nm. When $t_I > 0.6$ nm, as $V_{total}^P$ is increased, the MR ratio increases first and then decreases, which is clear for $t_I$ = 0.6, 0.7, and 0.8 nm. The increase of the MR ratio originates from the decrease of $R_{total}^P$, whereas the decrease of the MR ratio originates from the reduction of $P_{det}^{AP}(V_2^{AP})$ through SAS, as explained in Sec. V. When the barrier becomes thicker, the reduction of $P_{det}^{AP}(V_2^{AP})$ through SAS occurs in the higher $V_{total}^P$ range, because a higher $V_{total}^P$ is required to drive a large enough current for SAS in a device with thicker barriers. Hence, the decrease of the MR ratio for $t_I$ = 0.9 and 1.0 nm is not observed in Fig. S10, since $V_{total}^P$ (< 1.5 V) is not high enough to generate significant reduction of $P_{det}^{AP}(V_2^{AP})$ through SAS. The brown open dots in Fig. 15(b) shows the case of $t_I$ = 1.0 nm with $0 < V_{total}^P < 3$ V, where the decrease of the MR ratio is observed when $V_{total}^P > 2\,V$. On the other hand, when $t_I < 0.6$ nm, MR ratio monotonically decreases as $V_{total}^P$ is increased. This is because the reduction of $P_{det}^{AP}(V_2^{AP})$ through SAS occurs in the lower $V_{total}^P$ range; when the barriers become thinner, smaller $V_{total}^P$ can drive a large enough current for SAS in the device.

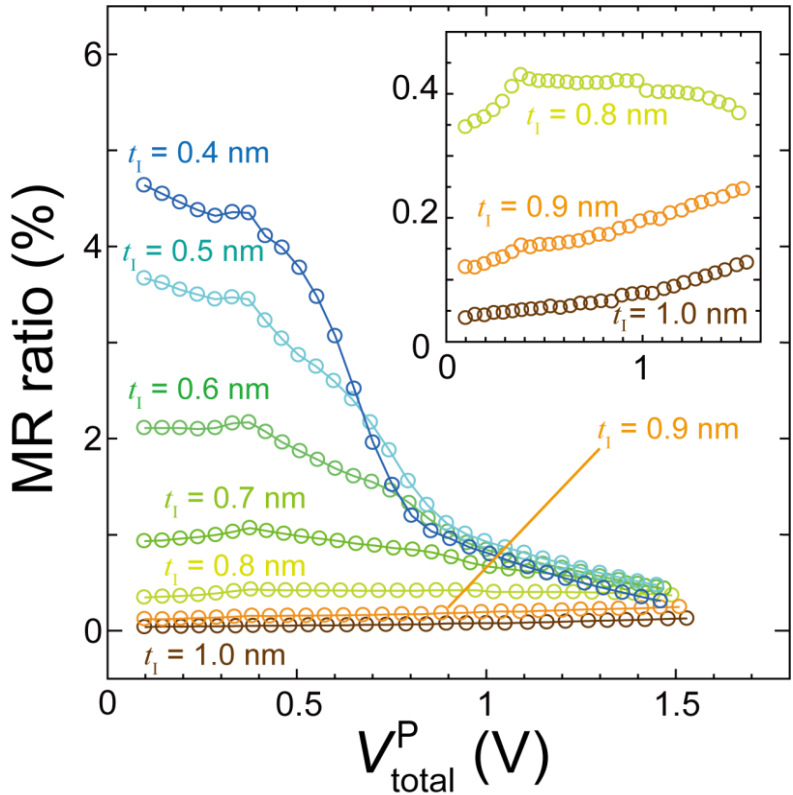

Fig. S10 Numerically calculated MR ratio plotted as functions of $V_{total}^P$ with various $t_I$ when ferromagnetic electrode properties are considered (spin-filter barrier is neglected). The data is extracted from Fig. 16(a). The inset is a magnified view of the MR ratio curves for $t_I$ = 0.8, 0.9 and 1.0 nm.